\documentclass[twocolumn]{aastex63}

\usepackage{times}
\usepackage{graphics,epsfig}
\usepackage{graphicx}
\usepackage{amsmath}
\usepackage{amssymb}
\usepackage{enumitem}
\usepackage{color}
\usepackage{wrapfig}
\DeclareGraphicsExtensions{.pdf,.png,.jpg}
\usepackage{color}
\usepackage{lineno}
\definecolor{darkcerulean}{rgb}{0.03, 0.27, 0.49}

\definecolor{bondiblue}{rgb}{0.0, 0.58, 0.71}

\definecolor{cerulean}{rgb}{0.0, 0.48, 0.65}

\definecolor{newsam}{rgb}{0.05, 0.10, 0.45}

\newcommand{\kms}{\rm km~s$^{-1}$}

\newcommand{\arcs}{$^{\prime\prime}$}
\newcommand{\arcm}{$^{\prime}$}

\newcommand{\mst}{\ensuremath{\rm M_{*}}}
\newcommand{\mhi}{\ensuremath{\rm M_{HI}}}

\newcommand{\hi}{{\rm H}{\sc i}}

\received{---}
\revised{---}
\accepted{---}
\submitjournal{ApJ}

\shorttitle{Gas fraction of groups and pairs}
\shortauthors{Roychowdhury et al.}

\begin{document}

\title{The variation of the gas content of galaxy groups and pairs compared to isolated galaxies}

\correspondingauthor{Sambit Roychowdhury}
\email{sambit.roychowdhury@icrar.org}

\author[0000-0002-5820-4589]{Sambit Roychowdhury}
\affiliation{International Centre for Radio Astronomy Research (ICRAR), University of Western Australia, 35 Stirling Highway, Crawley, WA 6009, Australia}
\affiliation{ARC Centre of Excellence for All Sky Astrophysics in 3 Dimensions (ASTRO 3D), Australia}

\author{Martin J. Meyer}
\affiliation{International Centre for Radio Astronomy Research (ICRAR), University of Western Australia, 35 Stirling Highway, Crawley, WA 6009, Australia}
\affiliation{ARC Centre of Excellence for All Sky Astrophysics in 3 Dimensions (ASTRO 3D), Australia}

\author{Jonghwan Rhee}
\affiliation{International Centre for Radio Astronomy Research (ICRAR), University of Western Australia, 35 Stirling Highway, Crawley, WA 6009, Australia}
\affiliation{ARC Centre of Excellence for All Sky Astrophysics in 3 Dimensions (ASTRO 3D), Australia}

\author{Martin A. Zwaan}
\affiliation{European Southern Observatory, Karl Schwarzschild Stra\ss e 2, D-85748 Garching, Germany}

\author{Garima Chauhan}
\affiliation{International Centre for Radio Astronomy Research (ICRAR), University of Western Australia, 35 Stirling Highway, Crawley, WA 6009, Australia}
\affiliation{ARC Centre of Excellence for All Sky Astrophysics in 3 Dimensions (ASTRO 3D), Australia}

\author{Luke J. M. Davies}
\affiliation{International Centre for Radio Astronomy Research (ICRAR), University of Western Australia, 35 Stirling Highway, Crawley, WA 6009, Australia}

\author{Sabine Bellstedt}
\affiliation{International Centre for Radio Astronomy Research (ICRAR), University of Western Australia, 35 Stirling Highway, Crawley, WA 6009, Australia}

\author{Simon P. Driver}
\affiliation{International Centre for Radio Astronomy Research (ICRAR), University of Western Australia, 35 Stirling Highway, Crawley, WA 6009, Australia}
\affiliation{SUPA, School of Physics \& Astronomy, University of St Andrews, North Haugh, St Andrews KY16 9SS, UK}

\author{Claudia del P. Lagos}
\affiliation{International Centre for Radio Astronomy Research (ICRAR), University of Western Australia, 35 Stirling Highway, Crawley, WA 6009, Australia}
\affiliation{ARC Centre of Excellence for All Sky Astrophysics in 3 Dimensions (ASTRO 3D), Australia}

\author{Aaron S. G. Robotham}
\affiliation{International Centre for Radio Astronomy Research (ICRAR), University of Western Australia, 35 Stirling Highway, Crawley, WA 6009, Australia}
\affiliation{ARC Centre of Excellence for All Sky Astrophysics in 3 Dimensions (ASTRO 3D), Australia}

\author{Joss Bland-Hawthorn}
\affiliation{Sydney Institute for Astronomy, School of Physics, University of Sydney, NSW 2006}
\affiliation{ARC Centre of Excellence for All Sky Astrophysics in 3 Dimensions (ASTRO 3D), Australia}

\author{Richard Dodson}
\affiliation{International Centre for Radio Astronomy Research (ICRAR), University of Western Australia, 35 Stirling Highway, Crawley, WA 6009, Australia}

\author{Benne W. Holwerda}
\affiliation{Physics and Astronomy Department, University of Louisville, Louisville KY 40292, USA}

\author{Andrew M. Hopkins}
\affiliation{Australian Astronomical Optics, Macquarie University, 105 Delhi Rd, North Ryde, NSW 2113, Australia}

\author{Maritza A. Lara-L{\'o}pez}
\affiliation{Armagh Observatory and Planetarium, College Hill, Armagh, BT61 DG, UK}

\author{{\'A}ngel R. L{\'o}pez-S{\'a}nchez}
\affiliation{Australian Astronomical Optics, Macquarie University, 105 Delhi Rd, North Ryde, NSW 2113, Australia}
\affiliation{Department of Physics and Astronomy, Macquarie University, NSW 2109, Australia}
\affiliation{ARC Centre of Excellence for All Sky Astrophysics in 3 Dimensions (ASTRO 3D), Australia}

\author{Danail Obreschkow}
\affiliation{International Centre for Radio Astronomy Research (ICRAR), University of Western Australia, 35 Stirling Highway, Crawley, WA 6009, Australia}

\author{Kristof Rozgonyi}
\affiliation{Faculty of Physics, Ludwig-Maximilians-Universit\"{a}t, Scheinerstr. 1, 81679 Munich, Germany}
\affiliation{International Centre for Radio Astronomy Research (ICRAR), University of Western Australia, 35 Stirling Highway, Crawley, WA 6009, Australia}
\affiliation{ARC Centre of Excellence for All Sky Astrophysics in 3 Dimensions (ASTRO 3D), Australia}

\author{Matthew T. Whiting}
\affiliation{ATNF, CSIRO Astronomy and Space Science, PO Box 76, Epping, NSW 1710, Australia}

\author{Angus H. Wright}
\affiliation{Ruhr University Bochum, Faculty of Physics and Astronomy, AIRUB, German Centre for Cosmological Lensing, 44780 Bochum, Germany}

\begin{abstract}
We measure how the atomic gas (\hi) fraction ($f_{HI}={\rm \frac{M_{HI}}{M_{*}}}$) of groups and pairs taken as single units vary with average stellar mass ($\langle {\rm M_*} \rangle$) and average star-formation rate ($\langle {\rm SFR} \rangle$), compared to isolated galaxies.
The \hi\ 21 cm emission observation are from (i) archival ALFALFA survey data covering three fields from the GAMA survey (provides environmental and galaxy properties), and (ii) DINGO pilot survey data of one of those fields. 
The mean $f_{HI}$ for different units (groups/pairs/isolated galaxies) are measured in regions of the log($\langle {\rm M_*} \rangle$) -- log($\langle {\rm SFR} \rangle$) plane, relative to the z $\sim 0$ star-forming main sequence (SFMS) of individual galaxies, by stacking $f_{HI}$ spectra of individual units.
For ALFALFA, $f_{HI}$ spectra of units are measured by extracting \hi\ spectra over the full groups/pair areas and dividing by the total stellar mass of member galaxies. For DINGO, $f_{HI}$ spectra of units are measured by co-adding \hi\ spectra of individual member galaxies, followed by division by their total stellar mass. 
For all units the mean $f_{HI}$ decreases as we move to higher $\langle {\rm M_*} \rangle$ along the SFMS, and as we move from above the SFMS to below it at any $\langle {\rm M_*} \rangle$.
From the DINGO-based study, mean $f_{HI}$ in groups appears to be lower compared to isolated galaxies for all $\langle {\rm M_*} \rangle$ along the SFMS.
From the ALFALFA-based study we find substantially higher mean $f_{HI}$ in groups compared to isolated galaxies (values for pairs being intermediate) for ${\langle{\rm M_*}\rangle}\lesssim10^{9.5}~{\rm M_{\odot}}$, indicating the presence of substantial amounts of \hi\ not associated with cataloged member galaxies in low mass groups.
\end{abstract}


\section{Introduction} 
\label{sec:intro}

Understanding the impact high density environments of groups, and their scaled-up versions: clusters, have on galaxies is crucial to understanding galaxy evolution.
There has been a long history of empirical studies aiming to parametrise the environmental effect on galaxy evolution.
Early work from \citet{1980ApJ...236..351D} showed a correlation between morphology and density, in that regions with higher density of galaxies have higher fraction of (morphologically classified) elliptical galaxies.
Following this, \citet{1984ApJ...285..426B} noticed that at the present epoch there are as many star-forming spiral galaxies in clusters as in the field, but that the spirals in clusters are much redder than in the field.
At the same time, they found their data to be consistent with a scenario where there has been an overall decline in star formation since z$\sim0.5$, but independent of galactic environment.
A potential avenue through which {\it morphological} evolution of galaxies can occur within cluster environments compared to isolated systems is `galaxy harassment' -- i.e. multiple high-speed encounters between galaxies through which gas is tidally stripped within group and cluster environments \citep{1996Natur.379..613M}.
The dense intra-group or intra-cluster medium can also remove the gas reservoirs of galaxies through `ram-pressure stripping', while `strangulation' prevents replenishment of cold gas by heating and stripping of hot gas from their halos \citep[e.g.][]{1972ApJ...176....1G,1980ApJ...237..692L,2000ApJ...540..113B,2006ApJ...647..910H}.

Meanwhile \citet{2004MNRAS.353..713K} showed that even though the star formation rate per unit stellar mass decreases significantly in high density environments for galaxies with stellar masses $>10^{10}~M_{\odot}$, this decrease in star formation activity occurs over long ($>$1 Gyr) timescales and therefore cannot be driven by rapid processes that alter galactic structure or by galaxy-galaxy mergers.
There is mounting evidence also that the evolution of the galaxy population towards quenched red galaxies, as well as the environmental relationships mentioned above, are both driven by quenching of satellite galaxies and not their centrals, which happens rapidly, and for a significant fraction before they were in a halo and became a satellite \citep{2008MNRAS.387...79V,2012MNRAS.424..232W,2019MNRAS.483.5444D}.
The star-formation rate (SFR) of galaxies residing in close pairs are also affected by their environment, with the SFR of the higher mass galaxy enhanced due to tidal turbulence and shocks in gas, while the SFR of the lower mass galaxy suppressed due to gas heating and stripping \citep{2015MNRAS.452..616D,2021MNRAS.501.2969G}. 

Most of the potential effects of dense environments on galaxies discussed above affect the gas in galaxies, which being the fuel for star formation, consequently affects the star formation.
Many observational studies therefore focus on the effect of dense environments on the gas in galaxies, specifically atomic hydrogen (\hi).
The deficiency of \hi\ in galaxies within dense environments like groups or clusters as compared to the field has long been noted \citep{1984ARA&A..22..445H}.
The degree of \hi\ depletion within clusters is related to morphology with early type galaxies affected more \citep{2001ApJ...548...97S}, though the proportion of gas-poor spirals increases continuously towards the centre of clusters \citep{2001ApJ...548...97S} and groups \citep{2009MNRAS.400.1962K}.
Morphology and distribution of \hi\ in galaxies nearer to the centres of clusters show special features like curtailed \hi\ disks compared to optical disks, gas displaced from disks, one-sided \hi\ tails pointing away from the centre \citep{2009AJ....138.1741C}.
Galaxies within groups have a significantly flatter \hi\ mass function compared to the field \citep{2001A&A...377..812V,2009MNRAS.400.1962K}.
The scaling relations of \hi\ with \mst, stellar mass surface density, and colour are significantly offset towards lower gas content in (dynamically young) clusters, but the difference between field and cluster galaxies gradually decreases as we move towards massive bulge-dominated systems \citep{2011MNRAS.415.1797C}.
More specifically, the gas content of satellite galaxies is significantly dependant on the environment they are in, and the systematic environmental suppression of gas content in satellite galaxies starts in the group regime well before they reach the cluster environment \citep{2017MNRAS.466.1275B}.
Therefore a picture of gas-depleted galaxies, specifically satellite galaxies, within high density regimes from groups to clusters is emerging.

Whether the depletion of gas in the satellite galaxies in dense environments affects their star formation properties and quenches them, or whether the depletion merely removes gas not directly involved in ongoing star-formation from the outer disk, implying that the quenching of galaxies is not necessarily environment dependant, remains an open question and an area of active research.
Even if the only effect of environment is the removal of gas from the outer disks of galaxies, this results in a depletion of the reservoir for future star formation and should result in a decrease the star formation \citep[e.g. see][]{2021arXiv210912078S}.
Such decrease though will only occur on very large timescales ($\sim$10 Gyrs) given relation between gas and star formation as determined in the \hi-dominated regime \citep{2015MNRAS.449.3700R}.

A complementary approach might improve our understanding of the above issues.
Instead of focussing on relating gas content to stellar and star formation properties of galaxies in different environments as in previous studies, this approach involves relating the overall gas content with the overall stellar and star formation properties when considering groups and pairs as individual {\it units}, with isolated galaxies acting as the control sample.
Such an approach will not only show whether the overall gas content in dense environments is less (or more) compared to isolated environments, but also verify whether the overall star formation in dense environments is measurably affected by any observed effect on their overall gas content. 

Based on existing studies, we can at best relate the overall gas content of groups to their stellar mass in a circuitous way.
The mass of the dark matter halo within which a group or pair of galaxies resides, is proportional to the stellar mass (\mst) of the brightest cluster/group galaxy (BCG) \citep{2010ApJ...717..379B}.
A recent study \citet{2020ApJ...894...92G} has now measured the \hi\ masses in dark matter halos of different masses at z~$\sim 0$ by stacking the spectra of entire groups.
They find that although \hi\ mass does in general increase with halo mass, the increase is not monotonic, and has a strong dependence on halo richness.

In this study we take up the complementary approach described above, and correlate the gas content of groups/pairs/isolated galaxies taken as individual units with their \mst\ and SFR.
In order to do so in a consistent way considering the considerable differences in total \mst, SFR and gas between groups/pairs/isolated galaxies, we work with average quantities.
The gas content is measured through the stacking of \hi\ spectra from entire groups/pairs/isolated galaxies.
Throughout this paper we adopt $\Omega_{\Lambda}~=~0.70$, $\Omega_m~=~0.30$, $H_0~=~70~km s^{-1} Mpc^{-1}$.

\section{Data}
\label{sec:dat}

\begin{table}
\begin{center}
\caption{Extent of the ALFALFA--GAMA overlap, in terms of the three equatorial GAMA fields.}
\label{tab:alfa}
\begin{tabular}{lccc}
\hline
\hline
& G09 & G12 & G15 \\
\hline
RA (J2000) & 129$^{\circ}$ to 141$^{\circ}$ & 174$^{\circ}$ to 186$^{\circ}$ & 211.5$^{\circ}$ to 223.5$^{\circ}$ \\
Dec (J2000) & 0$^{\circ}$ to $+$3$^{\circ}$ & 0$^{\circ}$ to $+$2$^{\circ}$ & 0$^{\circ}$ to $+$3$^{\circ}$ \\
\hline
\hline
\end{tabular}
\end{center}
\end{table}

\subsection{Atomic hydrogen data}
\label{sec:dathi}

The \hi\ 21 cm emission data used in this study comes from two different sources -- one single dish, and one interferometric.
The single dish data is ideal for sampling the \hi\ emission over the entire group/pair area (details in Section~\ref{sec:exalfa} below), while the interferometric data provides a comparison set where the \hi\ emission from the individual constituent galaxies in groups/pairs only are considered.

First, we use archival \hi\ 21 cm emission data cubes from the Arecibo Legacy Fast ALFA \citep[ALFALFA,][]{2018ApJ...861...49H} survey, which utilized the Arecibo L-Band Feed Array (ALFA) seven-beam receiver on the ({\it RIP}) 305-m Arecibo dish.
We use the data from ALFALFA to study the gas content of galaxies in the three equatorial Galaxy And Mass Assembly (GAMA) survey \citep{2011MNRAS.413..971D,2013MNRAS.430.2047H,2015MNRAS.452.2087L,2018MNRAS.474.3875B} fields viz. G09, G12, and G15.
We are restricted to $\sim$half the area covered by the three GAMA fields as ALFALFA only observed the sky with DEC $>0^{\circ}$.
ALFALFA also only covers the very nearby Universe upto a redshift of z~$\le0.06$ (1335 MHz to 1435 MHz), and we only consider galaxies in the region of spatial overlap between ALFALFA and GAMA (Table~\ref{tab:alfa}).
ALFALFA \hi\ data cubes have asymmetric Gaussian beams with full-width-at-half-maxima (FWHM) of 3.8\arcm$\times$3.3\arcm, channel spacing of $\sim$24.4 kHz, and average rms of 1.86 mJy beam$^{-1}$ channel$^{-1}$ \citep{2018ApJ...861...49H}.

Second, the Deep Investigation of Neutral Gas Origins \citep[DINGO,][]{2009pra..confE..15M,2012MNRAS.426.3385D} is an ongoing deep survey of GAMA fields in \hi\ 21 cm emission using the Australian Square Kilometre Array Pathfinder \citep[ASKAP,][]{2008ExA....22..151J,2009IEEEP..97.1507D,2021PASA...38....9H}.
The Phase I pilot observations of DINGO have been completed, and we use the combined \hi\ data cube from 17.7 hrs of observations of a $\sim$6.5$^{\circ} \times$ 6.5$^{\circ}$ area centred on 14h40m41.67s $+$00$^{\circ}$28$^{\prime}$35.6$^{\prime \prime}$, primarily to improve our understanding of the results obtained using the archival ALFALFA data.
The observed field's spatial overlap with the G15 GAMA equatorial field is 216.85$^{\circ}$ to 223.5$^{\circ}$ in RA (J2000), and $-$2$^{\circ}$ to $+$3$^{\circ}$ in Dec (J2000).
The observations were processed and the \hi\ data cube created using ASKAPsoft \citep{2019ascl.soft12003G}.
The \hi\ cube used has a lower frequency cutoff at 1295.5 MHz (upto a redshift of z~$\le0.096$), a higher frequency cutoff at 1433.5 MHz, a Gaussian beam with FWHM of 30\arcs, channel spacing of 18.5 kHz, and average rms varies between $\sim$1.4 to 1.7 mJy beam$^{-1}$ channel$^{-1}$ as we move from the low frequency end to the high frequency end of the cube. 

Note that when smoothed to ALFALFA spatial and velocity resolutions, the $\sim$18 hrs of data from the DINGO pilot observations used in this work is $\sim$50 times less sensitive than ALFALFA data.
This, in addition to the fact that the single dish ALFALFA data is anyway suited for our complementary approach of sampling the \hi\ emission over entire group/pair areas, means that we base our study on the ALFALFA data, with the DINGO pilot data providing a comparison set for our results.

\subsection{GAMA catalogs}

We use a number of catalogs available from the multi-wavelength GAMA survey\footnote{\url{http://www.gama-survey.org/dr3/}} to define our sample and draw the \mst\ and SFR of galaxies from.
We take care to only include in our study detections confirmed to be galaxies based on latest galaxy catalog from GAMA.

For our sample selection, we use the latest GAMA galaxy group catalog \citep{2011MNRAS.416.2640R} to define groups and pairs of galaxies in the ALFALFA-GAMA overlap volume.
This catalog is generated using a friend-of-friends based grouping algorithm.
In order to be consistent, we define pairs of galaxies based on the same catalog and not based on e.g. a combination of projected separation and velocity separation \citep{2014MNRAS.444.3986R,2015MNRAS.451.3249A}.
Thus GAMA galaxy group catalog entries with number of members $\ge$3 are considered as `groups', while those with number of members $=2$ are `pairs'.
Note that this therefore only selects `stand-alone' pairs, i.e. pairs of galaxies that are not within groups.
The stellar mass completeness limit of GAMA varies with redshift.
The group catalog does not enforce a stellar mass cutoff, in order to include all possible detected galaxies when defining groups.
For our study the stellar mass completeness varies within a modest range, starting from $\sim 10^7~M_{\odot}$ at $z=0.01$, to $\sim 10^8~M_{\odot}$ at $z=0.06$, and $\sim 2 \times 10^8~M_{\odot}$ at $z=0.1$.
We therefore use all groups from the catalog in our chosen redshift range and sky area, not to especially bias ourselves against groups consisting mostly of low stellar mass members.
The false positive and false negative rates for group membership is $\sim 20\%$ but depends on group multiplicity \citep[see][for details]{2011MNRAS.416.2640R}.

The \mst\ and SFR of the galaxies are based on {\sc magphys} \citep{2008MNRAS.388.1595D} fits to the full spectral energy distributions (SEDs) of GAMA galaxies \citep[see][]{2016MNRAS.461..458D,2018MNRAS.475.2891D}. 
 
\section{Analysis and results}
\label{sec:anares}

\subsection{Extracting \hi\ spectra from ALFALFA cubes}
\label{sec:exalfa}

\begin{figure}
\begin{center}
\end{center}
\includegraphics[width=3.5truein]{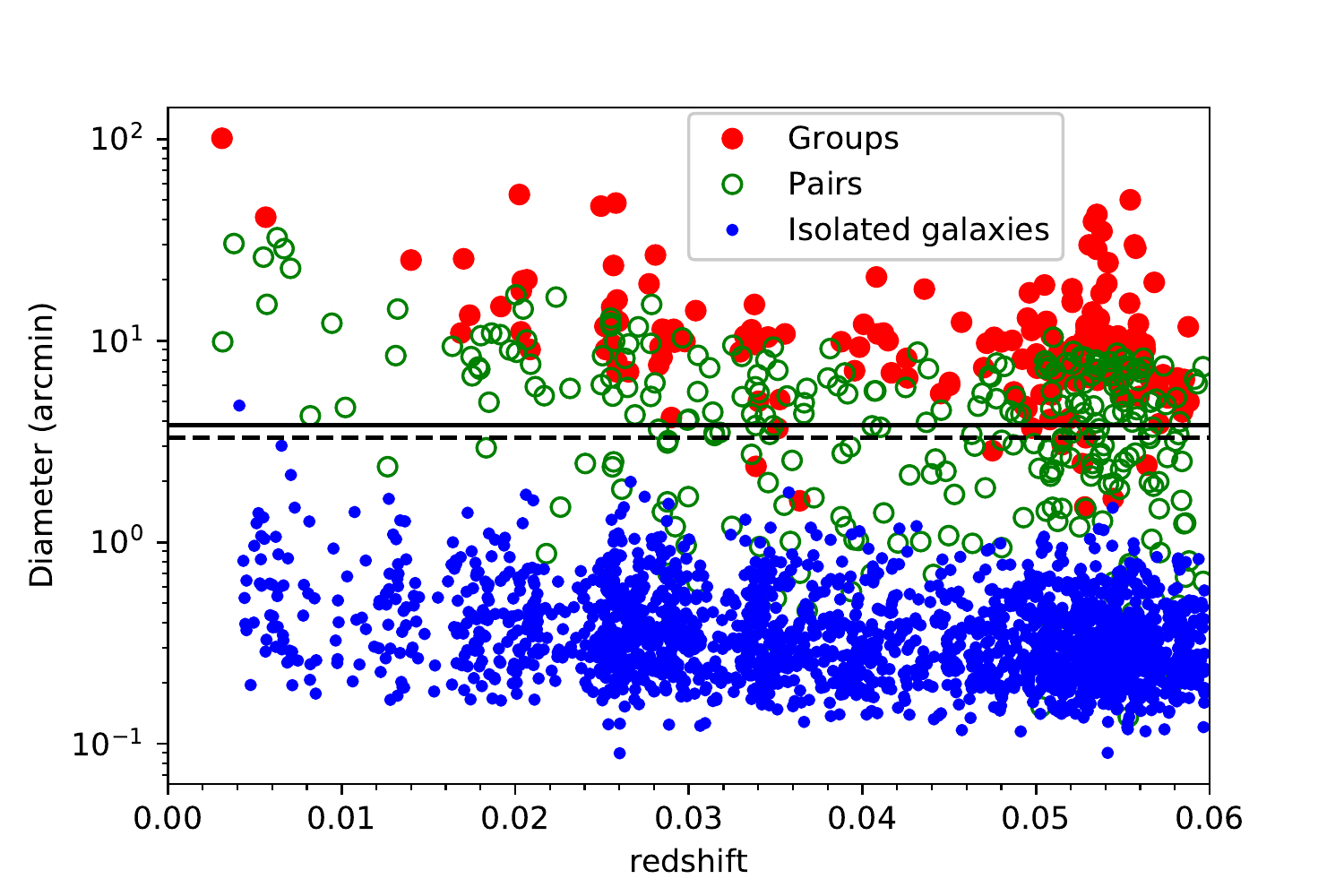}
\caption{The sizes of groups/pairs/isolated galaxies whose spectra are extracted from ALFALFA cubes. Note that the isolated galaxy diameters plotted are twice the R100 radii values from the latest GAMA galaxy catalog (see text for details). Also see text regarding group/pair sizes. The solid and dashed horizontal lines are the FWHMs of the ALFALFA major and minor axis respectively.}
\label{fig:siz}
\end{figure}

The ALFALFA--GAMA overlap sample with available \mst\ and SFR from GAMA, as decribed in Section~\ref{sec:dat}, includes 170 groups, 288 pairs, and 2066 isolated galaxies.
As mentioned in Section~\ref{sec:dat}, the beam size of the ALFALFA survey is large, such that at z~$=0.06$ it corresponds to a spatial scale of 265 kpc, and at  z~$=0.03$ the spatial scale is 137 kpc.
Figure~\ref{fig:siz} shows a comparison of the group (and pair) sizes with the ALFALFA beam size as a function of redshift.
Each group's (or pair's) radius is defined by the most distant group member based on the projected distance away from the iterative central galaxy from the GAMA galaxy group catalog.
Thus the data is ideal for sampling the \hi\ emission coming from the entire group/pair area, and we extract the spectra over the entire group/pair area in the manner described below.
Our novel approach also means we do not have to worry about spatial confusion, which can be a big issue when considering galaxies in group and pair environments.

For each group or pair in our ALFALFA--GAMA overlap sample we use the redshift of the respective iterative central galaxy (BCG) from the GAMA galaxy group catalog to ascertain the central frequency of the \hi\ spectrum for that group or pair.
For our control sample -- the isolated galaxies within the ALFALFA--GAMA overlap region, the same is done using their cataloged redshifts.
If the group/pair/galaxy is covered spatially and in frequency space by one of the ALFALFA cubes, we extract a spectrum by summing over the flux density over a certain area for each channel spanning $\pm$1500 \kms\ on either side of the central frequency.
The spectra are extracted over such a large velocity range in order to have enough `line-free' channels, defined below in Section~\ref{sec:stack1}, which are necessary to remove any residual baseline in the spectra and also to estimate the noise level of each spectrum.
The radius of the circular area over which the flux density if summed over for each groups/pair is the quadrature sum of the larger half-width-at-half-maxima of the ALFALFA beam ($\sim$1.9\arcm) and the respective group/pair radius.
For isolated galaxies, the radius of the circular area is simply the larger half-width-at-half-maxima of the ALFALFA beam ($\sim$1.9\arcm).
Assuming the maximum extent of an isolated galaxy's \hi\ emission to be twice the R100 radii values from the latest GAMA galaxy catalog \citep{2020MNRAS.496.3235B}, which is the approximate elliptical semi-major axis containing 100\% of the flux for all bands, we see from Fig.~\ref{fig:siz} that such a circular aperture ensures that we pick up all the \hi\ flux from any isolated galaxy.
For each extracted spectrum if there is a frequency channel for which 40\% or more of the pixels have weights $<$ 10 in the associated ALFALFA weight maps, we drop that channel from the spectrum. 

\subsection{The average \mst\ -- average SFR plane}
\label{sec:sfrms}

\begin{figure*}
\begin{center}
\begin{tabular}{cc}
{\mbox{\includegraphics[width=3.5truein]{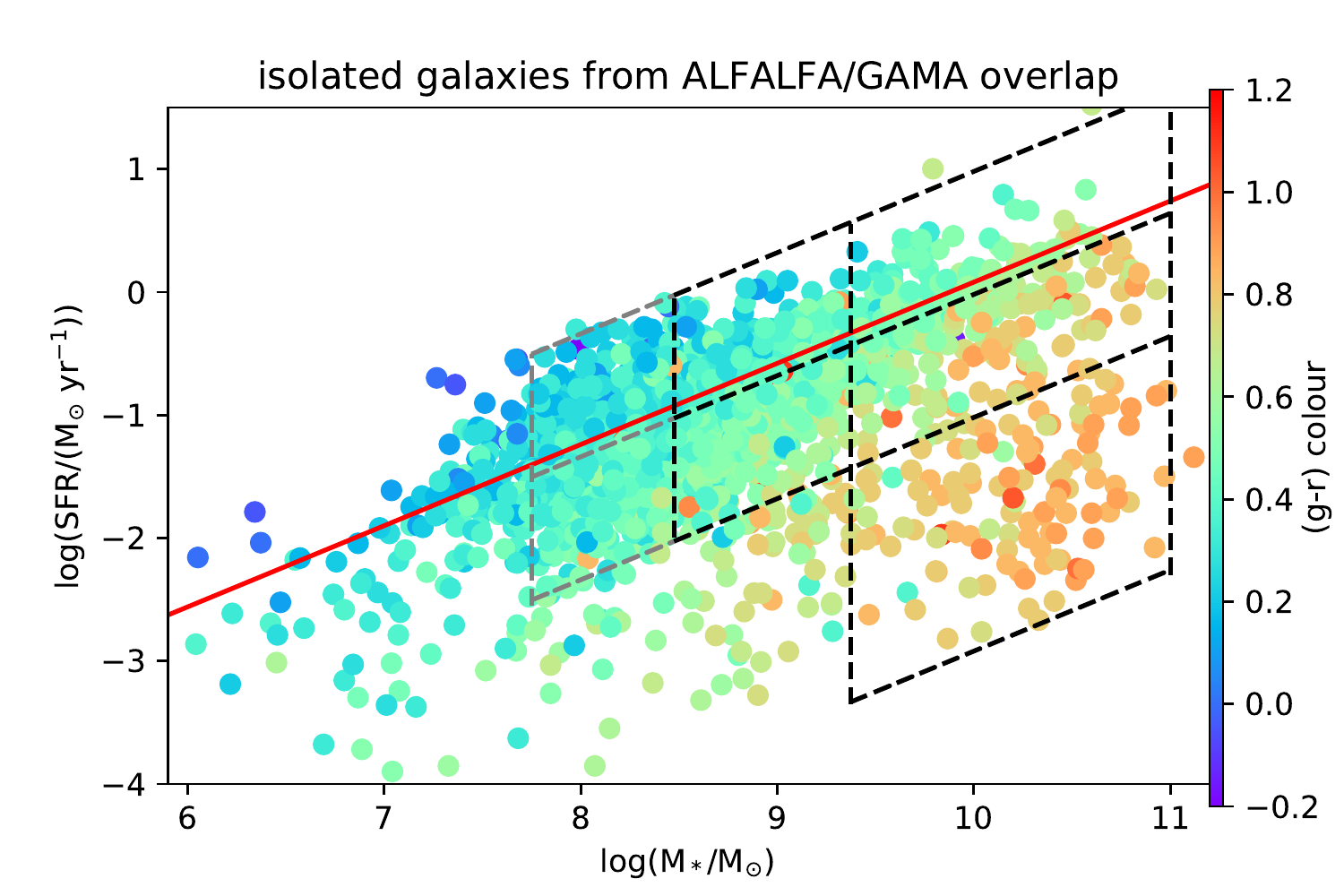}}}&
{\mbox{\includegraphics[width=3.5truein]{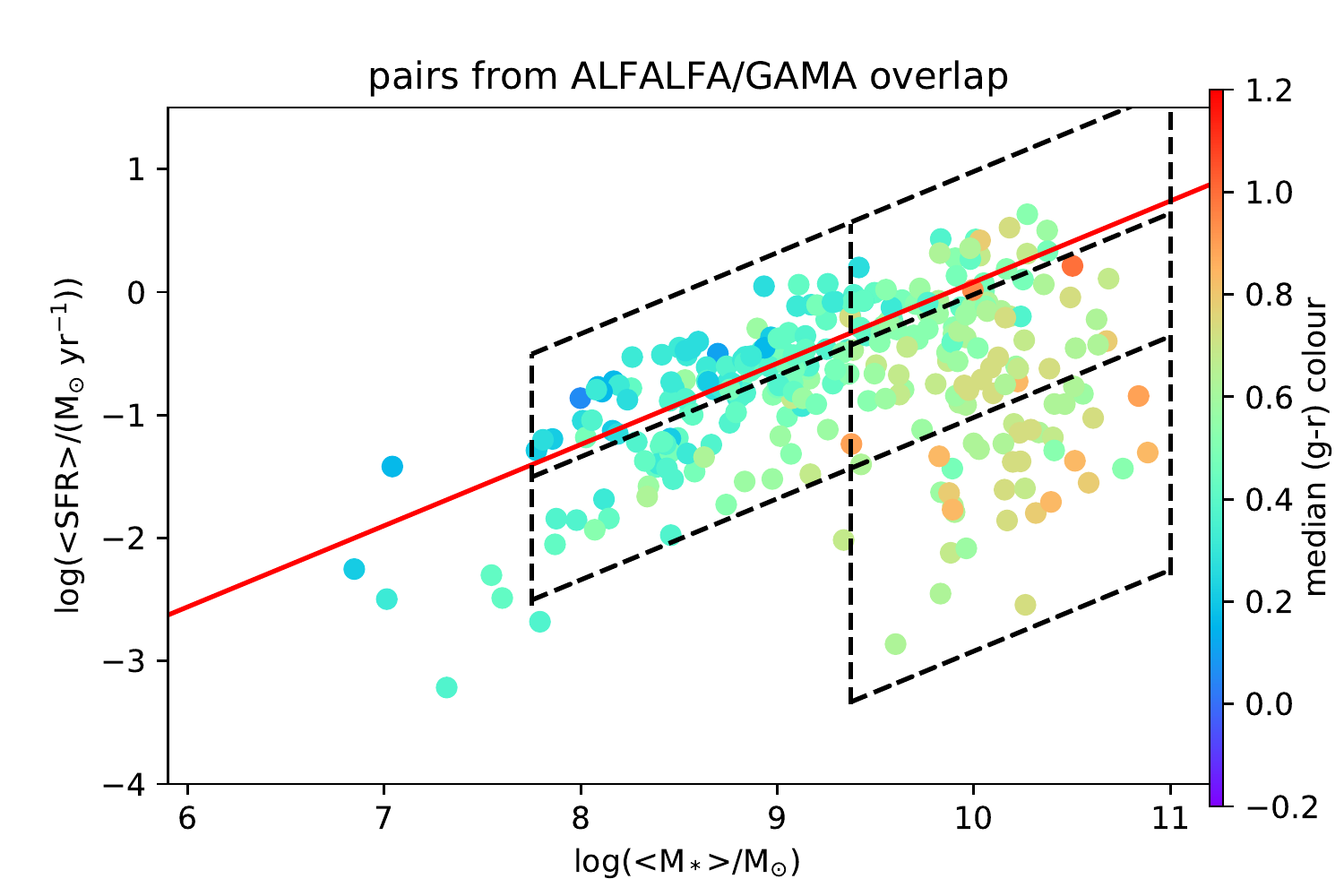}}}\\
{\mbox{\includegraphics[width=3.5truein]{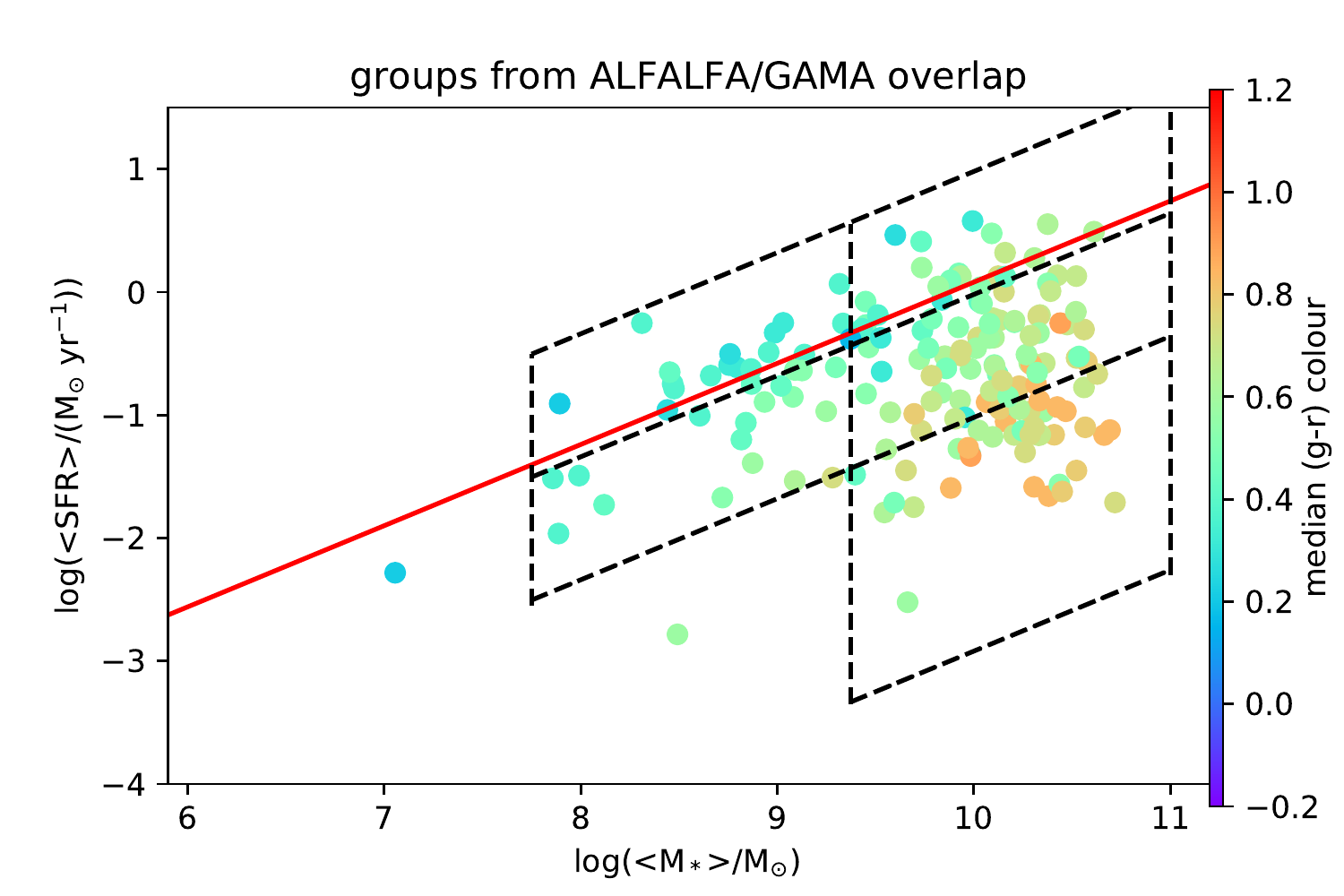}}}&
{\mbox{\includegraphics[width=3.5truein]{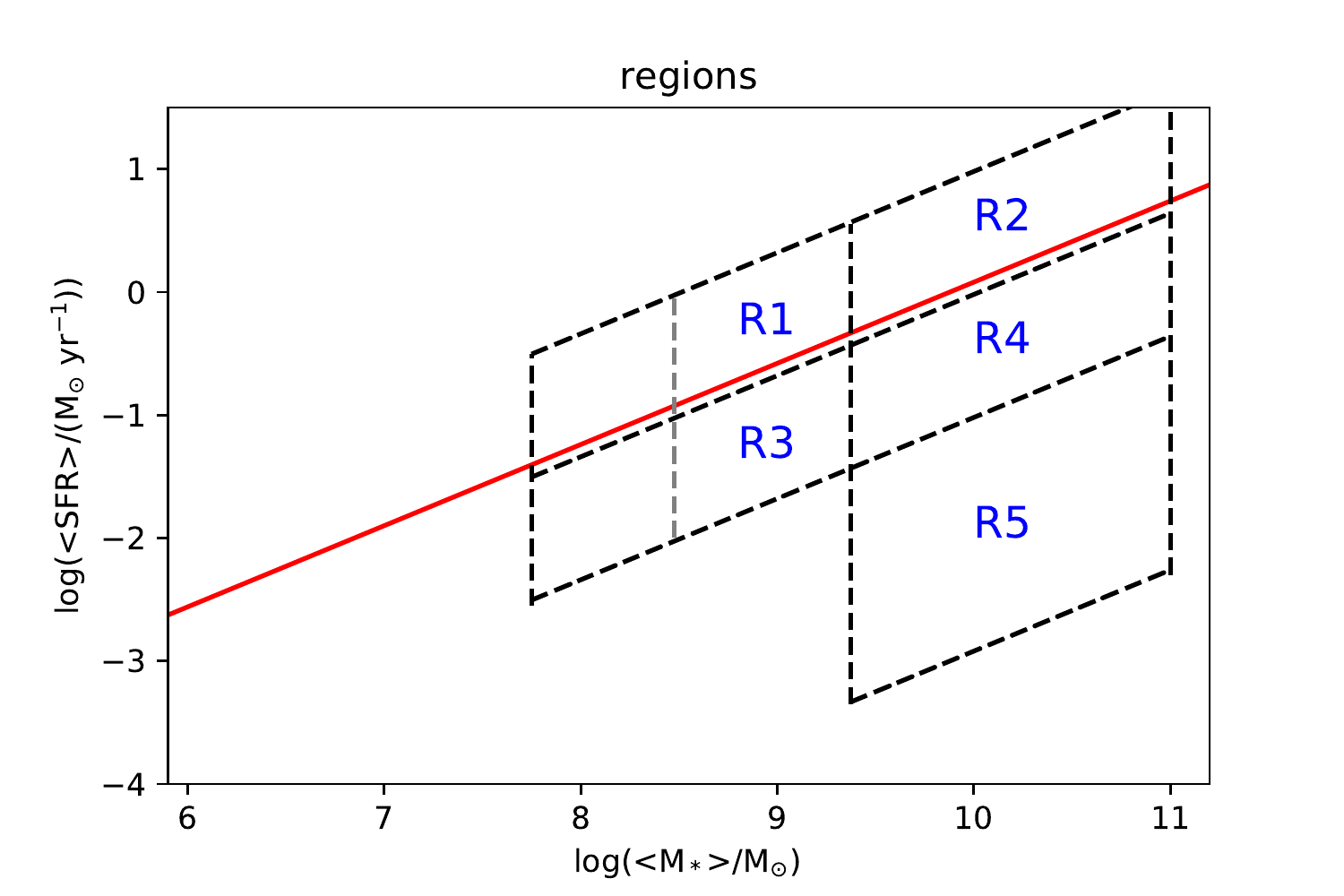}}}\\
\end{tabular}
\end{center}
\caption{log(SFR) plotted against log(\mst) for isolated galaxies from the ALFALFA--GAMA overlap sample (top left panel), and log($\langle$SFR$\rangle$) plotted against log($\langle$\mst$\rangle$) for galaxies in pairs (top right panel) and groups (bottom left panel). The points are colour-coded according to the median $g-r$ colour for groups/pairs, and g$-$r colour for isolated galaxies. The red bold line in each panel is the fit to the SFMS for z~$<0.1$ galaxies from \citet{2016MNRAS.461..458D}. The bottom right panel names the regions marked in each of the panels.}
\label{fig:sfrms}
\end{figure*}

\begin{table*}
\begin{center}
\caption{Measured values of different parameters for units within the five regions marked in Fig.~\ref{fig:sfrms}. For log($\langle$SFR$\rangle$) and log($\langle$\mst$\rangle$) the median value in each region is presented, while the error bars represent the standard deviation around the respective median value. The mean gas fractions ($f_{HI}~=~$M$_{HI}$/M$_{*}$) are measured using flux-weighted stacking in the different regions (see Section~\ref{sec:sfrms} for details). Error bars are based on standard deviation in line-channels measured using 1000 bootstrapped samples drawn randomly from the original spectra. The velocity widths are the 2$\sigma$ values from the Gaussian fits to the stacked spectra, whenever such a fit was possible.}
\label{tab:5reg}
\begin{tabular}{lccccc}
\hline
\hline
Environment 			& Region R1 			& Region R2 			& Region R3 			& Region R4 			& Region R5 \\
				& log(${\rm \frac{\langle M_* \rangle}{M_{\odot}}}$)	& log(${\rm \frac{\langle M_* \rangle}{M_{\odot}}}$)	& log(${\rm \frac{\langle M_* \rangle}{M_{\odot}}}$)	& log(${\rm \frac{\langle M_* \rangle}{M_{\odot}}}$)	& log(${\rm \frac{\langle M_* \rangle}{M_{\odot}}}$) \\
\hline
Isolated 			& 8.45$\pm$0.39			& 9.73$\pm$0.32			& 8.62$\pm$0.41			& 10.00$\pm$0.40		& 10.21$\pm$0.40 \\
Isolated -- smaller region	& 8.76$\pm$0.25			&				& 8.84$\pm$0.24			&				& \\
Pairs 				& 8.76$\pm$0.40			& 9.79$\pm$0.30			& 8.80$\pm$0.40			& 9.96$\pm$0.32			& 10.23$\pm$0.33 \\
Groups 				& 8.79$\pm$0.37			& 9.92$\pm$0.31			& 8.87$\pm$0.47			& 10.10$\pm$0.29		& 10.29$\pm$0.32 \\
\hline         
\hline                           
Environment 			& Region R1 				& Region R2 				& Region R3 				& Region R4 				& Region R5 \\
				& log(${\rm \frac{\langle SFR \rangle}{M_{\odot} yr^{-1}}}$)	& log(${\rm \frac{\langle SFR \rangle}{M_{\odot} yr^{-1}}}$)	& log(${\rm \frac{\langle SFR \rangle}{M_{\odot} yr^{-1}}}$)	& log(${\rm \frac{\langle SFR \rangle}{M_{\odot} yr^{-1}}}$)	& log(${\rm \frac{\langle SFR \rangle}{M_{\odot} yr^{-1}}}$) \\
\hline
Isolated 			& $-$0.76$\pm$0.31			& 0.06$\pm$0.27				& $-$1.26$\pm$0.38			& $-$0.38$\pm$0.33			& $-$1.65$\pm$0.49 \\
Isolated - smaller region	& $-$0.58$\pm$0.23			&					& $-$1.06$\pm$0.30			&					& \\
Pairs 				& $-$0.58$\pm$0.30			& 0.02$\pm$0.26				& $-$1.15$\pm$0.41			& $-$0.41$\pm$0.31			& $-$1.38$\pm$0.47 \\
Groups 				& $-$0.59$\pm$0.25			& 0.13$\pm$0.26				& $-$1.06$\pm$0.41			& $-$0.52$\pm$0.31			& $-$1.16$\pm$0.34 \\
\hline    
\hline 
Environment 			& Region R1 			& Region R2 			& Region R3 			& Region R4 			& Region R5 \\
				& ${\rm \langle \frac{M_{HI}}{M_{*}} \rangle}$	& ${\rm \langle \frac{M_{HI}}{M_{*}} \rangle}$	& ${\rm \langle \frac{M_{HI}}{M_{*}} \rangle}$	& ${\rm \langle \frac{M_{HI}}{M_{*}} \rangle}$	& ${\rm \langle \frac{M_{HI}}{M_{*}} \rangle}$ \\
\hline
Isolated 			& 1.84$\pm$0.17			& 0.38$\pm$0.04			& 0.83$\pm$0.07			& 0.15$\pm$0.02			& 0.02$\pm$0.01 \\
Isolated -- smaller region	& 1.09$\pm$0.10			&				& 0.55$\pm$0.07			&				& \\
Pairs 				& 2.39$\pm$0.26			& 0.41$\pm$0.08			& 1.66$\pm$0.26			& 0.11$\pm$0.02			& $\pm$0.01 \\
Groups 				& 4.77$\pm$0.87			& 0.51$\pm$0.07			& 1.59$\pm$0.41			& 0.19$\pm$0.03			& 0.02$\pm$0.01 \\    
\hline    
\hline   
Environment 			& Region R1 			& Region R2 			& Region R3 			& Region R4 			& Region R5 \\
				& $\Delta v_{HI,\pm 2 \sigma}$	& $\Delta v_{HI,\pm 2 \sigma}$	& $\Delta v_{HI,\pm 2 \sigma}$	& $\Delta v_{HI,\pm 2 \sigma}$	& $\Delta v_{HI,\pm 2 \sigma}$ \\
				& ${\rm (km~s^{-1})}$    		& ${\rm (km~s^{-1})}$			& ${\rm (km~s^{-1})}$			& ${\rm (km~s^{-1})}$			& ${\rm (km~s^{-1})}$ \\
\hline
Isolated			& 205				& 365				& 171				& 427				& (295) \\
Isolated -- smaller region	& 238				&                               & 210				&				& \\
Pairs				& 244				& 245				& 284				& 326				& \\
Groups				& 316				& 367				& 207				& 605				& (195) \\
\hline
\hline
\end{tabular}
\end{center}
\end{table*}

As our aim is to relate the gas content of groups and pairs to their star formation properties, we measure the {\it average} SFR and {\it average} \mst\ for each group/pair in our sample, denoted from this point as $\langle$SFR$\rangle$ and $\langle$\mst$\rangle$ respectively.
In Fig.~\ref{fig:sfrms} we compare these measurements with those for the isolated galaxies in our GAMA-ALFALFA overlap sample, which as mentioned before is effectively our control dataset.
The star-forming main sequence (SFMS) for GAMA galaxies for the z~$<0.1$ regime from \citet{2016MNRAS.461..458D} is plotted in all panels for reference.
Following expectations, from this figure we find that the $\langle$SFR$\rangle$ and $\langle$\mst$\rangle$ values of galaxies in groups and pairs lie within the extent of values spanned by isolated galaxies.
It is also evident that the $\langle$\mst$\rangle$ of pairs and groups tend to lie on the higher side of isolated galaxy stellar masses, more so for groups than pairs.
This works almost exclusively uses the average quantities $\langle$SFR$\rangle$ and $\langle$\mst$\rangle$, and it might be useful to conceptualize what these quantities signify.

In all panels, $g-r$ colours (median values for groups/pairs) become redder as one moves towards higher $\langle$\mst$\rangle$ values along the SFMS, or to lower $\langle$SFR$\rangle$ values away from the SFMS and towards the `quenched' galaxy regime. 
The similar (median) $g-r$ colours across the log($\langle$\mst$\rangle$) -- log($\langle$SFR$\rangle$) plane for groups/pairs/isolated galaxies might be indicative of the primacy of secular evolution of galaxies compared to environmental processing, if one considers that the $g-r$ colour is tied to the evolutionary stage of the galaxy/pair/group in and it traces the evolution from the star-forming to the quenched regime.
Exploring the details of the same is beyond the scope of the present study though, and in Fig.~\ref{fig:sfrms} the $g-r$ colours are presented for qualitative purposes only and we do not use them in our subsequent analysis.

We divide the log($\langle$\mst$\rangle$) -- log($\langle$SFR$\rangle$) plane into 5 distinct regions as marked and named in Fig.~\ref{fig:sfrms}, and subsequently stack the \hi\ in all galaxies/pairs/groups within each such region (see Sec.~\ref{sec:stack1}).
These regions were defined after some experimentation so as to cover the region of the log($\langle$\mst$\rangle$) -- log($\langle$SFR$\rangle$) plane where most of the groups and pairs in our sample lie, with each region having enough number of units (especially for groups) for a significant detection of their \hi\ content through stacking, while at the same time not being too extended in $\langle$\mst$\rangle$ or $\langle$SFR$\rangle$ for us not to able to draw any conclusions.
In the vertical direction, there are two regions on-or-above the SFMS (1 dex above a line running parallel to the SFMS but offset 0.1 dex below), two regions just below the SFMS (1 dex below the line running parallel to the SFMS but offset 0.1 dex below), and finally a region covering the quenched galaxy regime which goes a further 2 dex below in the vertical direction.
The regions are bound at the very low mass and the very high mass end by log(${\rm \frac{\langle M_* \rangle}{M_{\odot}}}$) values of 7.75 and 11 respectively.
The line dividing the low $\langle$\mst$\rangle$ regions (R1 and R3) from the high $\langle$\mst$\rangle$ regions (R2, R4 and R5) is exactly half-way between the bounding $\langle$\mst$\rangle$ values in log-space (log(${\rm \frac{\langle M_* \rangle}{M_{\odot}}}$) = 9.375).
The median and standard deviation of the log($\langle$\mst$\rangle$) and log($\langle$SFR$\rangle$) value for the units in these regions are listed in Table~\ref{tab:5reg}.

The median log($\langle$\mst$\rangle$) value for groups and pairs in the lower $\langle$\mst$\rangle$ regions (R1 and R3) is on the higher side compared to that for isolated galaxies.
This is because a densely populated portion of the SFMS of isolated galaxies (a region of log($\langle$\mst$\rangle$) -- log($\langle$SFR$\rangle$) plane where there aren't many groups/pairs) lies in the lower log($\langle$\mst$\rangle$) portion of these regions.
In order to ensure that we are comparing like with like, we also define smaller versions of R1 and R3 regions for isolated galaxies (marked in Fig.~\ref{fig:sfrms}) by shifting the lower bound of log($\frac{\langle M_* \rangle}{M_{\odot}}$) from 7.75 to 8.475.
As can be seen from Table~\ref{tab:5reg}, the median log(SFR) and log(\mst) values for isolated galaxies in the smaller R1 and R3 regions better match the median values of log($\langle$SFR$\rangle$) and log($\langle$\mst$\rangle$) for groups and pairs.


\subsection{$\langle f_{HI} \rangle$ in different regions of the $\langle$\mst$\rangle$ -- $\langle$SFR$\rangle$ plane}
\label{sec:stack1}

\begin{figure*}
\begin{center}
\begin{tabular}{cc}
{\mbox{\includegraphics[width=3.5truein]{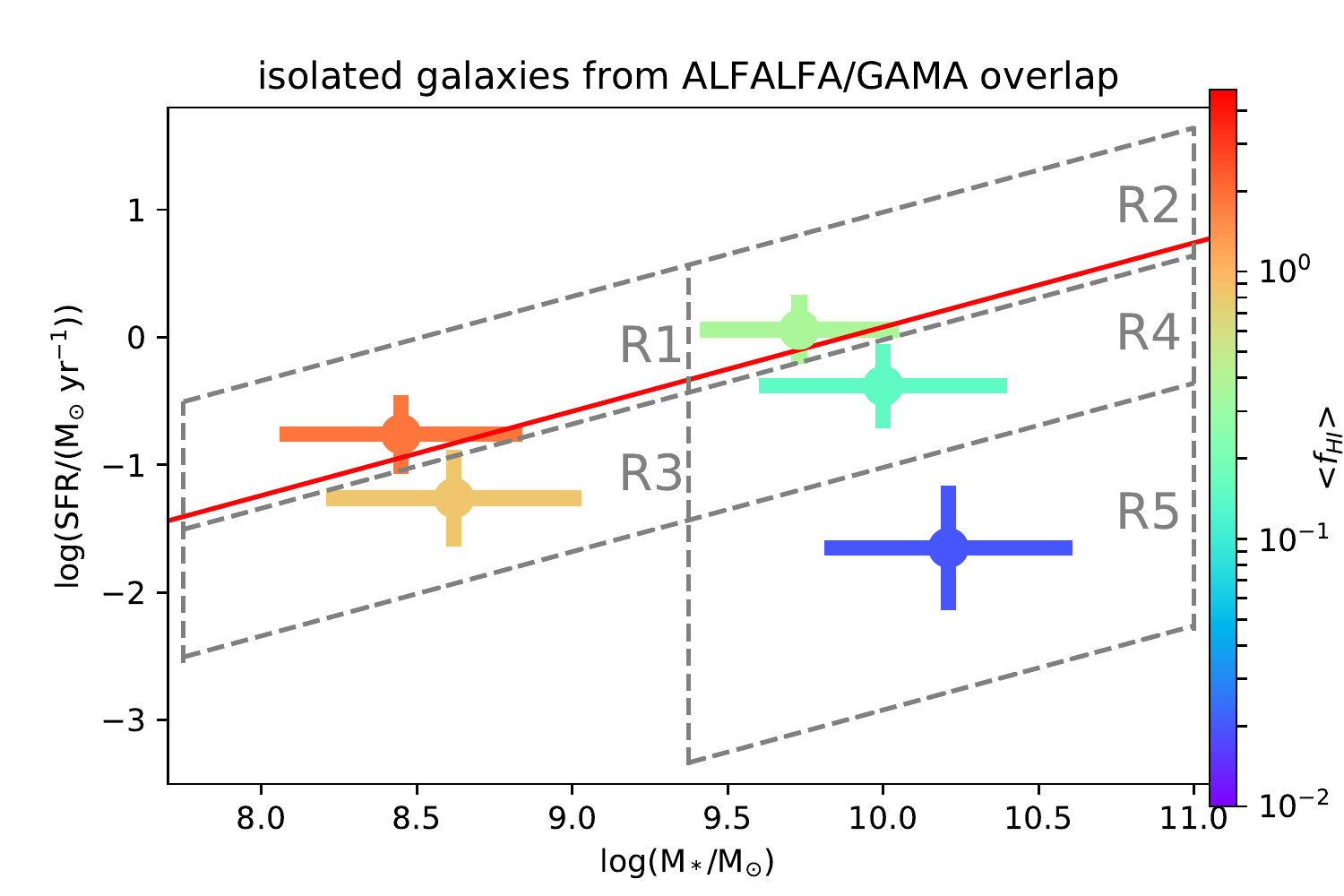}}}&
{\mbox{\includegraphics[width=3.5truein]{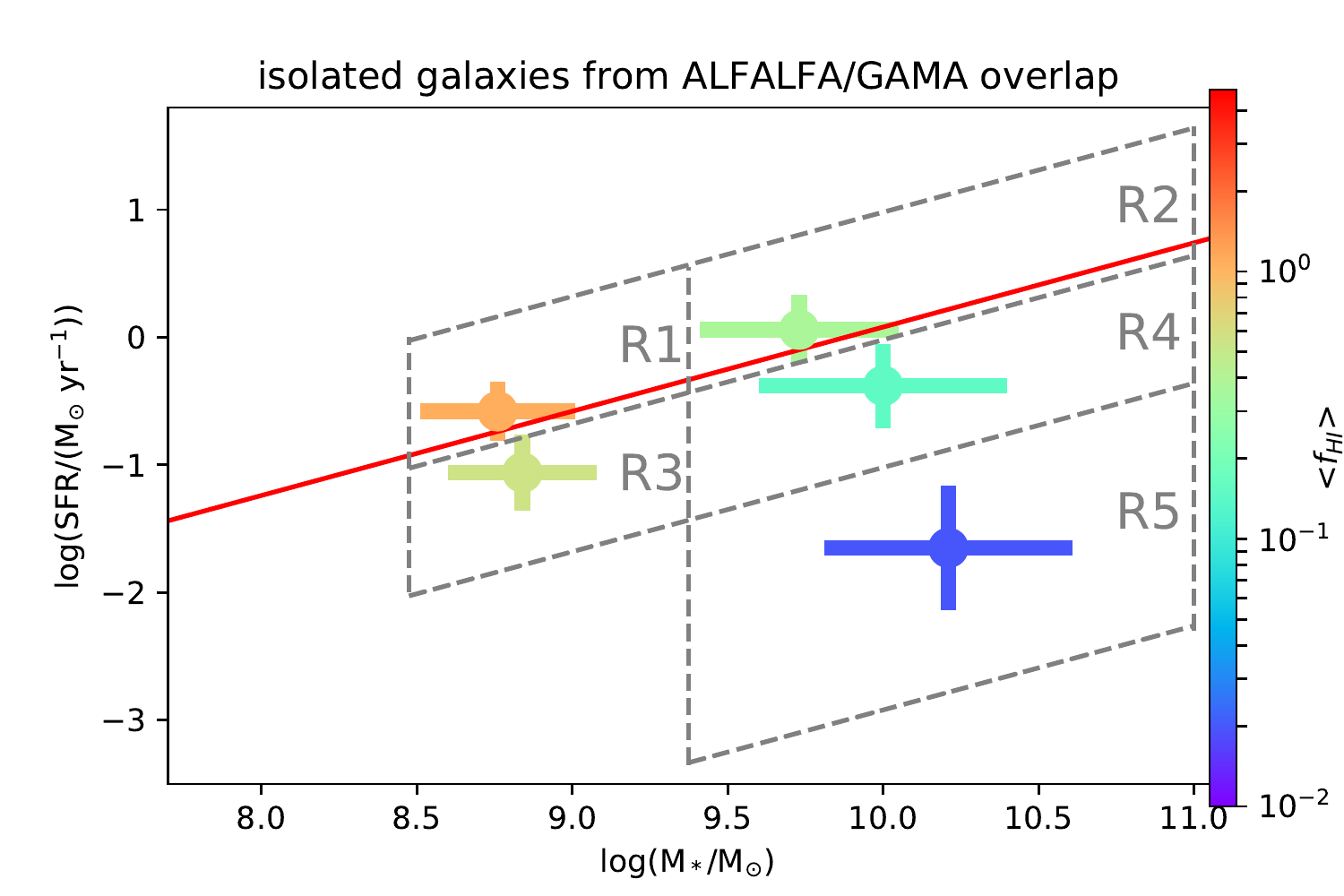}}}\\
{\mbox{\includegraphics[width=3.5truein]{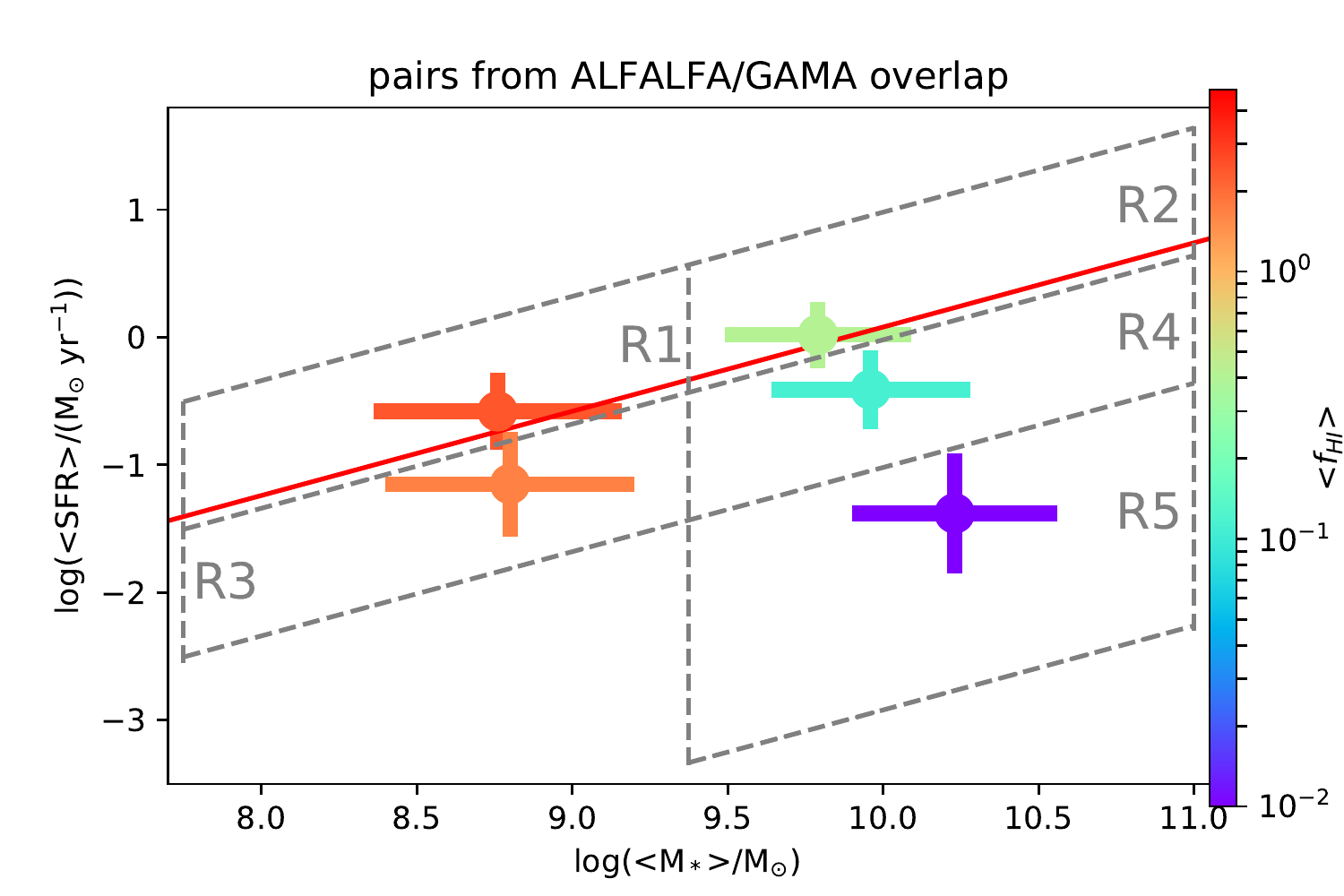}}}&
{\mbox{\includegraphics[width=3.5truein]{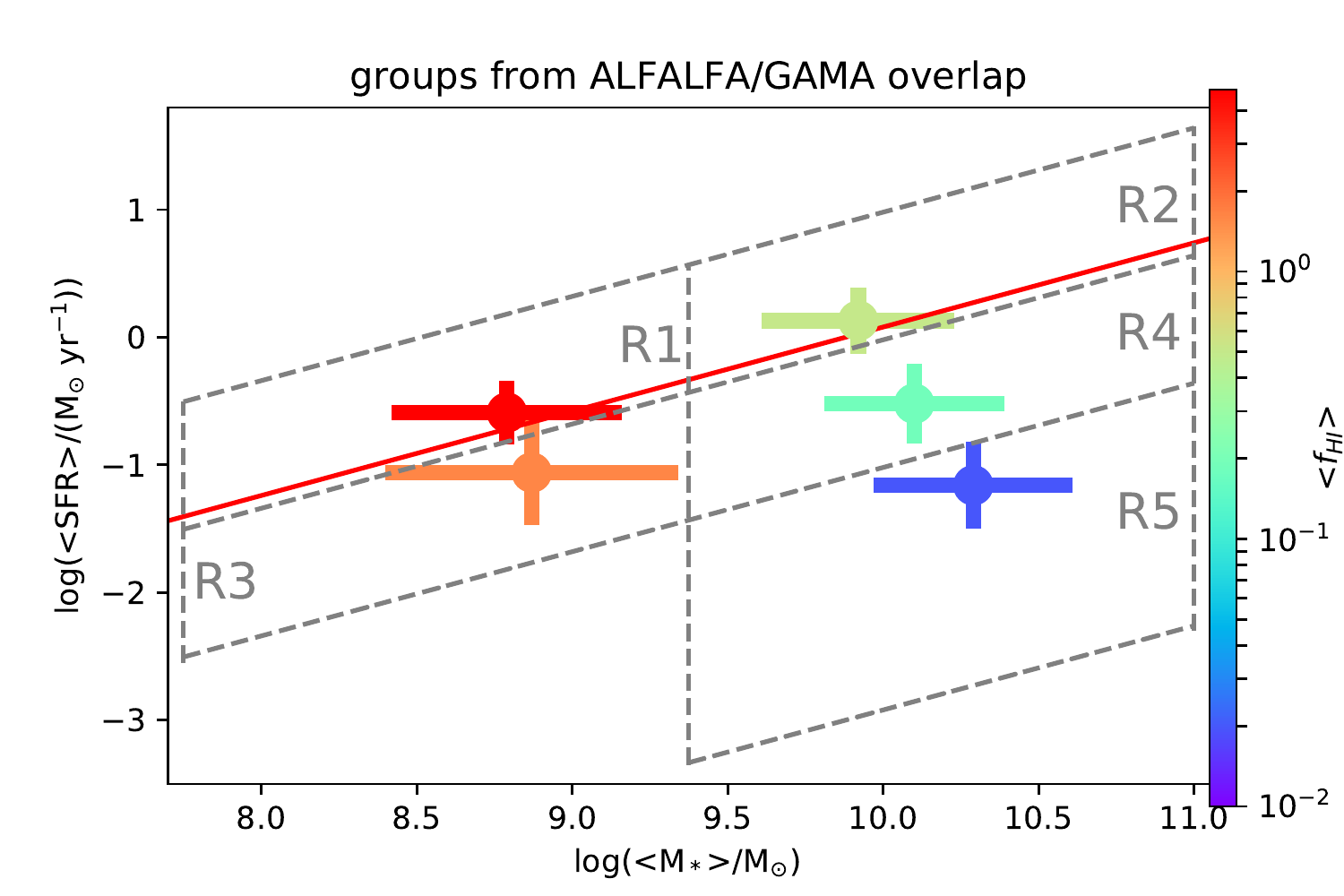}}}\\
\end{tabular}
\end{center}
\caption{The measured values of the mean gas fractions ($\langle f_{HI} \rangle$) shown in colour (logarithmic scale) for units within the five regions marked with dashed lines (same as in Fig.~\ref{fig:sfrms}) and as listed in Table~\ref{tab:5reg}. The points are plotted at the measured median values for $\langle$\mst$\rangle$ and $\langle$SFR$\rangle$ of units within each region, with the corresponding standard deviations shown as errorbars. The upper two panels for isolated galaxies differ only in regions R1 and R3, where the left panel shows the result for the full regions and the right panel shows the result for the smaller R1 and R3 regions. Note that for the smaller R1 and R3 regions the median values for $\langle$\mst$\rangle$ and $\langle$SFR$\rangle$ are similar to those for pairs and groups in regions R1 and R3.}
\label{fig:sfrmsv}
\end{figure*}

We stack the \hi\ 21 cm emission spectra for all the groups/pairs/galaxies within each of the five regions described in Section~\ref{sec:sfrms}.
As the number of groups (and pairs) in our sample is limited, we opt for stacking in order to get statistically significant detections of the \hi\ emission.
Since the extent of each of the regions in $\langle$\mst$\rangle$ is large, we do not measure the total \hi\ content of groups/pairs/galaxies within a given region as such a quantity will likely be dominated by the more massive groups/pairs/galaxies within each region.
Instead we stack on a quantity which is effectively normalized by \mst, i.e. \hi\ gas fraction, ${\ensuremath{f_{HI}~=~\frac{M_{HI}}{M_{*}}}}$.
Note that from this point we use the term `gas fraction' interchangeably with `\hi\ fraction', as on galactic scales for galaxies with any \mst, \hi\ dominates the gas fraction as opposed to molecular gas \citep{2017ApJS..233...22S}.

We also opt for flux-weighting instead of luminosity-weighting of our spectra, as each of our stacks span a large range in terms of luminosity distance and the latter method will bias our results against the nearer groups/pairs/galaxies in our stacks.
In order to implement this, for each of our spectra, we measure the root mean squared (RMS) noise in flux density units (Jy) at `line-free' frequencies.
For groups/pairs, the line-free frequencies are defined as those outside the group/pair velocity dispersion, and an additional 300 \kms\ buffer to account for galactic rotation, on either side of the central frequency of the \hi\ emission.
For isolated galaxies, the velocity range is simply 300 \kms\ on either side of the central frequency of the \hi\ emission.

The flux density spectra were converted to mass spectra using \citep[following][]{2013MNRAS.433.1398D}:
\begin{equation}
\frac{M_{HI,\nu_{obs}}}{M_{\odot} MHz^{-1}}~=~4.98 \times 10^7 \left( \frac{S_{\nu_{obs}}}{Jy} \right) \left( \frac{D_L}{Mpc} \right)^2,
\label{eqn:mhi}
\end{equation}
Where $S_{\nu_{obs}}$ is the observed flux density and $D_L$ is the luminosity distance.
Next, in order to enable stacking the frequency axis of all individual spectra were shifted to the rest frame using the redshift of the respective iterative central galaxy (for groups/pairs) or the galaxy itself (isolated).
If the respective redshift is z, the frequency shift is:
\begin{equation}
\nu_{em}~=~\nu_{obs}(1+z),
\label{eqn:nu}
\end{equation}
where $\nu_{obs}$ is the observed frame and $\nu_{em}$ is the rest frame emission frequency.
The mass spectra needs to be normalized to conserve the total mass during this frequency shift:
\begin{equation}
M_{HI,\nu_{em}}~=~\frac{M_{HI,\nu_{obs}}}{(1+z)}.
\label{eqn:massc}
\end{equation}
All individual spectra are re-gridded on to a uniformly spaced frequency axis with channel separation of 25 KHz. 
Next, all spectra are divided by the total \mst\ of the unit (group/pair/isolated galaxy), in order to convert them to $f_{HI}$ spectra.
Finally for each unit (group/pair/isolated galaxy), $f_{HI}$ spectra for units within any given region of the log($\langle$\mst$\rangle$) -- log($\langle$SFR$\rangle$) plane are then stacked using: 
\begin{equation}
\langle f_{HI} \rangle~=~\frac{\sum_{i}^{n} w_i f_{HI,em,i}}{\sum_{i}^{n} w_i},
\label{eqn:sum}
\end{equation}
where $f_{HI,em,i}$ is the $i^{th}$ spectrum with corresponding weight $w_i~=~\frac{1}{\sigma_i^2}$, where $\sigma_i$ is the RMS noise in flux density units (Jy) at `line-free' frequencies measured previously. 

Shifted to their rest frequencies, at this stage the velocity resolution of the spectra, corresponding to the frequency resolution of 25 KHz, are quite fine at $\sim$5 \kms.
We average each spectrum over six of the original (re-gridded) channels, so that each averaged channel is $\sim$30 \kms\ wide.
This velocity resolution is more than sufficient for sampling the stacked signals that are detected (Table~\ref{tab:5reg} onwards), while decreasing the RMS noise in the averaged channels -- which is useful for the polynomial fit that is done next.
To remove any residual baseline effects, we subtract a second order polynomial fit to channels $>~\pm$450 \kms\ from the rest frequency of the \hi\ 21 cm emission line.
The measured spectra are shown in the Appendix (Figure Set 17).
Whenever there is a detection of the stacked \hi\ emission signal (which is the case for most of the stacks), we fit a Gaussian with its centre, peak and standard deviation all being free parameters.
As can be seen from Fig. 17.1 and Figure Set 17, the rest frequency of the \hi\ 21 cm emission line is close to the centre of all fitted Gaussians, and in nowhere beyond 2$\sigma$ of the respective Gaussian.
We sum over all channels within $\pm$2$\sigma$ of the fitted Gaussian in each case to recover the mean $f_{HI}$ ($\langle f_{HI} \rangle$) for each region.
In order to estimate the errors on our measured $\langle f_{HI} \rangle$, for each region we create 1000 bootstrapped samples drawn randomly from the original spectra (with replacement), and measure the standard deviation around the measured value per channel using these sets of bootstrapped spectra.
The error on the measured $\langle f_{HI} \rangle$ is determined by summing over the standard deviations of all channels within $\pm$2$\sigma$ of the fitted Gaussian in each case, and dividing by the square-root of the total number of channels summed over (as the measurement in each channel is independent).
When there is no clear detection of the \hi\ emission, the error is measured by summing over $\pm$150 \kms\ channels on either side of the rest frequency of the \hi\ 21 cm emission line.

The measured $\langle f_{HI} \rangle$ are tabulated in Table~\ref{tab:5reg} along with the associated errors, and also shown in Fig.~\ref{fig:sfrmsv}.
For units from all three categories (groups/pairs/isolated galaxies), two clear trends exist in the $\langle f_{HI} \rangle$ values measured for the different regions.
First, as we move from above the SFMS to below it in the log($\langle$\mst$\rangle$) -- log($\langle$SFR$\rangle$) plane, i.e. R1$\rightarrow$R3 or R2$\rightarrow$R4$\rightarrow$R5, the gas fractions steadily decrease to being consistent with the units having no measurable \hi\ in the quenched galaxy region R5.
Second, as we move from the lower $\langle$\mst$\rangle$ to higher $\langle$\mst$\rangle$, i.e. R1$\rightarrow$R2 or R3$\rightarrow$R4, there is a sharp drop in the measured $\langle f_{HI} \rangle$ for all categories.
In terms of comparing the $\langle f_{HI} \rangle$ values in similar regions from the different categories, the  measured values are remarkably similar for the higher $\langle$\mst$\rangle$ regions (R2, R4, and R5).
In the lower $\langle$\mst$\rangle$ regions (R1 and R3) though, for either region the measured $\langle f_{HI} \rangle$ values steadily increase as we move from isolated galaxies to pairs to groups.
For isolated galaxies, given that a densely populated portion of the SFMS lies in the lower log(\mst) portion of these regions, the $\langle f_{HI} \rangle$ for the full regions is higher than the smaller versions of R1 and R3 which actually have similar median \mst\ and SFR properties to the corresponding regions for pairs and groups. 
In any case, the trend of increasing $\langle f_{HI} \rangle$ as we move from isolated galaxies$\rightarrow$pairs$\rightarrow$groups is evident whether one considers the full or the smaller version of regions R1 and R3.

The velocity widths over which the $\langle f_{HI} \rangle$ for each of the regions are determined, taken to be $\pm$2$\sigma$ of the fitted Gaussian in each case, are also tabulated in Table~\ref{tab:5reg}.
In general the velocity widths become larger as we move up the SFMS (R1$\rightarrow$R2 or R3$\rightarrow$R4), and as we move from isolated galaxies$\rightarrow$pairs$\rightarrow$groups.
But these trends are not clearly established possibly due to the moderate significance of some of the stacked detections, which results in $\pm$2$\sigma$ of the fitted Gaussian not tracing the true underlying spread in velocities.

\subsection{Variation of $\langle f_{HI} \rangle$ along the SFMS}
\label{sec:stack2}

\begin{figure*}
\begin{center}
\begin{tabular}{cc}
{\mbox{\includegraphics[width=3.5truein]{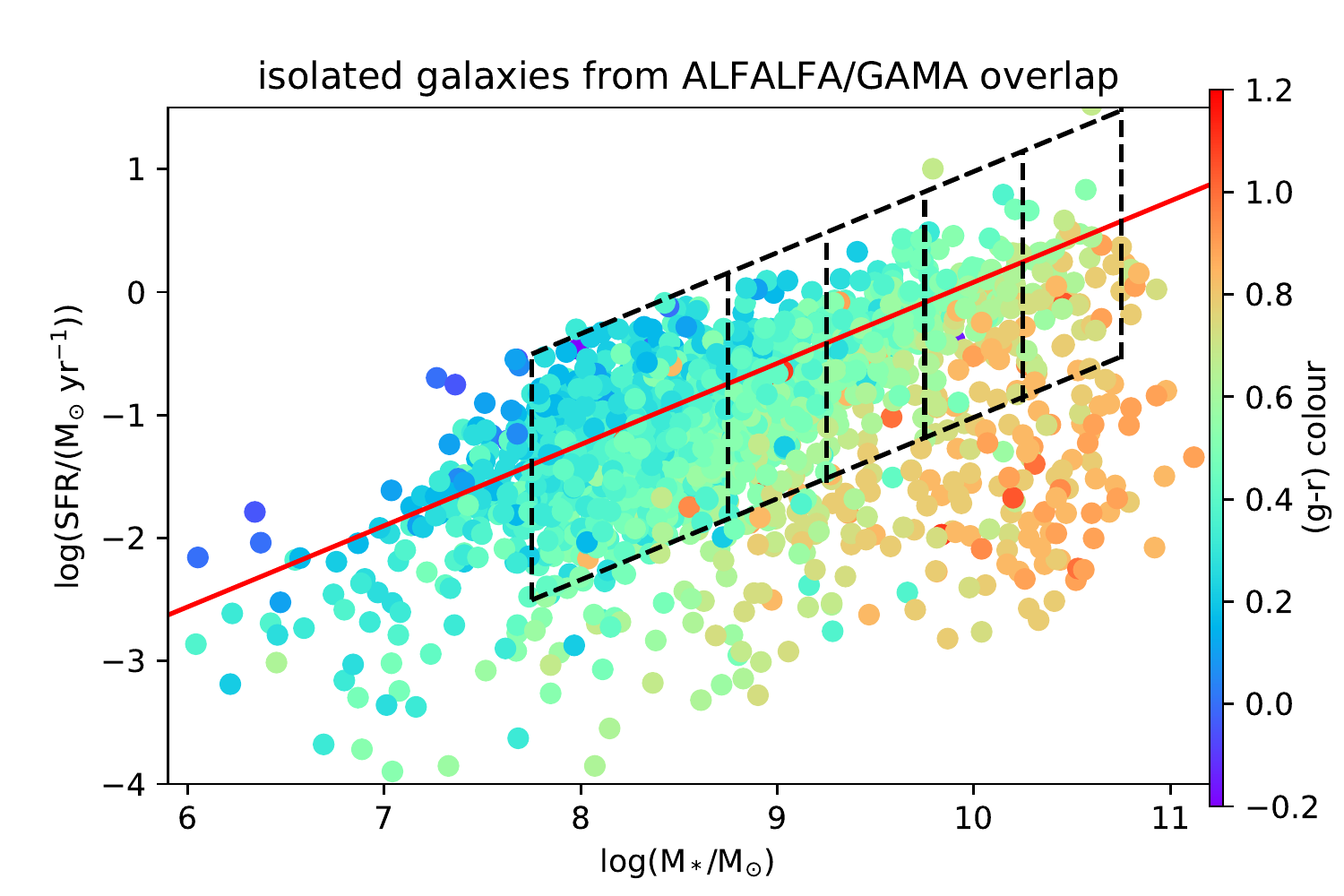}}}&
{\mbox{\includegraphics[width=3.5truein]{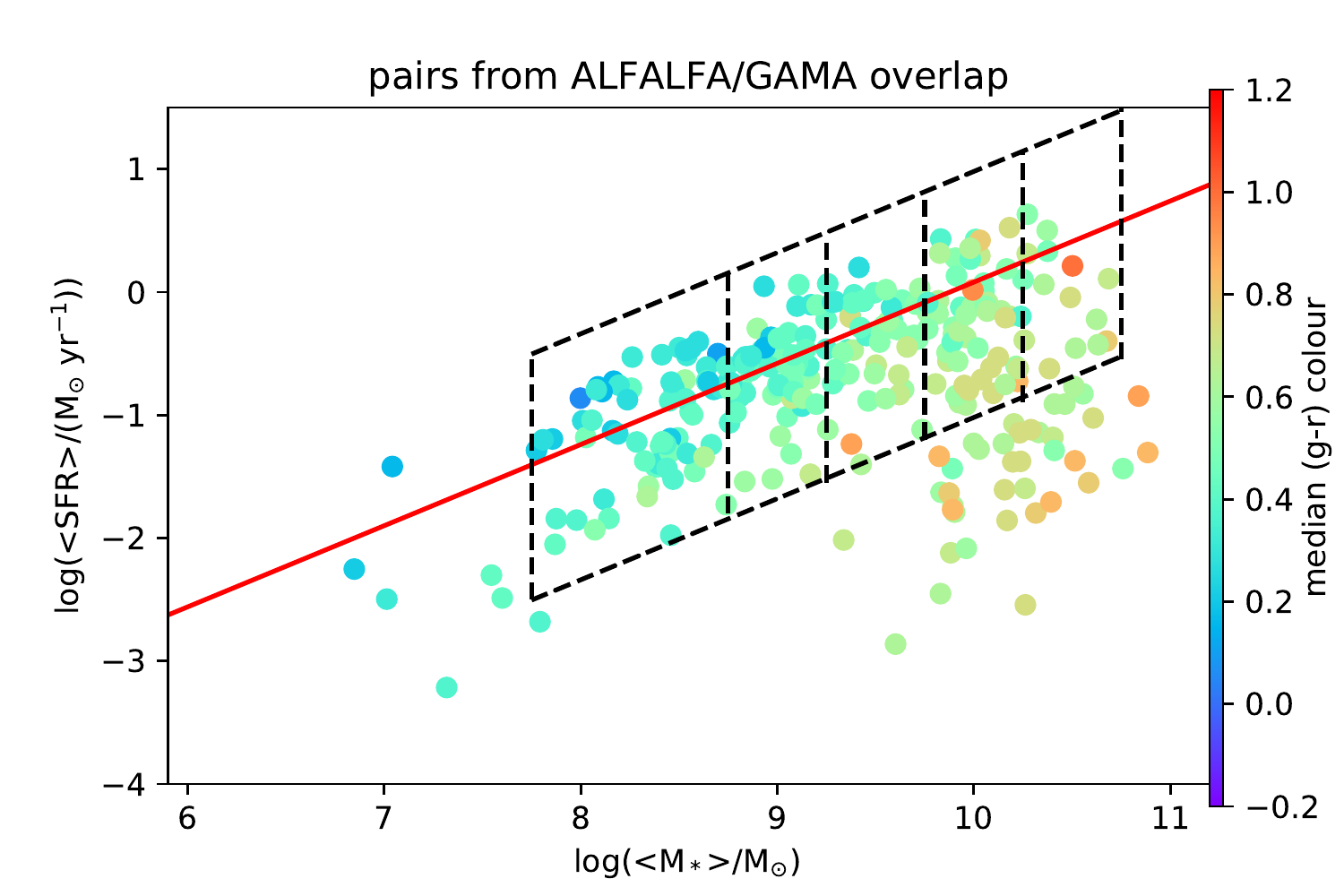}}}\\
{\mbox{\includegraphics[width=3.5truein]{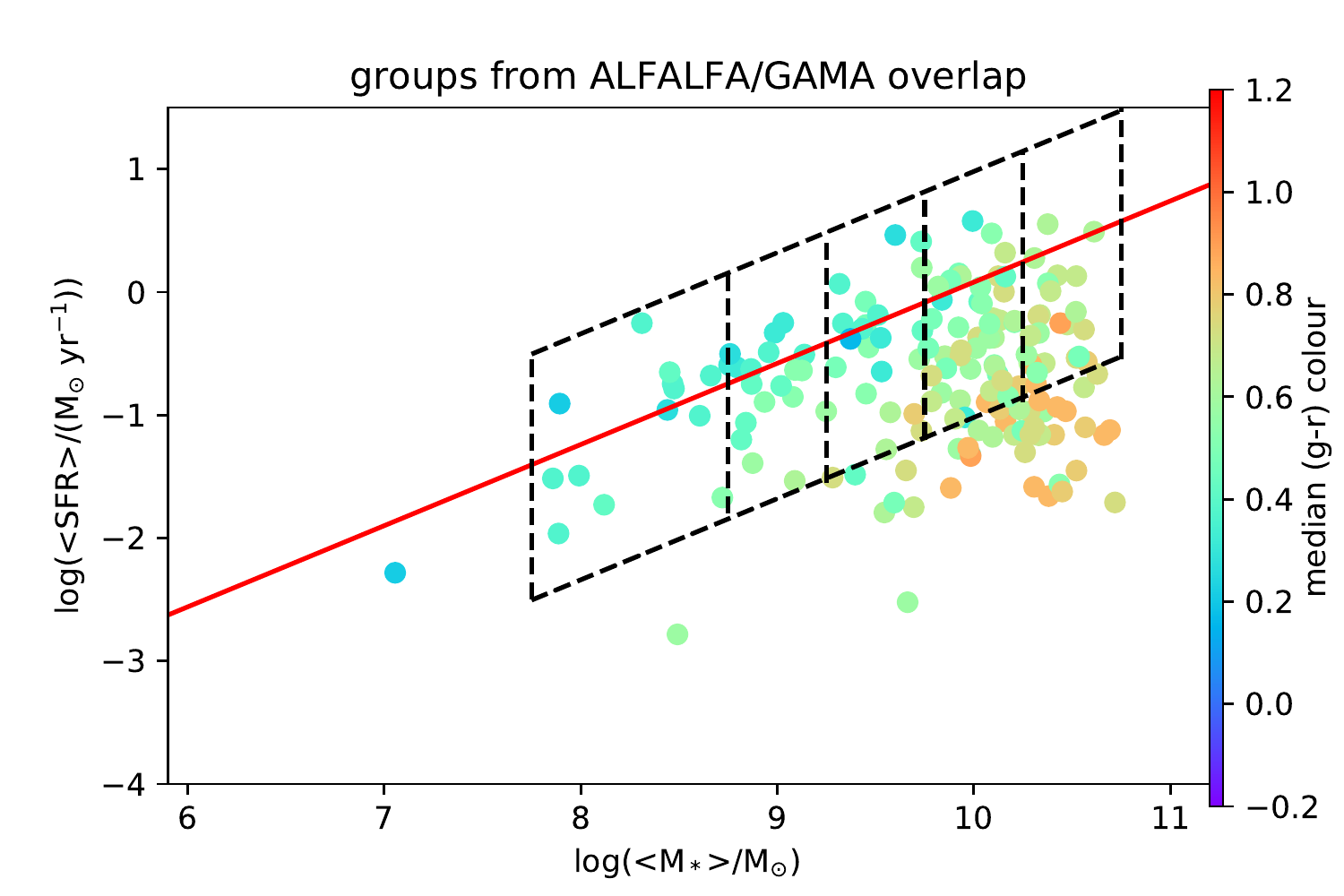}}}&
{\mbox{\includegraphics[width=3.5truein]{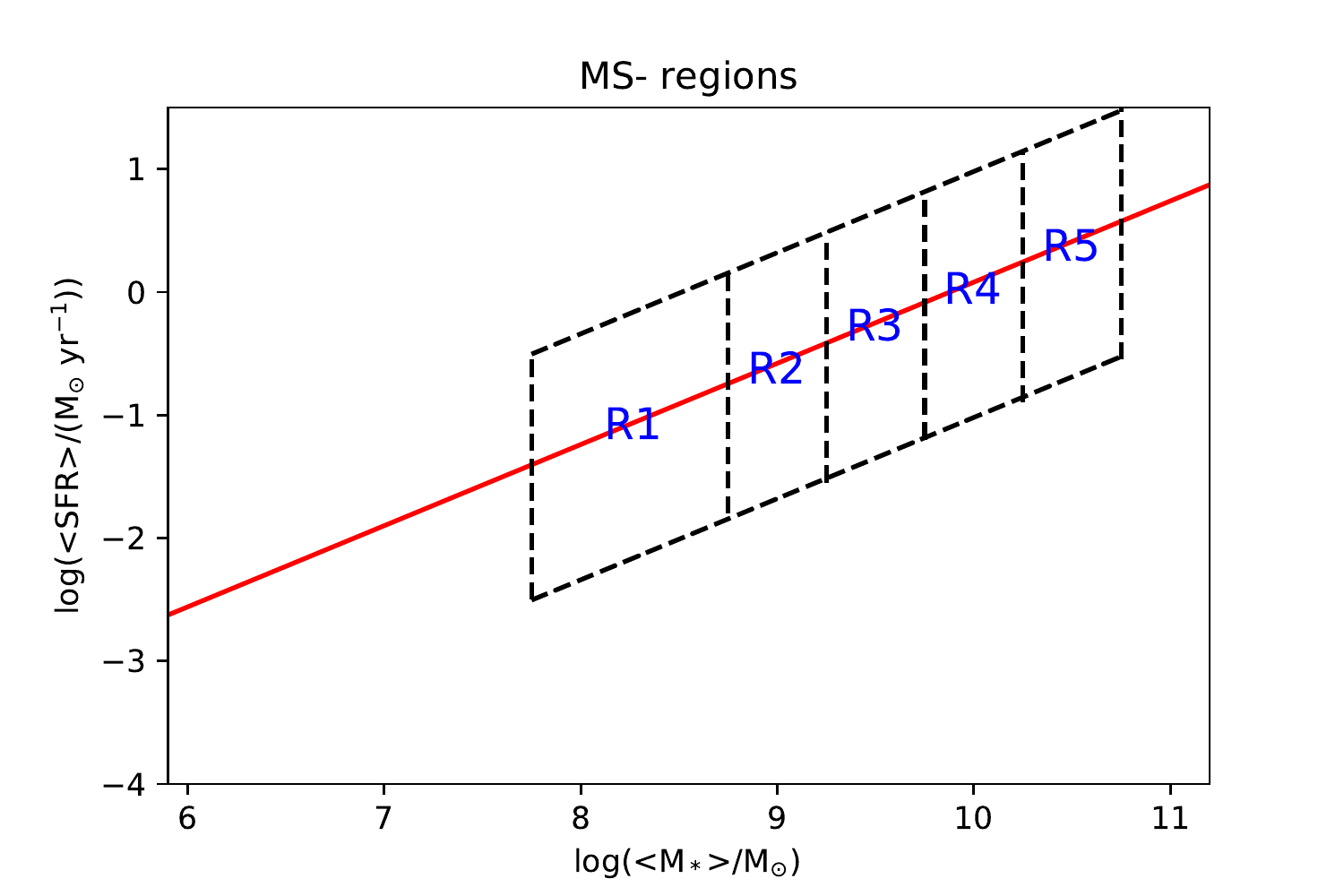}}}\\
\end{tabular}
\end{center}
\caption{log(SFR) plotted against log(\mst) for isolated galaxies from the ALFALFA--GAMA overlap sample (top left panel), and log($\langle$SFR$\rangle$) plotted against log($\langle$\mst$\rangle$) for galaxies in pair (top right panel) and group (bottom left panel), same as in Fig.~\ref{fig:sfrms}. The regions marked in this figure are defined to span the SFMS, and the bottom right panel names the regions.}
\label{fig:mssfrms}
\end{figure*}

\begin{table*}
\begin{center}
\caption{Measured values of different parameters for units within the five main-sequence regions marked in Fig.~\ref{fig:mssfrms}. See Table~\ref{tab:5reg} caption for description of the tabulated quantities.}
\label{tab:msreg}
\begin{tabular}{lccccc}
\hline
\hline
Environment 			& Region MS-R1 			& Region MS-R2 			& Region MS-R3 			& Region MS-R4 			& Region MS-R5 \\
				& log(${\rm \frac{\langle M_* \rangle}{M_{\odot}}}$)	& log(${\rm \frac{\langle M_* \rangle}{M_{\odot}}}$)	& log(${\rm \frac{\langle M_* \rangle}{M_{\odot}}}$)	& log(${\rm \frac{\langle M_* \rangle}{M_{\odot}}}$)	& log(${\rm \frac{\langle M_* \rangle}{M_{\odot}}}$) \\
\hline
Isolated 			& 8.35$\pm$0.27			& 8.96$\pm$0.14			& 9.47$\pm$0.15			& 10.01$\pm$0.15		& 10.42$\pm$0.12 \\
Pairs 				& 8.44$\pm$0.26			& 9.01$\pm$0.13			& 9.48$\pm$0.14			& 9.98$\pm$0.13			& 10.39$\pm$0.15 \\
Groups 				& 8.44$\pm$0.30			& 8.96$\pm$0.14			& 9.53$\pm$0.14			& 10.03$\pm$0.13		& 10.38$\pm$0.10 \\
\hline         
\hline                           
Environment 			& Region MS-R1 				& Region MS-R2 				& Region MS-R3 				& Region MS-R4 				& Region MS-R5 \\
				& log(${\rm \frac{\langle SFR \rangle}{M_{\odot} yr^{-1}}}$)	& log(${\rm \frac{\langle SFR \rangle}{M_{\odot} yr^{-1}}}$)	& log(${\rm \frac{\langle SFR \rangle}{M_{\odot} yr^{-1}}}$)	& log(${\rm \frac{\langle SFR \rangle}{M_{\odot} yr^{-1}}}$)	& log(${\rm \frac{\langle SFR \rangle}{M_{\odot} yr^{-1}}}$) \\
\hline
Isolated 			& $-$1.08$\pm$0.41			& $-$0.79$\pm$0.36			& $-$0.38$\pm$0.36			& $-$0.13$\pm$0.38			& $-$0.04$\pm$0.44 \\
Pairs 				& $-$1.04$\pm$0.43			& $-$0.63$\pm$0.33			& $-$0.35$\pm$0.34			& $-$0.20$\pm$0.20			& $-$0.06$\pm$0.37 \\
Groups 				& $-$0.96$\pm$0.49			& $-$0.64$\pm$0.33			& $-$0.37$\pm$0.46			& $-$0.37$\pm$0.41			& $-$0.30$\pm$0.35 \\
\hline    
\hline 
Environment 			& Region MS-R1 			& Region MS-R2 			& Region MS-R3 			& Region MS-R4 			& Region MS-R5 \\
				& ${\rm \langle \frac{M_{HI}}{M_{*}} \rangle}$	& ${\rm \langle \frac{M_{HI}}{M_{*}} \rangle}$	& ${\rm \langle \frac{M_{HI}}{M_{*}} \rangle}$	& ${\rm \langle \frac{M_{HI}}{M_{*}} \rangle}$	& ${\rm \langle \frac{M_{HI}}{M_{*}} \rangle}$ \\
\hline
Isolated 			& 1.59$\pm$0.11			& 0.62$\pm$0.07			& 0.38$\pm$0.04			& 0.17$\pm$0.02			& 0.09$\pm$0.01 \\
Pairs 				& 3.48$\pm$0.32			& 0.74$\pm$0.16			& 0.49$\pm$0.08			& 0.17$\pm$0.03			& 0.07$\pm$0.03 \\
Groups 				& 3.67$\pm$1.55			& 3.00$\pm$0.65			& 0.62$\pm$0.09			& 0.22$\pm$0.03			& 0.09$\pm$0.02 \\    
\hline    
\hline   
Environment 			& Region MS-R1 			& Region MS-R2 			& Region MS-R3 			& Region MS-R4 			& Region MS-R5 \\
				& $\Delta v_{HI,\pm 2 \sigma}$	& $\Delta v_{HI,\pm 2 \sigma}$	& $\Delta v_{HI,\pm 2 \sigma}$	& $\Delta v_{HI,\pm 2 \sigma}$	& $\Delta v_{HI,\pm 2 \sigma}$ \\
				& ${\rm (km~s^{-1})}$    		& ${\rm (km~s^{-1})}$			& ${\rm (km~s^{-1})}$			& ${\rm (km~s^{-1})}$			& ${\rm (km~s^{-1})}$ \\
\hline
Isolated			& 185				& 249				& 340				& 385				& 560 \\
Pairs				& 261				& 208				& 250				& 409				& (808) \\
Groups				& 348				& 436				& 375				& 503				& 550 \\
\hline
\hline
\end{tabular}
\end{center}
\end{table*}

\begin{figure*}
\begin{center}
\includegraphics[width=5truein]{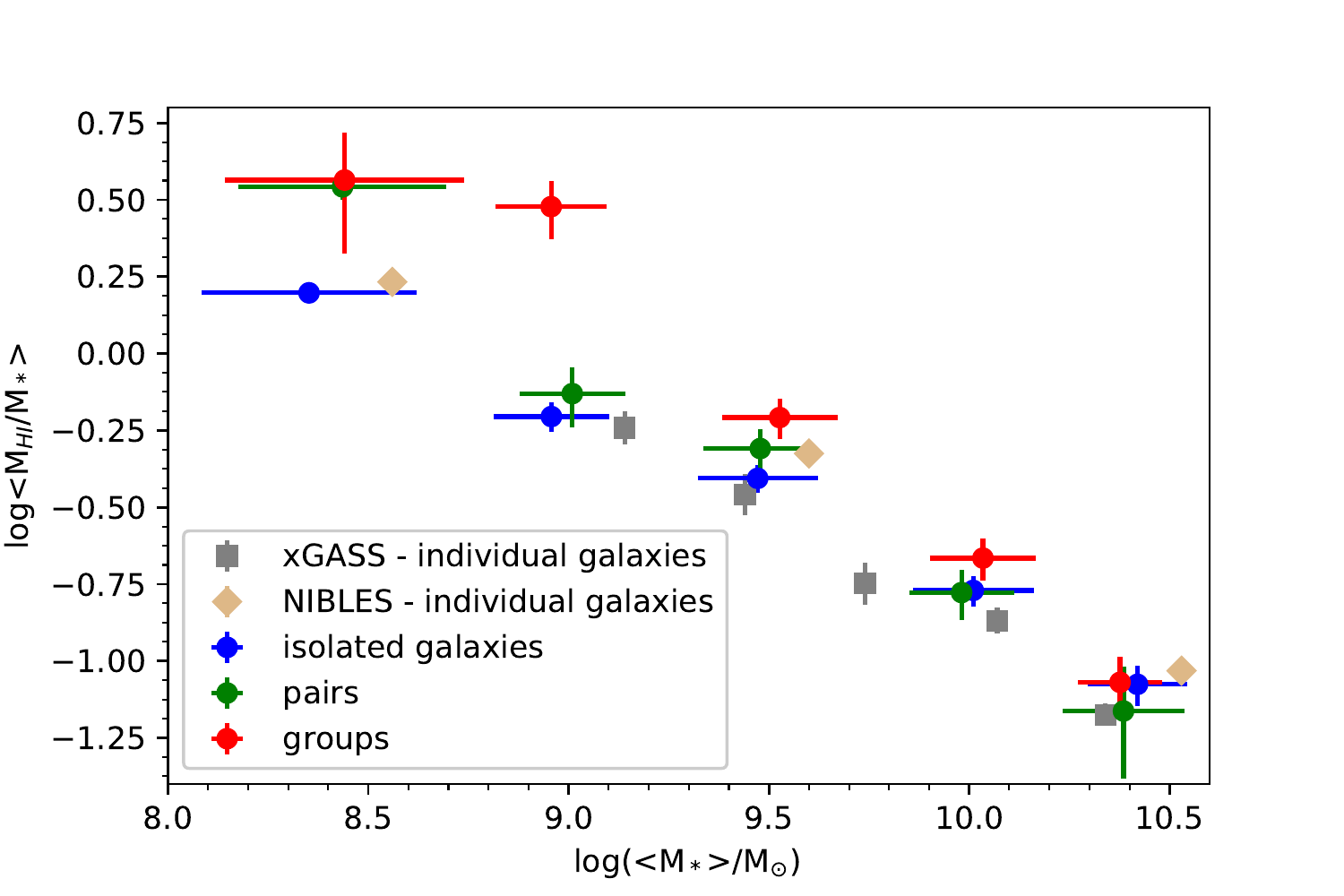}
\end{center}
\caption{The variation of gas fraction with log($\langle$\mst$\rangle$) along the SFMS, and with the unit category. The $\langle f_{HI} \rangle$ and the median log($\langle$\mst$\rangle$)s tabulated in Table~\ref{tab:msreg} are plotted against each other for the three categories of galaxies. Also plotted are values for individual galaxies from the xGASS survey \citep[][]{2018MNRAS.476..875C} and the NIBLES survey \citep{2019MNRAS.487.4901H}.}
\label{fig:fhic}
\end{figure*}

\begin{figure}
\begin{center}
\end{center}
\includegraphics[width=3.5truein]{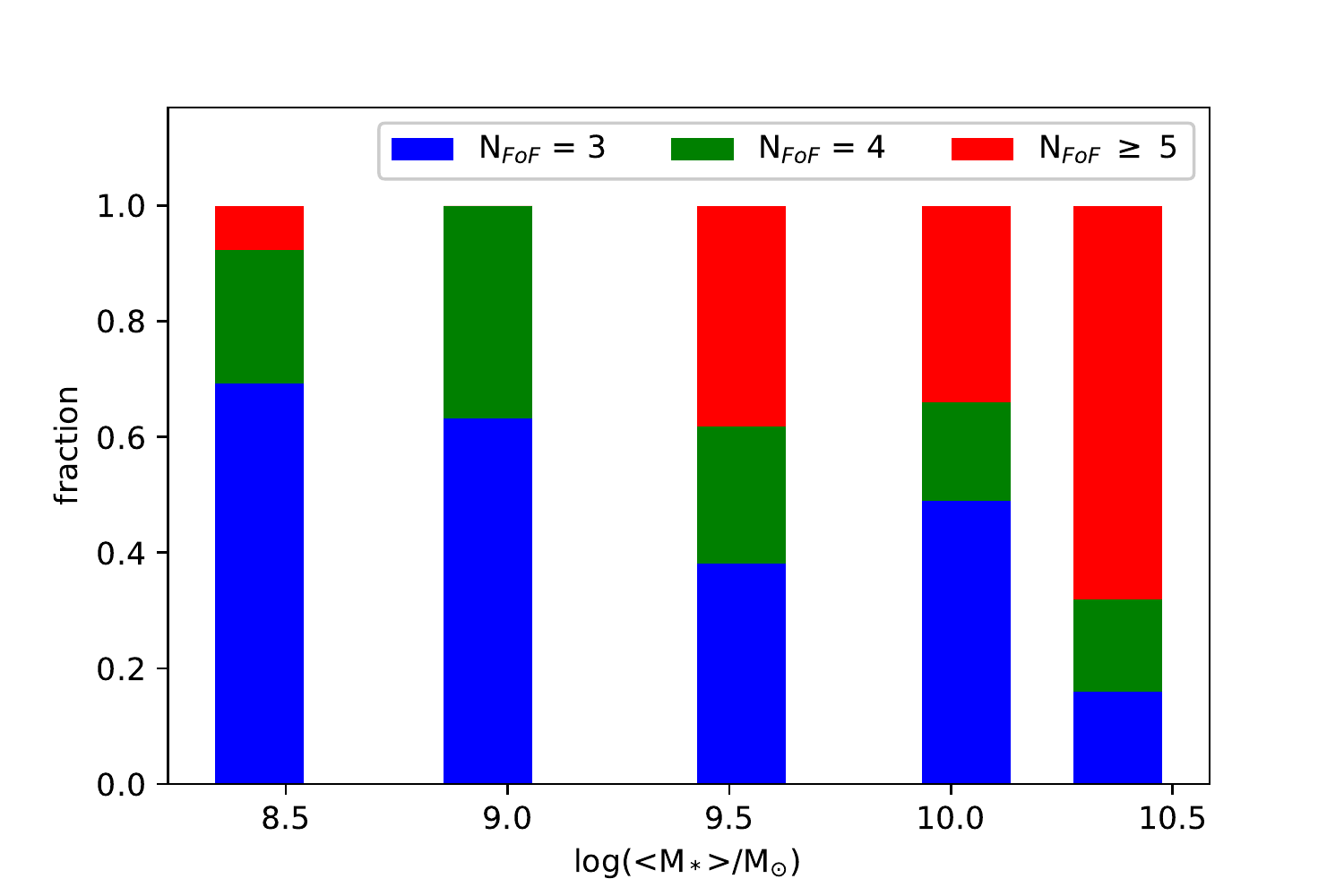}
\caption{Stacked histograms showing the fraction of groups with different number of member galaxies, for the five regions along the SFMS described in Section~\ref{sec:stack2}.}
\label{fig:histNFoF}
\end{figure}

\begin{figure}
\begin{center}
\end{center}
\includegraphics[width=3.5truein]{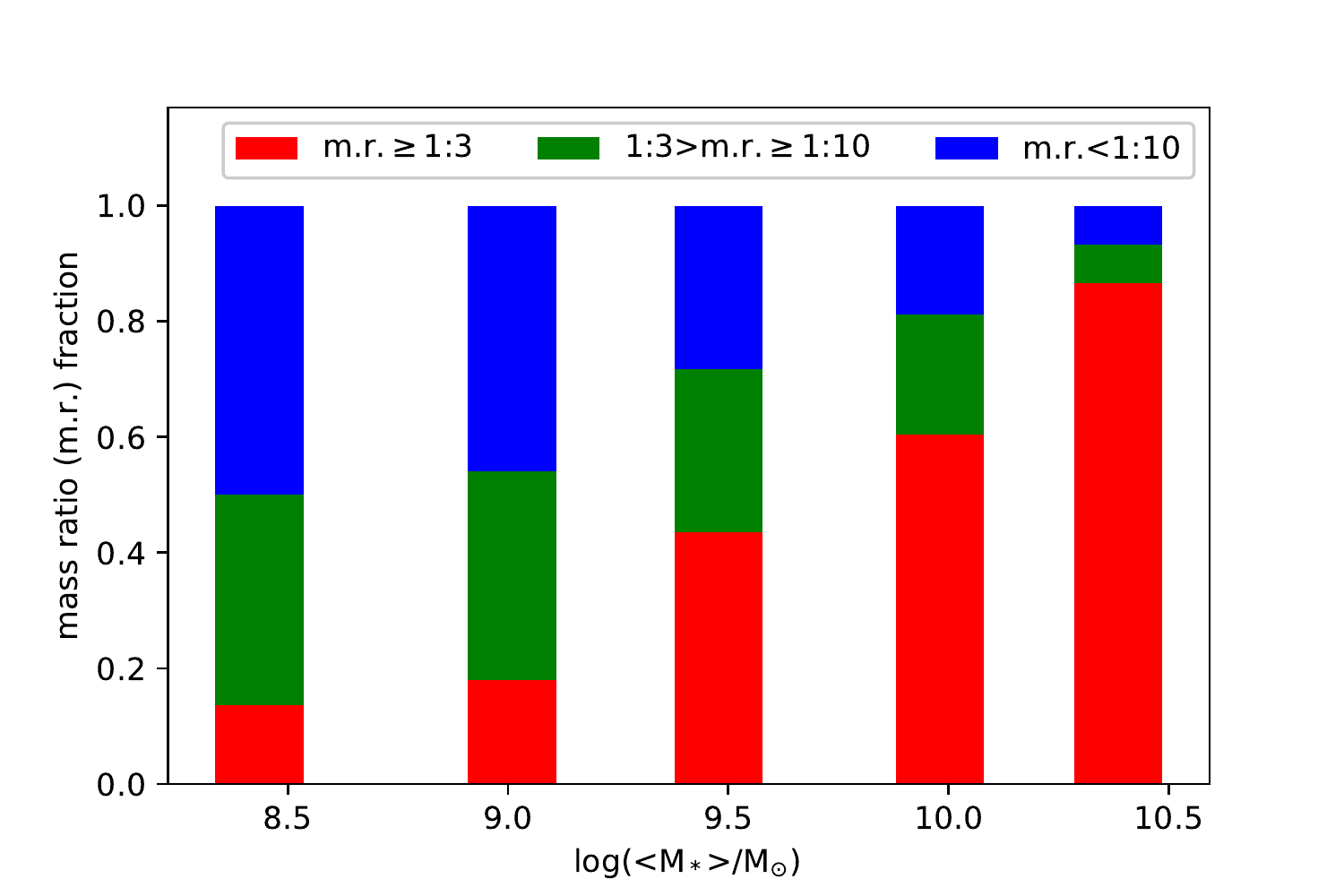}
\caption{Stacked histograms showing the fraction of pairs in which the ratio of the stellar masses of the companion galaxy to that of the base galaxy is $\geq 1:3$ (major merger candidate) or $< 1:3$ (minor merger candidate), with the latter further sub-divided into two categories having the ratio $\geq 1:10$, or $< 1:10$, for the five regions along the SFMS described in Section~\ref{sec:stack2}.}
\label{fig:histMR}
\end{figure}

Next we want to confirm with higher statistical significance the three important trends that emerge for star-forming groups/pairs/isolated galaxies in Section~\ref{sec:stack1}, viz., (i) decreasing $\langle f_{HI} \rangle$ with increasing $\langle$\mst$\rangle$, (ii) similar $\langle f_{HI} \rangle$ for the higher $\langle$\mst$\rangle$ regions of all three categories, and (iii) increasing $\langle f_{HI} \rangle$ in the lower $\langle$\mst$\rangle$ regions as we go from isolated galaxies$\rightarrow$pairs$\rightarrow$groups.
In order to do so, we divide the log($\langle$\mst$\rangle$) -- log($\langle$SFR$\rangle$) plane into five distinct regions along the SFMS, as marked and named in Fig.~\ref{fig:mssfrms}, and subsequently stack the \hi\ gas fraction in all galaxies/pairs/groups within each such region.
These regions have been defined so as to cover the densest population of groups and pairs in the log($\langle$\mst$\rangle$) -- log($\langle$SFR$\rangle$) plane along the SFMS. 
In the vertical direction, the region boundaries are parallel to the SFMS and encompass 2 dex, from 0.9 dex above the SFMS to 1.1 dex below it (see region definitions in Section~\ref{sec:stack1}).
Based on our results in Section~\ref{sec:stack1}, this would imply that for each region we are averaging over units with higher (above SFMS) and lower (below SFMS) gas fractions.
We need to do this as the regions defined for this section span much smaller ranges in log($\langle$\mst$\rangle$) compared to those in Section~\ref{sec:stack1}, and we need to have sufficient number of groups and pairs within each of our regions for detection, as our aim in this section is to specifically study the variation of $f_{HI}$ along the SFMS.
The regions MS-R1 through MS-R5 are centred on log($\langle$\mst$\rangle$) of 8.25, 9.0, 9.5, 10.0, and 10.5 respectively.
The median and standard deviation of the log($\langle$\mst$\rangle$) and log($\langle$SFR$\rangle$) value for the units in these regions are listed in Table~\ref{tab:msreg}.
Note that from this point we are ignoring the quenched regime (i.e. region R5 in Section~\ref{sec:stack1}), as for groups/pairs/isolated galaxies we do not find any detectable \hi\ emission in this regime.

We follow the same stacking procedure for units of all three categories (groups/pairs/isolated galaxies) contained in either region as outlined in Section~\ref{sec:stack1}.
The measured $\langle f_{HI} \rangle$ with errors (using bootstrapping) are tabulated in Table~\ref{tab:msreg}, and the stacked spectra are shown in Figure Set 17 in the Appendix.
With respect to the velocity widths over which the $\langle f_{HI} \rangle$ for each of the regions are determined, in general we see similar overall trends as in Section~\ref{sec:stack1}: they become larger as we move up the SFMS, and as we move from isolated galaxies$\rightarrow$pairs$\rightarrow$groups.
Again as in Section~\ref{sec:stack1}, these trends are not clearly established possibly due to the moderate significance of some of the stacked detections.

We plot the variation of the measured $\langle f_{HI} \rangle$ with the median log($\langle$\mst$\rangle$) of each region for the three categories in Fig.~\ref{fig:fhic}, given that the regions follow the SFMS.
Overall, we find a clear trend of $f_{HI}$ decreasing with \mst\ -- be it the galaxy stellar mass for isolated galaxies or the average stellar mass of a galaxy in a group or pair.
Also plotted are in Fig.~\ref{fig:fhic} are the $\langle f_{HI} \rangle$ values measured within different \mst\ bins for individual galaxies from the extended {\it GALEX} Arecibo SDSS Survey \citep[xGASS, data from][]{2018MNRAS.476..875C}, and the Nancay Interstellar Baryon Legacy Extragalactic Survey \citep[NIBLES, data from][]{2019MNRAS.487.4901H}.
We note that the measurements from both xGASS and NIBLES match those from isolated galaxies.
Thanks to our GAMA-based sample, in principle we can extend our study to \mst s as low as $10^6~M_{\odot}$ for isolated galaxies, but choose a higher \mst\ as the lower limit for our study to restrict ourselves to a region of the log($\langle$\mst$\rangle$) -- log($\langle$SFR$\rangle$) plane which has groups and pairs too (see Fig.~\ref{fig:mssfrms}).

More importantly, from Fig.~\ref{fig:fhic} and Table~\ref{tab:msreg} we find that at all (average) stellar masses except the highest $\langle$\mst$\rangle$ region/bin, the $\langle f_{HI} \rangle$ for groups is higher than that for an isolated galaxy, or galaxies from the xGASS survey.
The $\langle f_{HI} \rangle$ for pairs is intermediate to that for groups or isolated galaxies, but closer to the value for isolated galaxies in all regions/bins except for the lowest $\langle$\mst$\rangle$ one.
The difference in $\langle f_{HI} \rangle$ for groups with that for an isolated galaxy progressively decreases with increasing (average) stellar mass, till they become comparable for the highest (average) stellar mass region/bin.
The difference in $\langle f_{HI} \rangle$ values between groups (and pairs, see paragraph below) and isolated galaxies are further investigated in  Section~\ref{sec:pilot}, and what the observed trends might imply is discussed in Section~\ref{sec:dis} below.

The $\langle f_{HI} \rangle$ for pairs shows a peculiar behaviour.
For the lowest log($\langle$\mst$\rangle$) region/bin, the $\langle f_{HI} \rangle$ value for pairs is consistent with that for groups, but for all the other bins the $\langle f_{HI} \rangle$ value for pairs is consistent with that for isolated galaxies.
As we can see from Fig.~\ref{fig:histNFoF}, for the lowest log($\langle$\mst$\rangle$) region/bin, most of the groups have n$=$3 members, which implies pairs and groups are very similar entities for this particular region/bin. 
From region MS-R3 onwards for higher log($\langle$\mst$\rangle$) regions/bins, there are more groups with n$>$3 members, making the category `pairs' closer to isolated galaxies.
The above might explain why the $\langle f_{HI} \rangle$ for pairs is comparable to that of groups for the lowest log($\langle$\mst$\rangle$) regions/bins and their $\langle f_{HI} \rangle$ are comparable to those for isolated galaxies in higher log($\langle$\mst$\rangle$) regions/bins, though the $\langle f_{HI} \rangle$ for pairs being similar to that for isolated galaxies in region MS-R2 is harder to explain.
Note that the potential dependence of the $\langle f_{HI} \rangle$ trend for groups seen in Fig.~\ref{fig:fhic} on the number of galaxies in groups in explored further in Section~\ref{ssec:nfof}.

In Fig.~\ref{fig:histMR} we plot the histogram of pairs with different stellar mass ratios of the companion galaxy to that of the base galaxy.
The pairs with ratios $>1:3$ can be considered to be major merger candidates, while those with ratios smaller than this value can be considered to be candidates for minor meger.
We see that for the two lowest log($\langle$\mst$\rangle$) regions, minor merger candidates dominate the fraction of pairs, while the fraction of major merger candidates steadily increase with log($\langle$\mst$\rangle$) and their fraction dominates the highest log($\langle$\mst$\rangle$) regions.
For the two lowest log($\langle$\mst$\rangle$) regions, a substantial fraction of the pairs in fact have ratios less than $1:10$.
Minor mergers are known to be \hi-rich, and the trends seen in Fig.~\ref{fig:histMR} are completely consistent with the variation $\langle f_{HI} \rangle$ for pairs as seen in Fig.~\ref{fig:fhic}.
Nevertheless, the apparent excess of \hi\ for pairs in the lowest log($\langle$\mst$\rangle$) bin possibly has the same origin as that for groups, something we investigate further in the following sections.

\subsection{Analysis using DINGO pilot data from the G15 region}
\label{sec:pilot}

\begin{figure}
\begin{center}
\end{center}
\includegraphics[width=3.5truein]{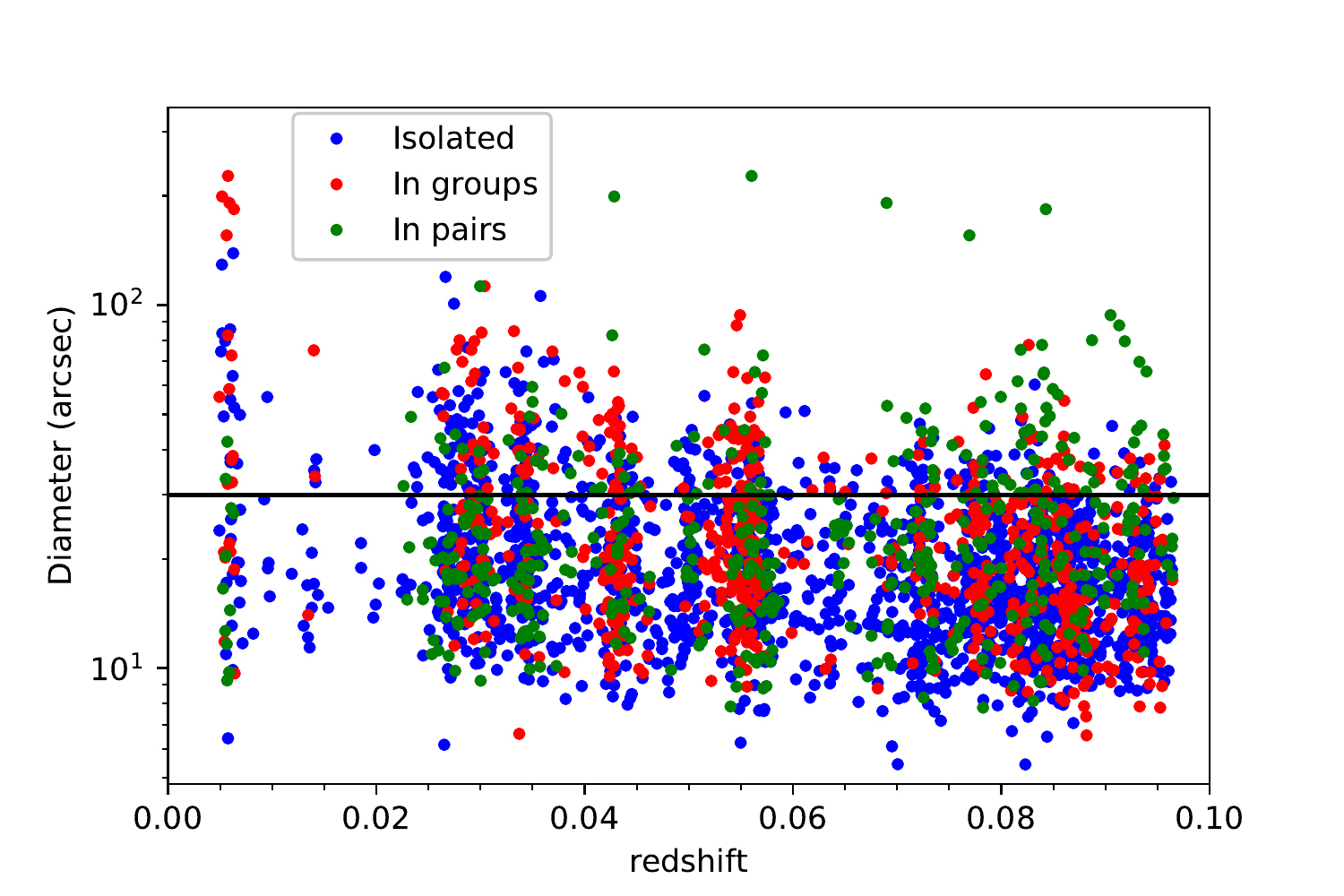}
\caption{The sizes of galaxies in groups and pairs, and isolated galaxies whose spectra are extracted from DINGO pilot data. Note that the galaxy diameters plotted are twice the R100 optical radii values from the latest GAMA galaxy catalog.}
\label{fig:sizeG15}
\end{figure}

\begin{figure*}
\begin{center}
\begin{tabular}{cc}
{\mbox{\includegraphics[width=3.5truein]{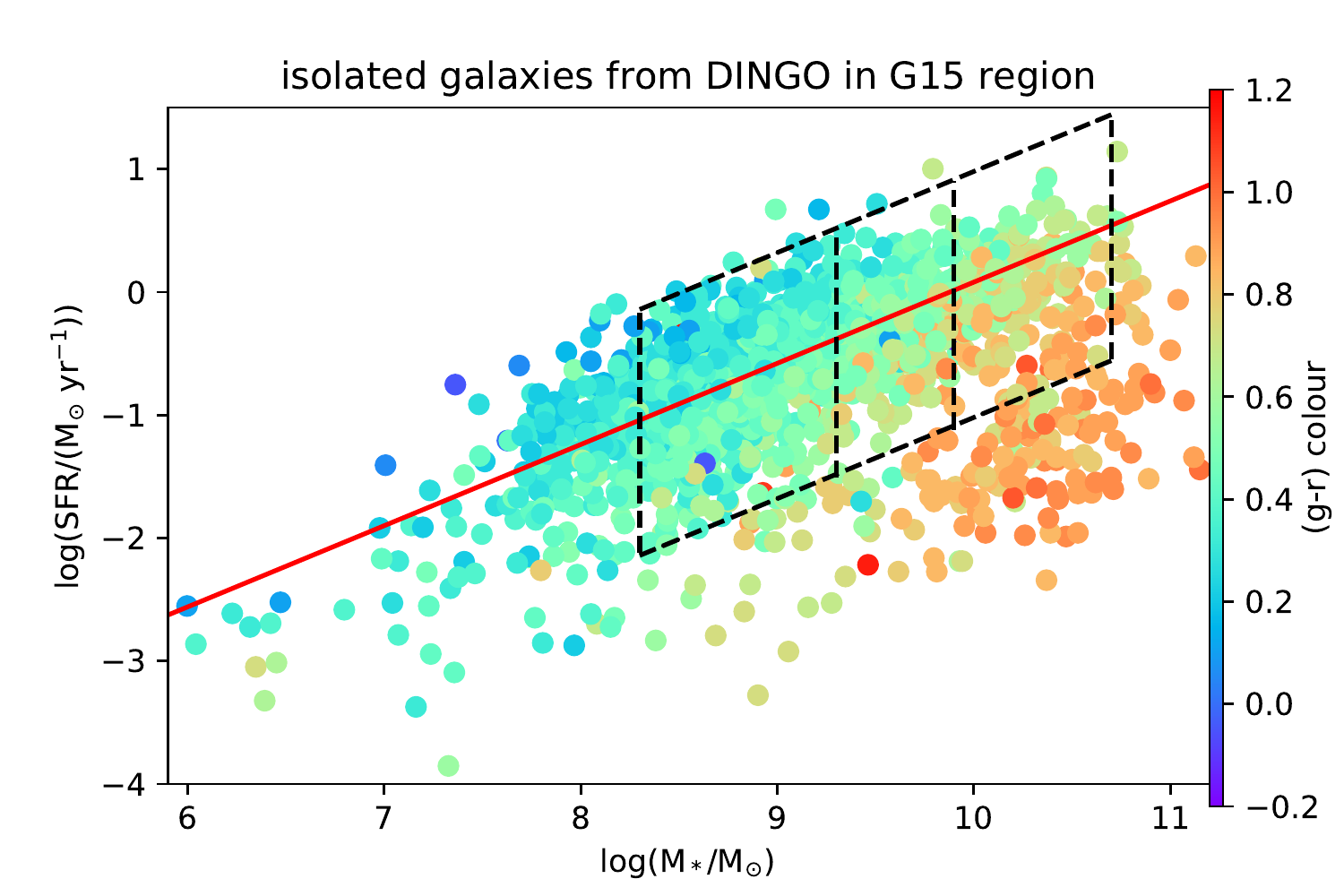}}}&
{\mbox{\includegraphics[width=3.5truein]{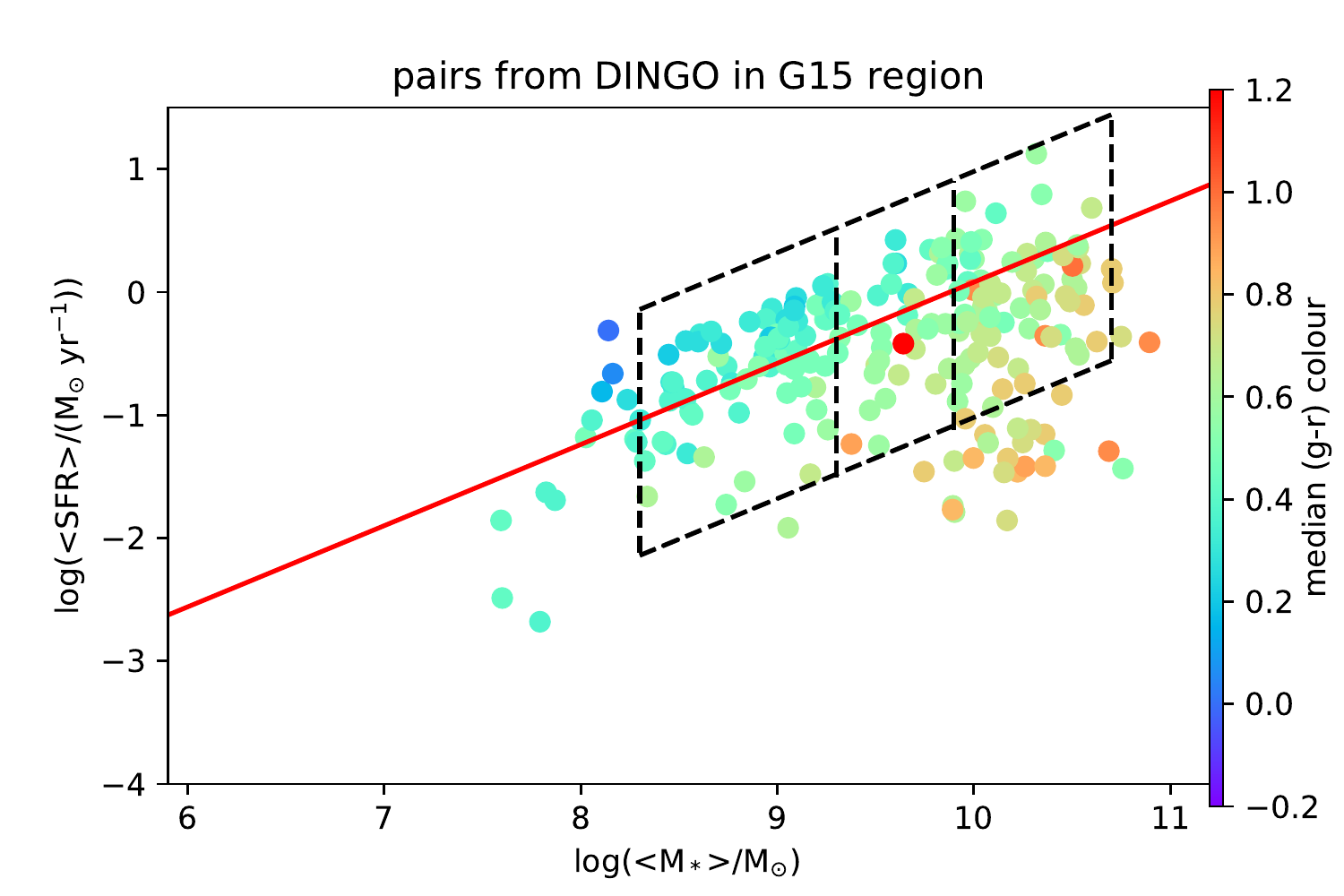}}}\\
{\mbox{\includegraphics[width=3.5truein]{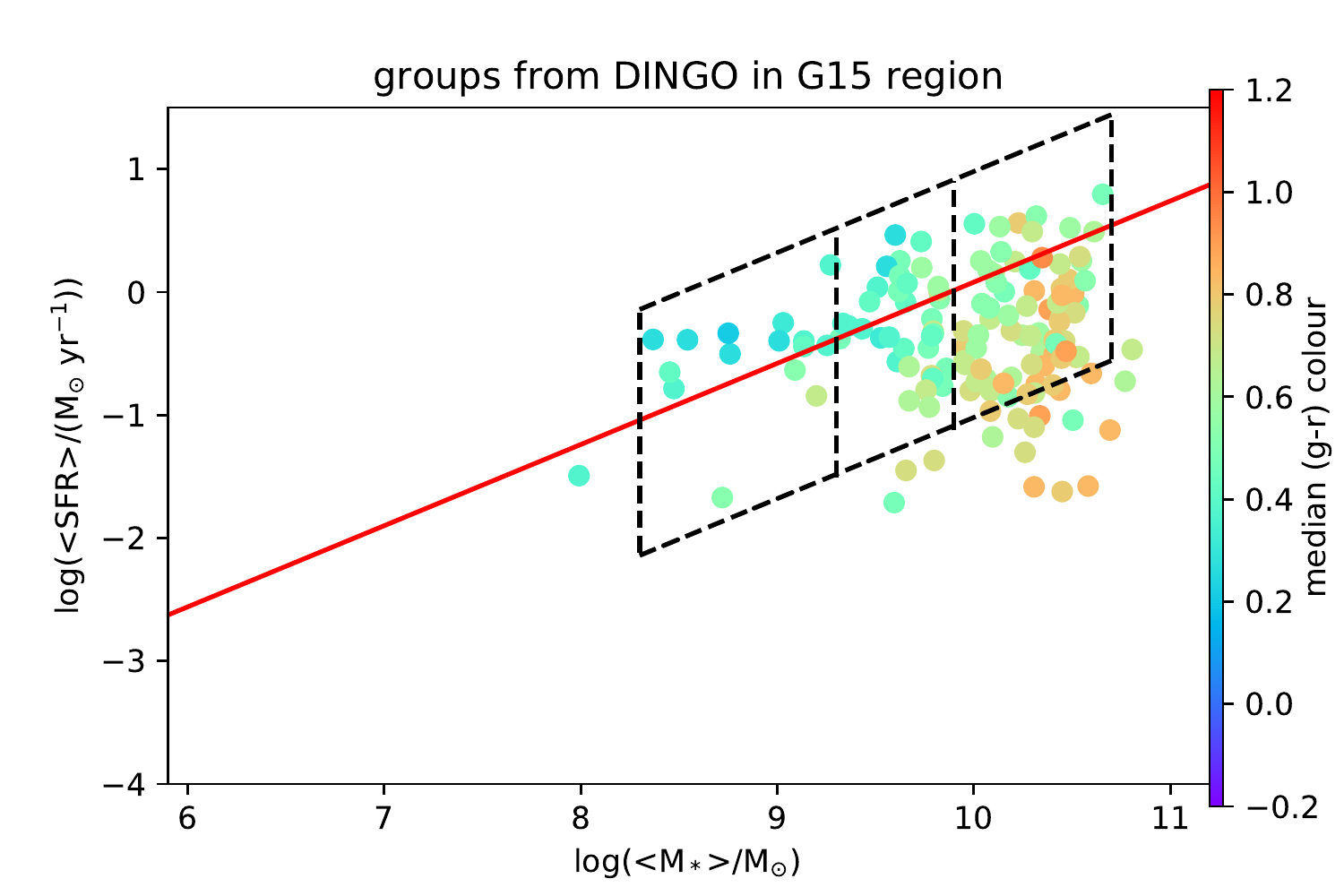}}}&
{\mbox{\includegraphics[width=3.5truein]{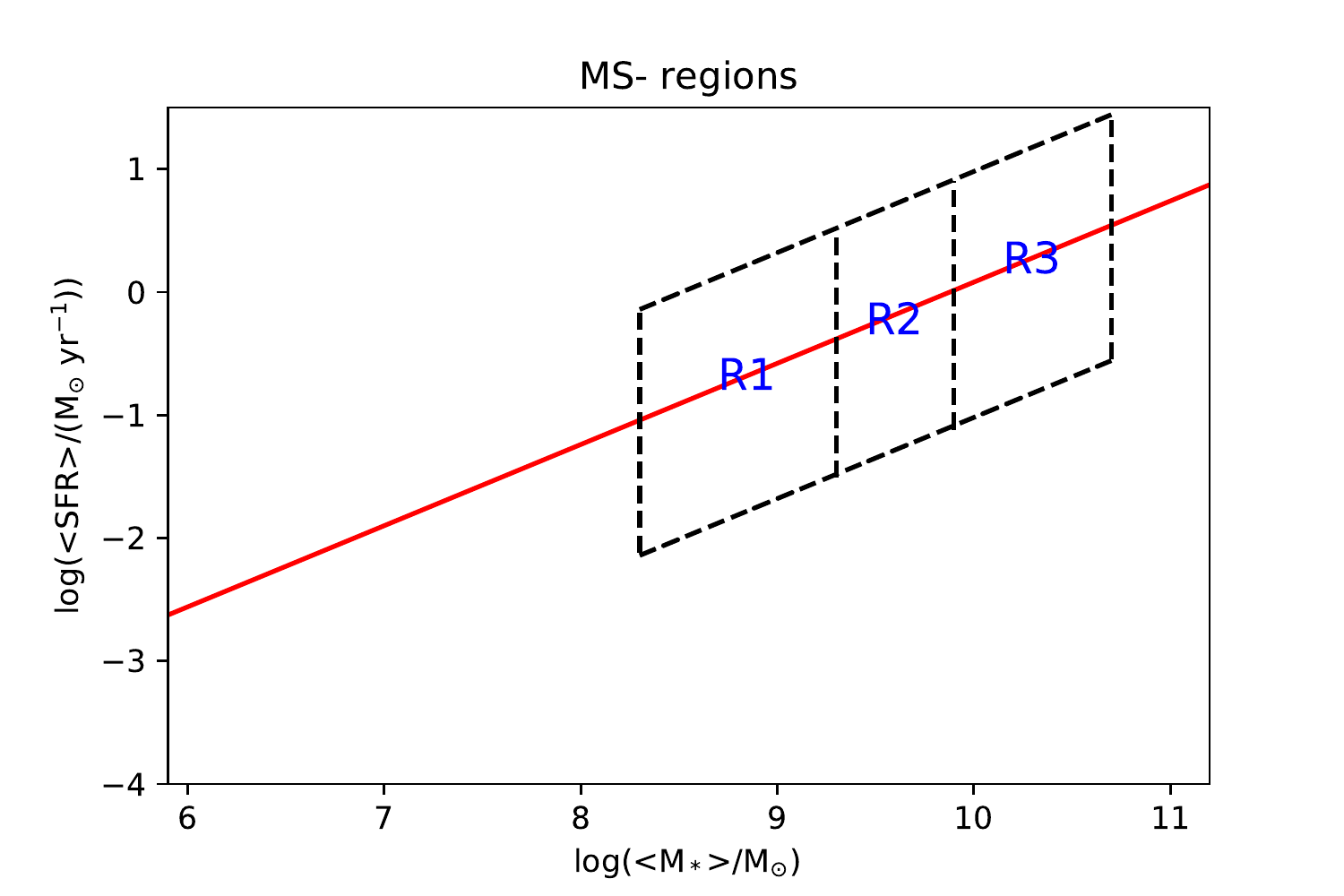}}}\\
\end{tabular}
\end{center}
\caption{log(SFR) plotted against log(\mst) for isolated galaxies from the DINGO G15 pilot survey (top left panel), and log($\langle$SFR$\rangle$) plotted against log($\langle$\mst$\rangle$) for galaxies in pair (top right panel) and group (bottom left panel). The points are colour-coded according to the median g$-$r colour for groups/pairs, and g$-$r colour for isolated galaxies. The red bold line in each panel is the fit to the SFMS for z~$< 0.1$ galaxies from \citet{2016MNRAS.461..458D}. The regions marked in this figure are defined to span the SFMS, and the bottom right panel names the regions.}
\label{fig:G15mssfrms}
\end{figure*}

\begin{table}
\begin{center}
\caption{Measured values of different parameters for units within the three main-sequence regions marked in Fig.~\ref{fig:G15mssfrms} (DINGO G15 pilot survey).
See Table~\ref{tab:5reg} caption for description of the tabulated quantities.}
\label{tab:G15msreg}
\begin{tabular}{lccccc}
\hline
\hline
Unit			& Region MS-R1 			& Region MS-R2 			& Region MS-R3 \\
				& log(${\rm \frac{\langle M_* \rangle}{M_{\odot}}}$)	& log(${\rm \frac{\langle M_* \rangle}{M_{\odot}}}$)	& log(${\rm \frac{\langle M_* \rangle}{M_{\odot}}}$) \\
\hline
Isolated 			& 8.82$\pm$0.27			& 9.55$\pm$0.17			& 10.20$\pm$0.20 \\
Pairs 		 		& 8.97$\pm$0.29			& 9.61$\pm$0.17			& 10.14$\pm$0.21 \\
Groups 		 		& 9.01$\pm$0.30			& 9.67$\pm$0.15			& 10.23$\pm$0.19 \\
\hline         
\hline                           
Unit 			& Region MS-R1 				& Region MS-R2 				& Region MS-R3 \\
				& log(${\rm \frac{\langle SFR \rangle}{M_{\odot} yr^{-1}}}$)	& log(${\rm \frac{\langle SFR \rangle}{M_{\odot} yr^{-1}}}$)	& log(${\rm \frac{\langle SFR \rangle}{M_{\odot} yr^{-1}}}$) \\
\hline
Isolated 			& $-$0.62$\pm$0.44			& $-$0.24$\pm$0.36			& $-$0.01$\pm$0.37 \\
Pairs 				& $-$0.55$\pm$0.42			& $-$0.27$\pm$0.43			& $-$0.04$\pm$0.46 \\
Groups 				& $-$0.43$\pm$0.39			& $-$0.30$\pm$0.36			& $-$0.21$\pm$0.42 \\
\hline    
\hline 
Unit 			& Region MS-R1 			& Region MS-R2 			& Region MS-R3 \\
				& ${\rm \langle \frac{M_{HI}}{M_{*}} \rangle}$	& ${\rm \langle \frac{M_{HI}}{M_{*}} \rangle}$	& ${\rm \langle \frac{M_{HI}}{M_{*}} \rangle}$ \\
\hline
Isolated 			& 0.92$\pm$0.05			& 0.36$\pm$0.03			& 0.13$\pm$0.02	 \\
Pairs 				& 1.02$\pm$0.09			& 0.25$\pm$0.04			& 0.11$\pm$0.02	 \\
Groups 				& 0.72$\pm$0.16			& 0.30$\pm$0.04			& 0.08$\pm$0.01	 \\    
\hline    
\hline   
Unit 			& Region MS-R1 			& Region MS-R2 			& Region MS-R3 \\
				& $\Delta v_{HI,\pm 2 \sigma}$	& $\Delta v_{HI,\pm 2 \sigma}$	& $\Delta v_{HI,\pm 2 \sigma}$\\
				& ${\rm (km~s^{-1})}$    		& ${\rm (km~s^{-1})}$			& ${\rm (km~s^{-1})}$ \\
\hline
Isolated			& 238				& 352				& 536 \\
Pairs				& 239				& 333				& 528 \\
Groups				& 170				& 335				& 441 \\
\hline
\hline
\end{tabular}
\end{center}
\end{table}

\begin{figure*}
\begin{center}
\includegraphics[width=5truein]{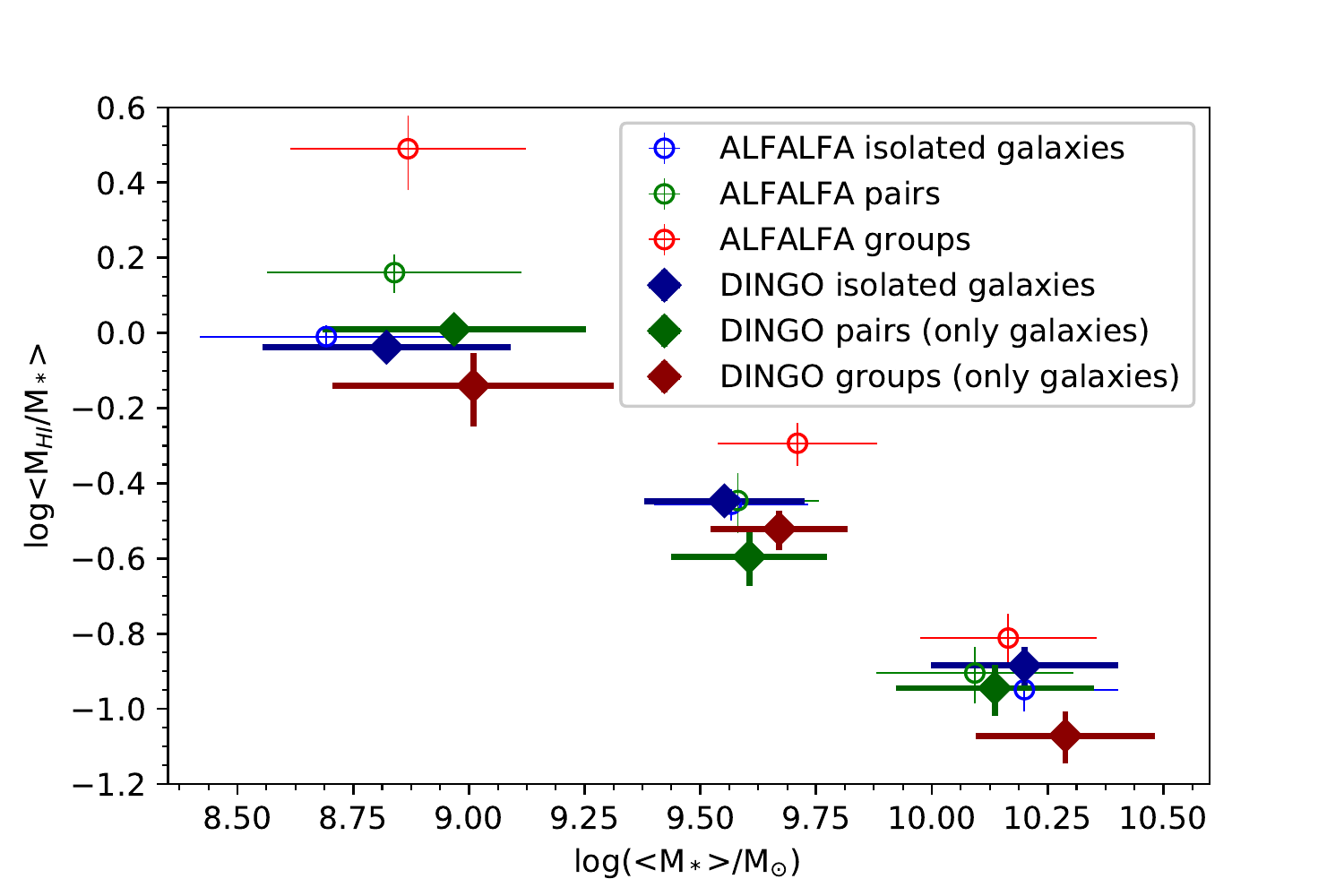}
\end{center}
\caption{The variation of gas fraction with log($\langle$\mst$\rangle$) along the SFMS, and with the unit category from the DINGO G15 pilot survey. The corresponding values from the same regions for the ALFALFA-GAMA overlap sample (Table~\ref{tab:apval} in the Appendix) are also plotted for comparison. The $\langle f_{HI} \rangle$ and the median log($\langle$\mst$\rangle$)s tabulated in Table~\ref{tab:G15msreg} are plotted against each other for the three categories of galaxies as filled symbols, while the values from Table~\ref{tab:apval} are plotted as open symbols.}
\label{fig:G15fhic}
\end{figure*}

\begin{figure}
\begin{center}
\end{center}
\includegraphics[width=3.5truein]{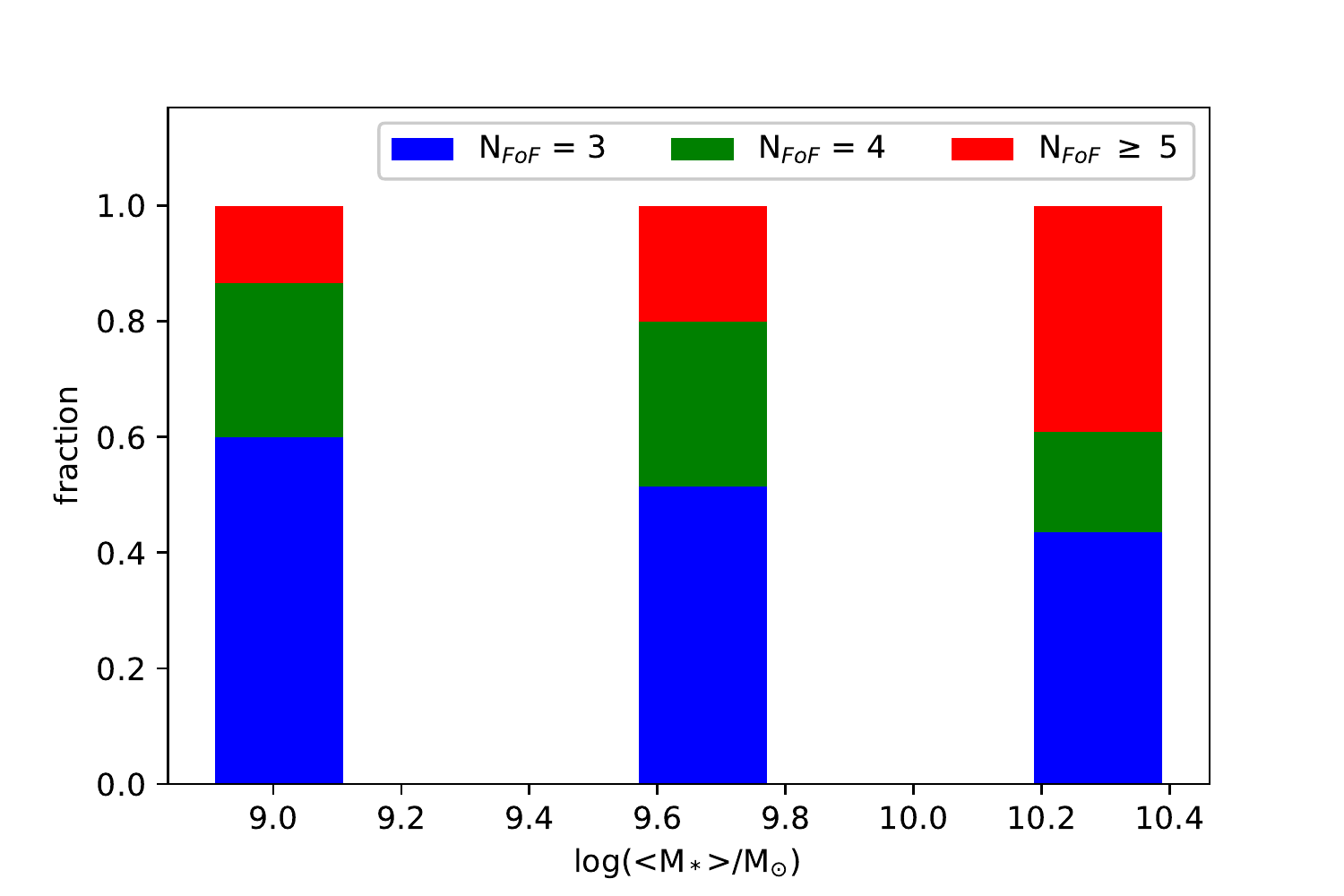}
\caption{Stacked histograms showing the fraction of groups with different number of member galaxies, for the three regions along the SFMS described in Section~\ref{sec:pilot}.}
\label{fig:G15histNFoF}
\end{figure}

\begin{table}
\begin{center}
\caption{The difference in the average \hi\ content of units lying on or around the SFMS on the log($\langle$\mst$\rangle$) -- log($\langle$SFR$\rangle$) plane, for three different log($\langle$\mst$\rangle$) ranges, between measurements from ALFALFA-GAMA and DINGO G15 samples. The difference is expressed as the percentage of the average \hi\ content for the corresponding unit in the DINGO G15 sample. See text for details.}
\label{tab:fhidiff}
\begin{tabular}{cccc}
\hline
\hline
log(${\rm \frac{\langle M_* \rangle}{M_{\odot}}}$)& Groups & Pairs & Isolated \\
range & & & galaxies \\
\hline
8.3--9.3 & 211$^{+179}_{-114}$\% & 5$^{+24}_{-20}$\% & $-$21$^{+11}_{-9}$\% \\
9.3--9.9 & 86$^{+58}_{-45}$\% & 34$^{+52}_{-38}$\% & 2$^{+19}_{-16}$\% \\
9.9--10.7 & 63$^{+48}_{-37}$\% & $-$3$^{+41}_{-29}$\% & $-$15$^{+24}_{-18}$\% \\
\hline
\hline
\end{tabular}
\end{center}
\end{table}

In order to understand the varying trends in $\langle f_{HI} \rangle$ for groups/pairs/isolated galaxies seen in Section~\ref{sec:stack2} it will be very useful to compare those values with the $\langle f_{HI} \rangle$ values when we co-add the spectra from individual galaxies that make up the groups and pairs.
ALFALFA's spatial resolution is far too coarse for us to be able to extract the \hi\ spectra for individual member galaxies of groups and pairs which have member galaxies in close proximity to each other.
Here the \hi\ data from the DINGO pilot Phase I observations of the G15 field described in Section~\ref{sec:dathi} becomes a valuable addition to our study, and this is the dataset we use for this part of the study.
The sample with available \mst\ and SFR from GAMA includes 140 groups, 223 pairs, and 1719 isolated galaxies.
Note that this data is $\sim$7 times better in terms of spatial resolution compared to the ALFALFA \hi\ data. 
At z~$= 0.1$ the angular size of the DINGO pilot survey beam corresponds to spatial size of 56 kpc, and midway at  z~$= 0.05$ the spatial size is 29 kpc.
As mentioned in Section~\ref{sec:dathi} though, when smoothed to the same spatial resolutions the DINGO pilot data we are using is much less sensitive than ALFALFA data.
Consequently we do not get any detectable emission when we stack the \hi\ spectra extracted over entire areas covered by groups/pairs aka Section~\ref{sec:exalfa}.
Thus this part of the study is exclusively based on spectra extracted in the direction of individual galaxies in group/pair/isolated environments, see Fig.~\ref{fig:sizeG15} for a comparison of the sizes of individual galaxies compared with the FWHM of the DINGO pilot \hi\ data.

For all the galaxies covered by the DINGO pilot observations in the G15 region, the redshift from the latest GAMA optical galaxy catalog is used to ascertain the central frequency of the \hi\ spectrum for that galaxy.
For each galaxy we extract a spectrum by summing over the flux density over a certain area for each channel spanning $\pm$1500 \kms\ on either side of the central frequency.
The spatial area over which the flux density is summed over is a circle whose radius is determined by summing in quadrature the half-width-at-half-maxima of the DINGO beam ($\sim$15\arcs) and galaxy's radius as defined by the R100 optical radius value from the latest GAMA galaxy catalog.
Note that we use the R100 radius as a conservative choice so as not to miss any \hi\ flux, as \hi\ disk of galaxies, particularly star-forming galaxies, can be much more extended than their optical disks \citep{2002A&A...390..829S,2005A&A...442..137N,2008MNRAS.386.1667B}.

For each of our galaxy spectra, we measure the root RMS noise in flux density units (Jy) at `line-free' frequencies, i.e. channels not within 300 \kms\ on either side of the central frequency of the \hi\ emission.
The flux density spectra were converted to mass spectra, the frequency axis of all individual spectra were shifted to the rest frame using the redshift of the respective galaxy, and the mass spectra normalized to conserve the total mass during this frequency shift, following Equations~\ref{eqn:mhi},~\ref{eqn:nu}, and \ref{eqn:massc} described in Section~\ref{sec:stack1}. 
All individual spectra are re-gridded on to a uniformly spaced frequency axis with channel separation of 18.6 KHz. 
At this point, for groups and pairs the \hi\ mass spectra of constituent galaxies are co-added together (without applying any weights), and divided by the total \mst\ of the group/pair (considering the stellar masses of the constituent galaxies whose spectra were coadded) in order to convert them to $f_{HI}$ spectra for groups/pairs as units.
The combined weight for each such group/pair spectrum is $\sqrt{\sum_{j}~\frac{1}{{\sigma_j}^2}}$ where $\sigma_j$ is the RMS noise in flux density units (Jy) at `line-free' frequencies measured previously for the $j$th galaxy in the group/pair.
For isolated galaxies, the re-gridded mass spectra are simply divided by their stellar masses to convert to $f_{HI}$ spectra, with a weight of $\frac{1}{\sigma}$ where $\sigma$ is the RMS noise in flux density units (Jy) at `line-free' frequencies measured previously for the corresponding isolated galaxy.

In this section our aim is to further explore the trends seen in measured $\langle f_{HI} \rangle$ in Section~\ref{sec:stack2}. 
In order to do so, we divide the log($\langle$\mst$\rangle$) -- log($\langle$SFR$\rangle$) plane into three distinct regions along the SFMS, as marked and named in Fig.~\ref{fig:G15mssfrms}, and subsequently stack the \hi\ gas fraction in all galaxies/pairs/groups within each such region.
These regions have been defined so as to cover the densest population of groups and pairs in the log($\langle$\mst$\rangle$) -- log($\langle$SFR$\rangle$) plane along the SFMS as in Section~\ref{sec:stack2}, but the number of regions have been reduced owing to the smaller number of galaxies/pairs/groups covered due to the limited sky coverage of the DINGO pilot G15 data.
The regions MS-R1 through MS-R3 are defined to be centred on log($\langle$\mst$\rangle$) of 8.8, 9.6 and 10.3 respectively.
In the vertical direction, the region boundaries are parallel to the SFMS and encompass 2 dex, from 0.9 dex above the SFMS to 1.1 dex below it.
The median and standard deviation of the log($\langle$\mst$\rangle$) and log($\langle$SFR$\rangle$) value for the units in these regions are listed in Table~\ref{tab:G15msreg}.
Note that again we are ignoring the quenched regime, as in region R5 from Section~\ref{sec:stack1} we do not find any detectable \hi\ emission for groups/pairs/isolated galaxies.

We follow the flux-weighted stacking procedure for groups/pairs/isolated galaxies contained in each region as outlined in Section~\ref{sec:stack1}, using the weights as described above where groups/pairs weights are determined slightly differently given that for them individual member galaxy spectra are co-added first in order to create a single group/pair spectrum.
Given that we do not observe any residual baseline effects in the continuum-subtracted \hi\ datacubes from the DINGO pilot observations, we do away with subtracting a second order polynomial fit to the stacked spectra as is done for the stacks described in Sections~\ref{sec:stack1} and \ref{sec:stack2}.
The measured $\langle f_{HI} \rangle$ with errors (using bootstrapping) are tabulated in Table~\ref{tab:G15msreg}, and the stacked spectra are shown in  Figure Set 17 in the Appendix.
From Table~\ref{tab:G15msreg} wee see two general trends with respect to the velocity widths over which the $\langle f_{HI} \rangle$ for each of the regions are determined.
The first trend, which is similar to trends as in Sections~\ref{sec:stack1} and \ref{sec:stack2}, is that the velocity widths become larger as we move up the SFMS.
But the second trend is actually opposite to what is seen in Sections~\ref{sec:stack1} and \ref{sec:stack2}: the velocity widths become smaller as we move from isolated galaxies$\rightarrow$groups, while once again the values for pairs are not easy to fit in to a general trend.
The difference in how we stack in this section compared to Sections~\ref{sec:stack1} and \ref{sec:stack2} is that in this section we align the spectra of all member galaxies in a group/pair at their central frequencies. 
Therefore the consistent trend in measured velocity widths of the stacked spectra we move from isolated galaxies$\rightarrow$group implies that isolated galaxies have larger velocity widths than galaxies in groups (and pairs at least for the two higher \mst\ regions).
We emphasize that as in Sections~\ref{sec:stack1} and \ref{sec:stack2}, the trends in velocity widths are not clearly established possibly due to the moderate significance of some of the stacked detections.

In order to better compare the measured $\langle f_{HI} \rangle$ with those for the ALFALFA-GAMA sample, we measure $\langle f_{HI} \rangle$ for units in the ALFALFA-GAMA sample in the same three regions as used here for the DINGO G15 sample.
The stacking procedure is exactly as described in Section~\ref{sec:stack1}.
The measured $\langle f_{HI} \rangle$ with errors (using bootstrapping) are tabulated in Table~\ref{tab:apval} in the Appendix, while the stacked spectra are shown in Figure Set 17 in the Appendix.

We plot the variation of the measured $\langle f_{HI} \rangle$ with the median log($\langle$\mst$\rangle$) of each region for the three categories in Fig.~\ref{fig:G15fhic}, along with the values for the ALFALFA-GAMA overlap sample for the same three regions as described in the previous paragraph.
We see that the trend of $\langle f_{HI} \rangle$/$f_{HI}$ decreasing with $\langle$\mst$\rangle$/\mst\ is reproduced in a consistent manner by the DINGO pilot observations, especially considering isolated galaxies from both the samples and the values from xGASS discussed in the context of Fig.~\ref{fig:fhic} in Section~\ref{sec:stack2}.
The more interesting finding from Fig.~\ref{fig:G15fhic} is that when considering the spectra taken at the positions of member galaxies of groups/pairs only as in this section, the  $\langle f_{HI} \rangle$ for groups and pairs is consistent with that for isolated galaxies at similar $\langle$\mst$\rangle$ s.
Unlike for the ALFALFA-GAMA overlap sample we do not find the  $\langle f_{HI} \rangle$ for groups (and pairs) to be higher than that for isolated galaxies, not even in the lowest $\langle$\mst$\rangle$ bin where this difference is considerable for the ALFALFA-GAMA overlap sample.
In Fig.~\ref{fig:G15histNFoF} we also confirm that in the lowest \mst\ bin the distribution of group memberships is very similar to the low \mst\ bins of the ALFALFA-GAMA overlap sample (Fig.~\ref{fig:histNFoF}).
If anything it appears that the $\langle f_{HI} \rangle$ for isolated galaxies are slightly larger than those for group galaxies at all \mst s from the stacking done using DINGO G15 data, though these differences are not statistically significant.

What these results signify are further elaborated in Section~\ref{sec:dis} below.
In Table~\ref{tab:fhidiff} though we have quantified the observed difference between the results when using the ALFALFA-GAMA data or DINGO G15 data.
The excess average \hi\ mass detected using the ALFALFA-GAMA data in different $\langle$\mst$\rangle$ bins is tabulated as a percentage of the average \hi\ mass measured using DINGO G15 data.
The average \hi\ masses in each in are calculated by multiplying the $\langle f_{HI} \rangle$ of the corresponding bin with the median $\langle$\mst$\rangle$ in the same bin, and the errors are from the corresponding errors in our $\langle f_{HI} \rangle$ measurements.
For groups, we detect a clear excess in \hi\ mass when comparing our results from ALFALFA-GAMA data to those from DINGO G15 data.
For the lowest $\langle$\mst$\rangle$ bin, the excess is consistent with being twice the amount detected when stacking on the member galaxies only.
Note that the clear excess in $\langle f_{HI} \rangle$ measured for pairs compared to isolated galaxies in the lowest $\langle$\mst$\rangle$ bin used in Section~\ref{sec:stack2} is not reflected in our results in this section, as for the DINGO G15 data we exclude most of the low $\langle$\mst$\rangle$ region covered by the lowest $\langle$\mst$\rangle$ bin used in Section~\ref{sec:stack2}.

\section{Discussion}
\label{sec:dis}

Following are the main deductions that can be made based on the results presented in Section~\ref{sec:anares}.

First, in the regime where the average stellar mass of a unit is low (log$(\frac{\langle M_* \rangle}{M_{\odot}}) \lesssim$ 9.5), the average \hi\ per unit \mst\ increases as one moves to denser environments, i.e. pairs and then groups, compared to isolated galaxies.
This result is derived using ALFALFA data which, given its large beam, for groups and pairs covers both the identified member galaxies as well as the intergalactic space between them.
When considering cumulative (co-added without weighting) stacks of galaxies in groups/pairs though, using high spatial resolution \hi\ data from DINGO pilot observations, the \hi\ per unit \mst\ becomes consistent in this regime for groups/pairs/isolated galaxies.
Therefore the excess \hi\ in pair and group environments in this low $\langle$\mst$\rangle$ regime cannot be associated to the member galaxies themselves, but must be in the intergalactic space between the already identified member galaxies either as dark non-star forming clouds or associated with faint galaxies lying below the GAMA detection limit.

Second, at higher $\langle$\mst$\rangle$, galaxies whether isolated or in pairs and groups have very similar gas fractions, which are also lower that the respective $\langle f_{HI} \rangle$ values in the low $\langle$\mst$\rangle$ regime.
This result is unchanged whether the \hi\ spectra for groups/pairs are extracted over the entire group/pair are or whether they are arrived at by co-adding \hi\ spectra from individual member galaxies.
Our results are consistent e.g. with a study of absorbers in groups with $\langle$\mst$\rangle$s in this regime by \citet{2019ApJS..240...15S}, which showed that warm and cool gas detected in the intra-group medium of such groups is neither massive nor volume-filling.
What is interesting is that even though the $\langle$\mst$\rangle$ values, having similar $\langle f_{HI} \rangle$ values, are comparable for the three categories, in reality their total stellar masses occupy vastly different ranges. 
Regarding how different the masses of the halos that the units of each category reside in are, please see Sections~\ref{ssec:halo} and \ref{sec:sim} below.
The lower $\langle f_{HI} \rangle$ values compared to the low $\langle$\mst$\rangle$ regime can be understood in terms of redder and more early type galaxies becoming prevalent in the higher $\langle$\mst$\rangle$ regime.
For example, \citet{2018ApJ...864...40P} showed that $f_{HI}$ varies with galaxy type, with the \hi\ content increasing with T-type and the dispersion in \hi\ content at a given \mst\ also becoming narrower with increasing T-type.
Ours being an averaged result though, we are insensitive to the variation in gas consumption and decrease in star-formation efficiency between late type galaxies of low and high mass, as they move from the high gas fraction to the moderate gas fraction regime \citep[e.g. see][]{2013MNRAS.433L..35L}.
The relevant part of our result though, is that in terms of these galaxies residing in different environments, this result is reminiscent of the findings from \citet{2015MNRAS.451.3249A}: when controlled for mass distribution, galaxies in different environments have similar properties. 
It is also noteworthy that the stellar mass cutoff for the galaxies used in their study was log$(\frac{M_*}{M_{\odot}}) \sim$ 9.5.

Third, when considering spectra for groups/pairs measured by co-adding \hi\ spectra from individual member galaxies, we find isolated galaxies to have higher \hi\ per unit \mst\ compared to galaxies in groups (and pairs for higher \mst s).
The velocity widths of isolated galaxies also appear to be larger than galaxies in groups at all \mst s.
Therefore isolated galaxies must be more gas rich than an average group (or pair) galaxy of the same stellar mass.
Though this particular trend is not statistically significant in our study, it has been noticed in earlier studies \citep[e.g.][]{2013A&A...550A.115H}, and in a stacking study using deeper DINGO Early Science observations we can clearly see that isolated galaxies have higher $\langle f_{HI} \rangle$ compared to galaxies in groups, and within groups themselves group satellites have higher $\langle f_{HI} \rangle$ compared to group centrals (Rhee et al., in preparation).
These results are consistent with the finding from \citet{2016MNRAS.455.4013D} that passive fractions of galaxies at all stellar masses is higher in interacting pairs and groups.

Fourth, as both the $\langle$\mst$\rangle$ and $\langle$SFR$\rangle$ for units move below the SFMS (which in turn is defined based on observations of individual galaxies residing in all kinds of environments) in the log($\langle$\mst$\rangle$) -- log($\langle$SFR$\rangle$) plane, their gas fraction also decreases.
We therefore find that the $\langle$\mst$\rangle$ and $\langle$SFR$\rangle$ values for pairs and groups show a similar evolutionary trend as normal galaxies, ending with the pair or group having almost no gas when its $\langle$\mst$\rangle$ and $\langle$SFR$\rangle$ values lies in the regime of galaxies undergoing quenching -- given that there is still measurable star formation in the units, this region is what is referred to as the `green valley'.
Note that the measured gas fractions are near zero on account of little or no \hi\ being detected and not due to any inordinate increase in the total stellar masses of units in this regime.
Therefore, on average galaxies in a group or pair are likely to be depleted of gas if the group or pair has {\it average} \mst\ and {\it average} SFR values in a certain regime, and this fact does not seem to de dictated by environmental factors or by the halo mass as isolated galaxies with \mst\ and SFR in the same regime are also depleted of gas.
What is interesting about this result is that there is evidence that environment does play a role in galaxy quenching, at least though mechanisms that are correlated with the group potential \citep[e.g.][]{2017MNRAS.469.3670S}.
Again, what we show here is an averaged result and we are insensitive to the details of the various quenching pathways through which galaxies traverse the green valley, pathways many recent studies are showing differ according to their morphologies \citep[][]{2015MNRAS.450..435S,2018MNRAS.476...12B,2018MNRAS.477.4116K}, or depending on whether they are satellites or centrals \citep[e.g.][]{2019MNRAS.483.5444D}.

Taken together, the observed trends are consistent with a scenario where in groups/pairs comprising of low \mst\ constituents there is a lot of excess gas in the intergalactic space, even though the galaxies making up the groups/pairs themselves on average have less gas per unit stellar mass compared to isolated galaxies with \mst\ similar to the $\langle$\mst$\rangle$ of the group/pair.
This extra \hi\ in the intergalactic space seems not to be present as the average stellar mass of the group/pair increases.
Whether the excess gas in the intergalactic space exists as dark non-star forming clouds, tidal debries, etc. is an open question.
There is of course the possibility of this \hi\ being associated with faint dwarf galaxies undetected by GAMA.
As stated earlier, the stellar mass completeness limit varies from just below $10^7 M_{\odot}$ at z~$= 0$ to $\sim 10^8 M_{\odot}$ at z~$= 0.06$.
Given the substantial excess of \hi\ within low $\langle$\mst$\rangle$ groups and pairs (Table~\ref{tab:fhidiff}), such faint undetected galaxies will have to be heavily gas dominated almost-dark galaxies with little star formation, as they are also not detected at wavelengths/bands tracing SFR.
Let us try to estimate how many such undetected galaxies are needed to account for the observed excess \hi.
Using the results presented in Tables~\ref{tab:G15msreg}, \ref{tab:apval}, and \ref{tab:fhidiff}, let us consider a hypothetical group of galaxies with multiplicity $3$ and $\langle$\mst$\rangle~=~10^{8.8}~M_{\odot}$.
The galaxies themselves have a total \mhi$~=~0.92 \times \ 3 \times 10^{8.8}~M_{\odot}~=~1.74 \times 10^{9}~M_{\odot}$.
The total mass of \hi\ in the group though is \mhi$~=~3.09 \times \ 3 \times 10^{8.8}~M_{\odot}~=~5.85 \times 10^{9}~M_{\odot}$
Assuming that all the undetected galaxies just avoided detection by GAMA and have a stellar mass of $10^{6.5} M_{\odot}$, the extrapolated average $f_{HI}$ at this \mst\ from the scaling relation determined by \citet{2011MNRAS.411..993F} \citep[which was shown to be valid for much lower stellar masses in][]{2019MNRAS.487.4901H} is $\sim$120.
\citet{2019MNRAS.487.4901H} show that the scaling relations seems to hold even at \mst$~=~10^{8.5}~M_{\odot}$, but whether the scaling relation would continue two further orders of magnitude below is another open question.
Nevertheless, given this extravagantly high gas fraction, we will need around $11$ such gas-rich dwarfs with \mst$~=~10^{6.5} M_{\odot}$ to be interspersed between the three group galaxies.
The corresponding number for \mst$~=~10^{6} M_{\odot}$ galaxies is $\sim 14$. 
Such a scenario is therefore definitely not beyond the realms of possibility.

Whether the gas in the intergalactic space of low $\langle$\mst$\rangle$ groups is in faint undetected galaxies or not, evidence has been accumulating which indicates the presence of substantial amounts of gas in the intergalactic space of groups, low multiplicity groups in particular.
\citet{2010ApJ...710..385B,2015ApJ...812...78B} found significant amounts of diffuse \hi\ in Hickson Compact groups by comparing single dish observations with interferometric observations, sometimes as high as 80\%. 
They concluded that the diffuse gas was likely of tidal origin.
Multiple recent searches for galaxies around the sightlines to quasars at the redshifts of intervening absorption by dense \hi\ (Damped Lyman-$\alpha$ systems, MgII absorbers, as well as Lyman limit systems, at redshifts ranging from z~$= 0.19$ to z~$= 1.5$), have found the absorbers to be associated with diffuse gas in between galaxies which are in groups or pairs \citep{2019MNRAS.485.1595P,2020MNRAS.492.2347H,2020MNRAS.499.5022D,2021MNRAS.505..738N,2021arXiv210910875W}.
As for the possibility of presence of molecular gas associated with this excess \hi\ in the intergalactic space, a recent study of CO emission in host galaxies of ${\ensuremath {\rm H_2}}$ absorption selected systems at intermediate redshifts (ranging from z~$= 0.19$ to z~$=1.15$) interestingly also finds that most of the molecular gas resides in dense gas pockets within the intergalactic medium of groups of galaxies \citep{2021MNRAS.506..514K}. 

There is a strong plausibility that the $\langle$\mst$\rangle$ -- $\langle$SFR$\rangle$ plane is capturing the evolution not only of galaxies but also of entire units like groups/pairs, given how groups/pairs isolated galaxies occupy overlapping regions of the plane including the region along the SFMS, and how their (median) g$-$r colours become redder as one moves towards higher $\langle$\mst$\rangle$ regime along the SFMS and then towards lower $\langle$SFR$\rangle$ (i.e. green valley) regime (Figs.~\ref{fig:sfrms}, \ref{fig:mssfrms} and \ref{fig:G15mssfrms}).
At the time of assembly of groups \hi\ in infalling dwarfs, heated by previous epochs of vigorous star formation, can be removed by the groups's coronal hot gas even from the very outskirts of the group \citep{2011ApJ...732...17N}.
\citet{2013ApJ...775...97N} with the help of the above model reproduced the higher observed fraction of gas rich dwarfs, and them being distributed closer to the group centre, in the M81 group compared to the Local Group \citep{2016ApJ...824..151G}, when they considered that the M81 group started accreting $\sim$2 Gyr later than the Local Group. 
The stripped \hi\ in recently assembled groups might well explain the excess \hi\ we find in groups in the low $\langle$\mst$\rangle$ regime.
 
In the sections that follow, we explore what effects hidden factors not explicitly considered in our analysis yet might have on our results, and whether the above picture can be reconciled with predictions from simulations which utilize our present understanding of galaxy formation and evolution.

\subsection{The effect of group multiplicity}
\label{ssec:nfof}

\begin{figure}
\begin{center}
\begin{tabular}{c}
{\mbox{\includegraphics[width=3.5truein]{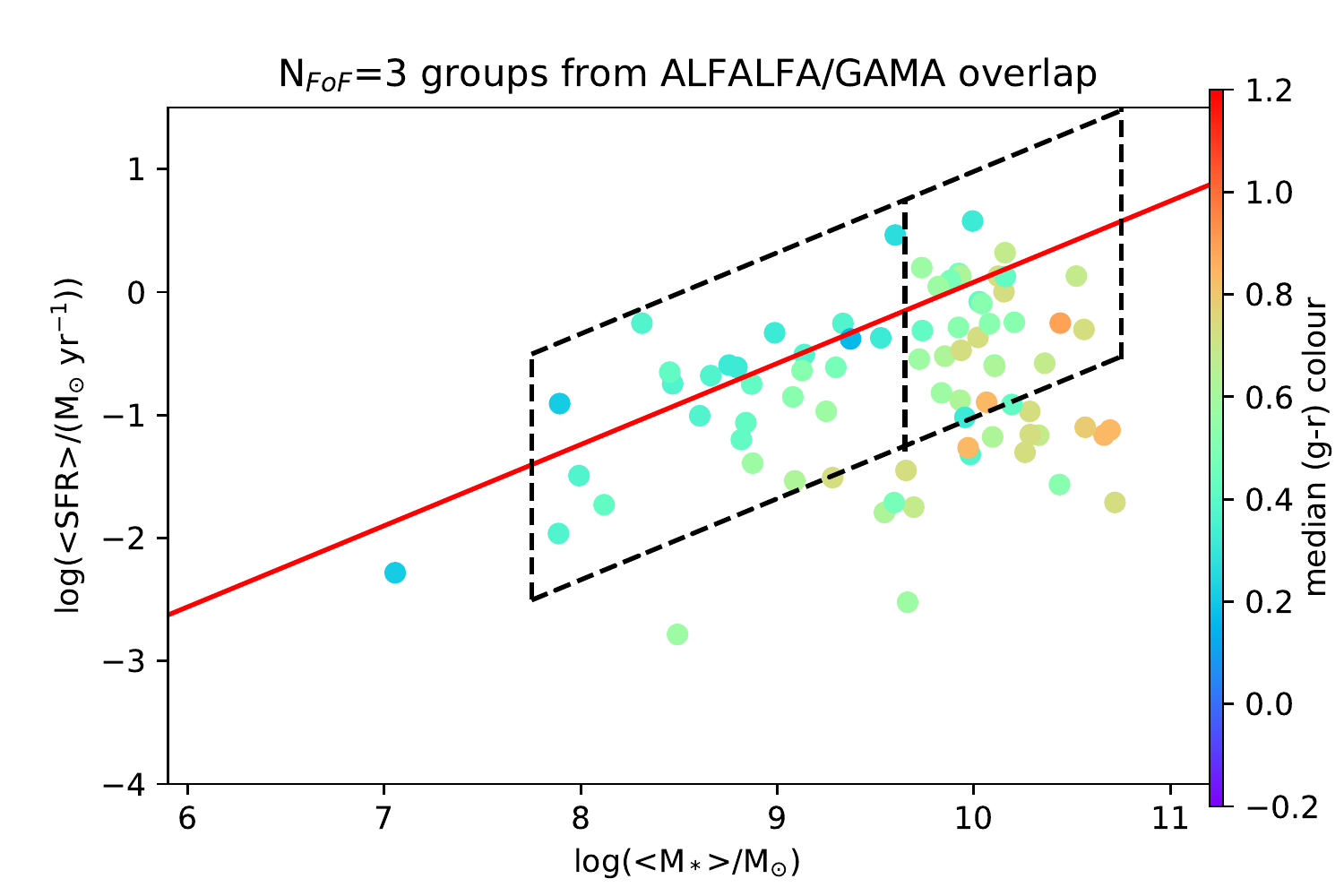}}}\\
{\mbox{\includegraphics[width=3.5truein]{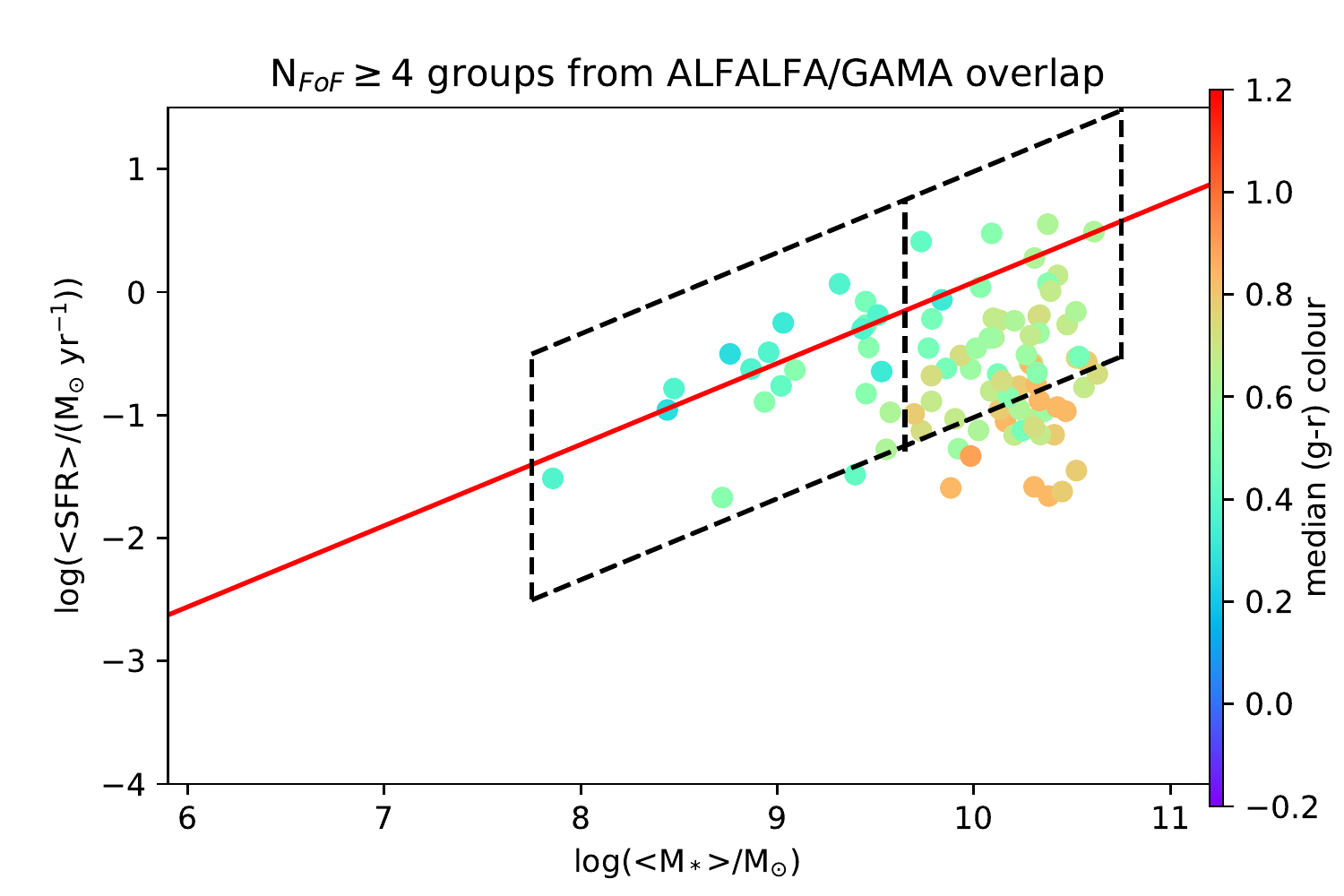}}}\\
{\mbox{\includegraphics[width=3.5truein]{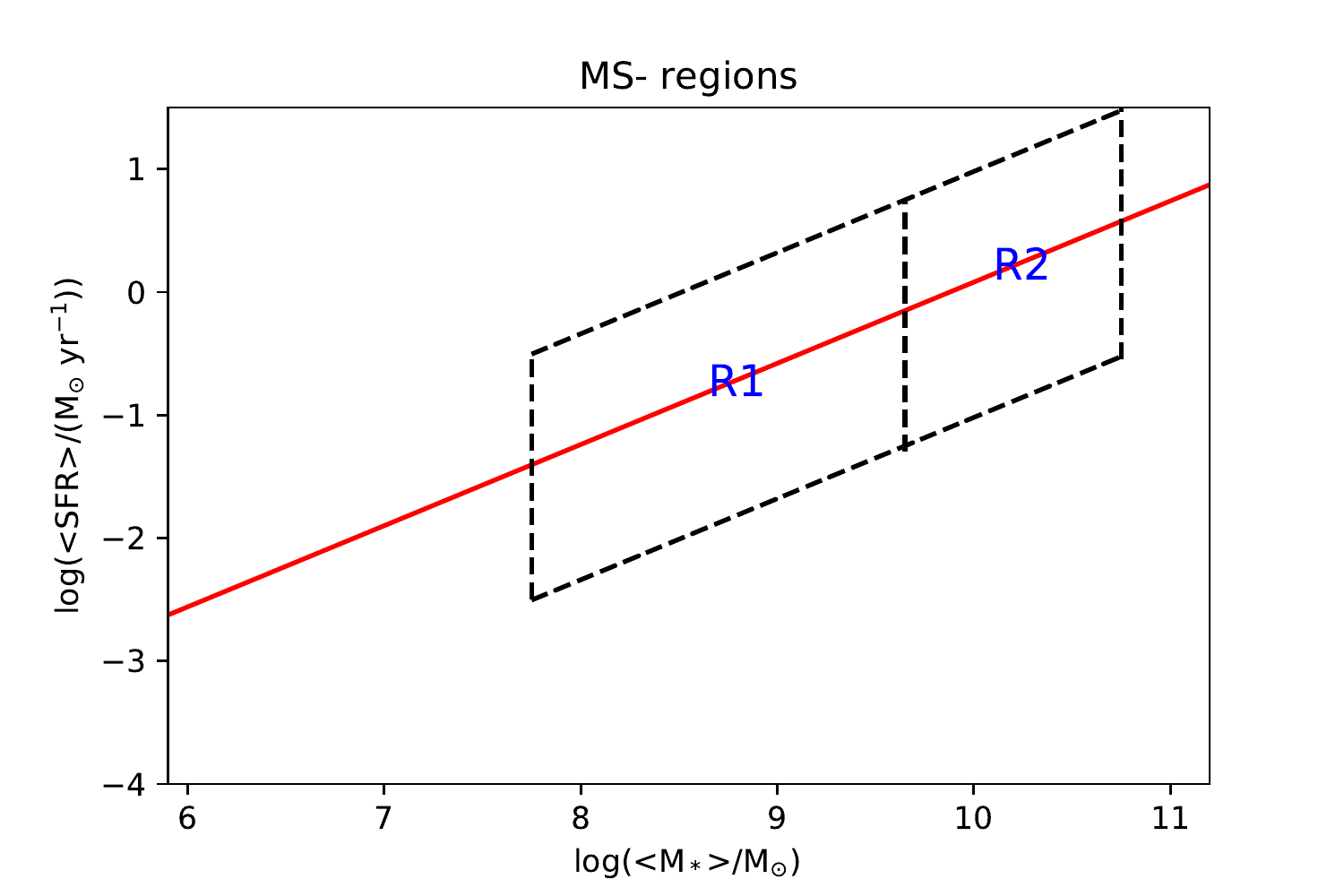}}}\\
\end{tabular}
\end{center}
\caption{log(SFR) plotted against log(\mst) from the ALFALFA--GAMA overlap sample for the two different categories of groups defined by their number of constituent galaxies (top two panels), colour coding same as in Fig.~\ref{fig:sfrms}. The regions marked in this figure are defined to span the SFMS, and the bottom panel names the regions.}
\label{fig:NFoFmssfrms}
\end{figure}

\begin{table}
\begin{center}
\caption{Measured values of different parameters for the two different categories of groups defined by their number of constituent galaxies, within the two main-sequence regions marked in Fig.~\ref{fig:NFoFmssfrms}. 
See Table~\ref{tab:5reg} caption for description of the tabulated quantities.}
\label{tab:NFoFmsreg}
\begin{tabular}{lccccc}
\hline
\hline
Unit			& Region MS-R1 			& Region MS-R2  \\
				& log(${\rm \frac{\langle M_* \rangle}{M_{\odot}}}$)	& log(${\rm \frac{\langle M_* \rangle}{M_{\odot}}}$) \\
\hline
Groups (N$_{FoF}=$3)	& 8.86$\pm$0.48			& 10.02$\pm$0.21 \\
Groups (N$_{FoF} \ge$4)	& 9.09$\pm$0.45			& 10.19$\pm$0.25 \\
\hline         
\hline                           
Unit 			& Region MS-R1 				& Region MS-R2 \\
				& log(${\rm \frac{\langle SFR \rangle}{M_{\odot} yr^{-1}}}$)	& log(${\rm \frac{\langle SFR \rangle}{M_{\odot} yr^{-1}}}$) \\
\hline
Groups (N$_{FoF}=$3) 	& $-$0.71$\pm$0.52			& $-$0.26$\pm$0.39 \\
Groups (N$_{FoF} \ge$4) & $-$0.63$\pm$0.44			& $-$0.48$\pm$0.41 \\
\hline    
\hline 
Unit 			& Region MS-R1 			& Region MS-R2  \\
				& ${\rm \langle \frac{M_{HI}}{M_{*}} \rangle}$	& ${\rm \langle \frac{M_{HI}}{M_{*}} \rangle}$ \\
\hline
Groups (N$_{FoF}=$3)	& 3.01$\pm$0.70			& 0.25$\pm$0.04	 \\  
Groups (N$_{FoF} \ge$4)	& 4.38$\pm$1.93			& 0.17$\pm$0.03	 \\  
\hline    
\hline   
Unit 			& Region MS-R1 			& Region MS-R2 \\
				& $\Delta v_{HI,\pm 2 \sigma}$	& $\Delta v_{HI,\pm 2 \sigma}$ \\
				& ${\rm (km~s^{-1})}$    		& ${\rm (km~s^{-1})}$ \\
\hline
Groups (N$_{FoF}=$3)	& 314				& 514 \\
Groups (N$_{FoF} \ge$4)	& (951)				& 539 \\
\hline
\hline
\end{tabular}
\end{center}
\end{table}

\begin{figure}
\begin{center}
\end{center}
\includegraphics[width=3.5truein]{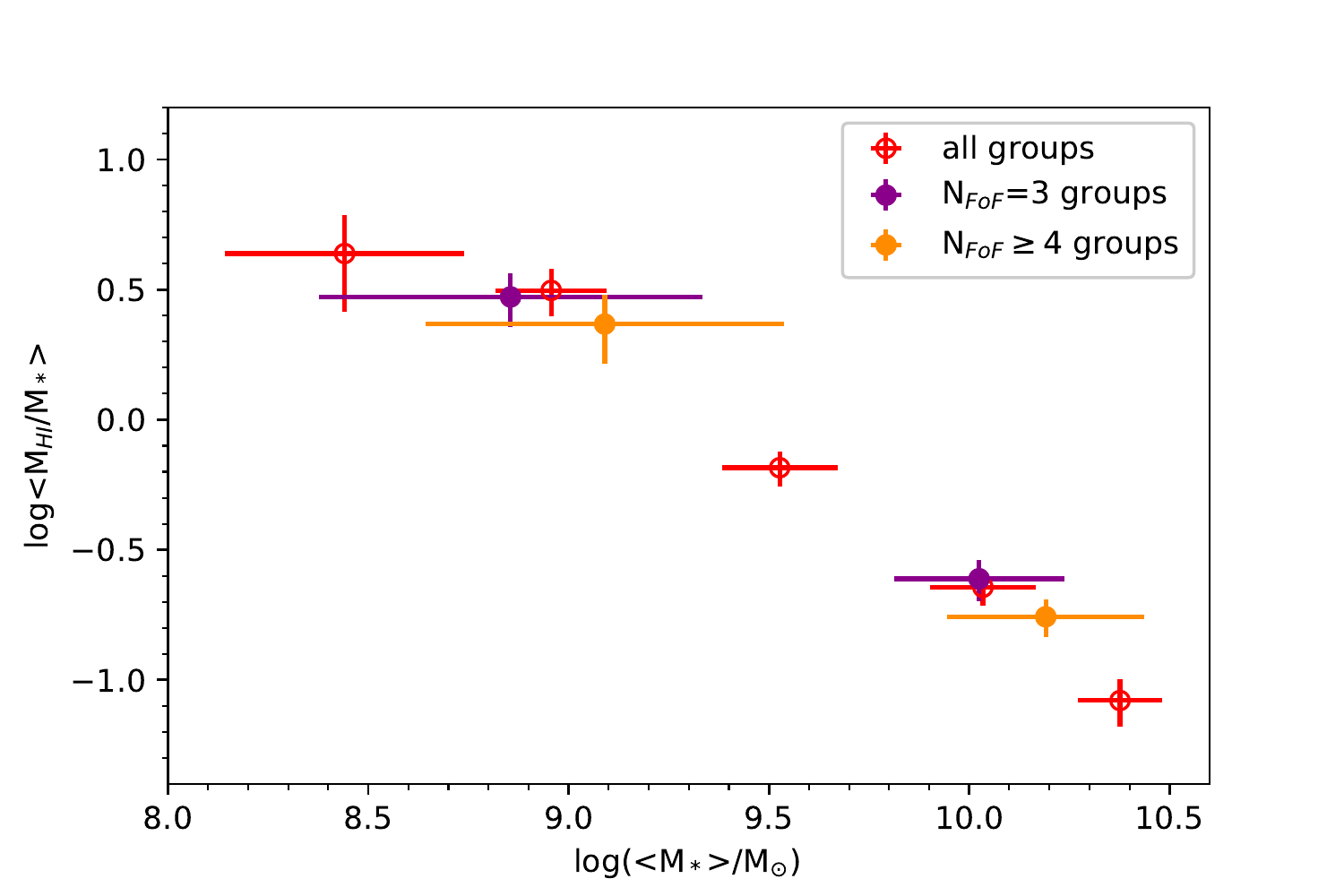}
\caption{The variation of gas fraction with log($\langle$\mst$\rangle$) along the SFMS as measured using ALFALFA spectra, for the two different categories of groups defined by their number of constituent galaxies. The corresponding values for groups from Figure~\ref{fig:fhic} are also plotted for comparison. The $\langle f_{HI} \rangle$ and the median log($\langle$\mst$\rangle$)s tabulated in Table~\ref{tab:NFoFmsreg} are plotted against each other as filled symbols, while the values for groups from Table~\ref{tab:msreg} are plotted as open symbols.}
\label{fig:fhicNFoF}
\end{figure}

The result from Section~\ref{sec:stack2} contained in Table~\ref{tab:msreg} and Fig.~\ref{fig:fhic}, that the $\langle f_{HI} \rangle$ for a group is higher than that for an isolated galaxy in the low $\langle$\mst$\rangle$ regime might potentially be biased by the fact that in the log$(\frac{\langle M_* \rangle}{M_{\odot}}) \lesssim$ 9.5 regime the groups are dominated by groups with only three members (Fig.~\ref{fig:histNFoF}).
Therefore we re-derive the results for groups from Section~\ref{sec:stack2} by dividing the groups into two parts: ones with 3 member galaxies, and others with 4 or more member galaxies.
We have to keep the limit for dividing our groups sample into two parts at N$_{FoF}~=~$3, as we hardly have any groups with 5 or more members in the low $\langle$\mst$\rangle$ regime.
Note that N$_{FoF}$ refers to the number of members in a group as deduced using the Friend-of-Friends algorithm used to create the GAMA Group catalog.

We divide the log($\langle$\mst$\rangle$) -- log($\langle$SFR$\rangle$) plane into two distinct regions along the SFMS, as marked and named in Fig.~\ref{fig:NFoFmssfrms}, and subsequently stack the \hi\ gas fraction in the two categories of groups within each such region.
In the vertical direction, the region boundaries are parallel to the SFMS and encompass 2 dex, from 0.9 dex above the SFMS to 1.1 dex below it.
The regions MS-R1 and MS-R2 are defined to be centred on log($\langle$\mst$\rangle$) of 8.7 and 9.2 respectively.
The median and standard deviation of the log($\langle$\mst$\rangle$) and log($\langle$SFR$\rangle$) value for the two categories of groups in these regions are listed in Table~\ref{tab:NFoFmsreg}.
We follow the same flux-weighted stacking procedure for the two categories of groups contained in each region as outlined in Section~\ref{sec:stack1}.
The measured $\langle f_{HI} \rangle$ with errors (using bootstrapping) are tabulated in Table~\ref{tab:NFoFmsreg}, and the stacked spectra are shown in Figure Set 17 in the Appendix.
We plot the variation of the measured $\langle f_{HI} \rangle$ with the median log($\langle$\mst$\rangle$) of each region for the two categories of groups in Fig.~\ref{fig:fhicNFoF}, along with the values for all groups from Fig.~\ref{fig:fhic}.
We find that, within errors, the $\langle f_{HI} \rangle$ measurements for N$_{FoF}~=~$3 and N$_{FoF}~\geq~$4 groups are consistent with each other.
In fact, it appears that the increase in $\langle f_{HI} \rangle$ as we move from higher $\langle$\mst$\rangle$ to lower $\langle$\mst$\rangle$ is even more pronounced for N$_{FoF}~\geq~$4 groups as compared to that for N$_{FoF}~=~$3 groups.
Therefore even though groups are dominated by N$_{FoF}~=~$3 groups at lower $\langle$\mst$\rangle$, there is no evidence that the much higher $\langle f_{HI} \rangle$ in groups in this $\langle$\mst$\rangle$ regime compared to isolated galaxies is linked to the increase in the relative fraction of N$_{FoF}~=~$3 groups.

\subsection{The effect of halo mass}
\label{ssec:halo}

\begin{figure*}
\begin{center}
\begin{tabular}{cc}
{\mbox{\includegraphics[width=3.5truein]{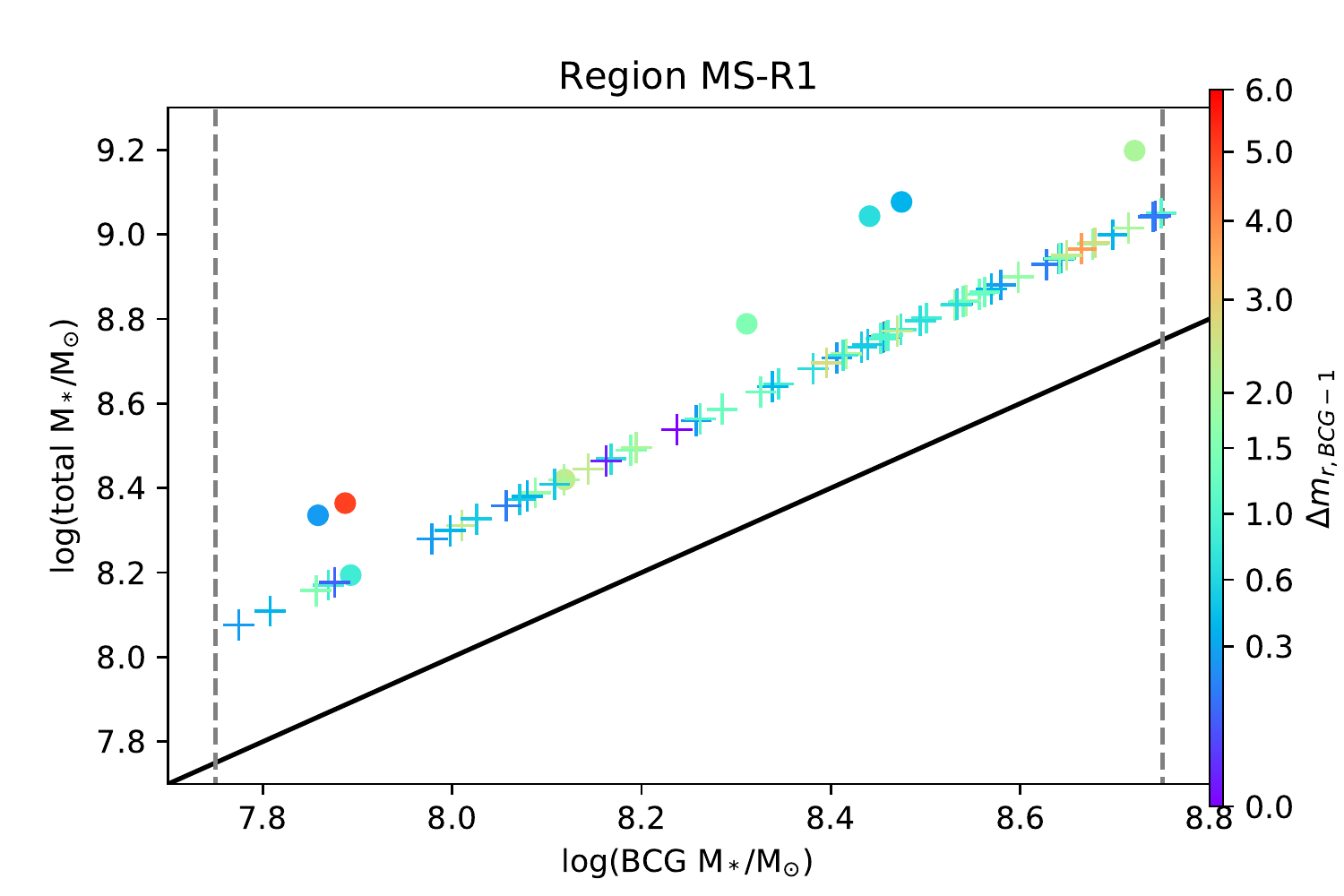}}}&
{\mbox{\includegraphics[width=3.5truein]{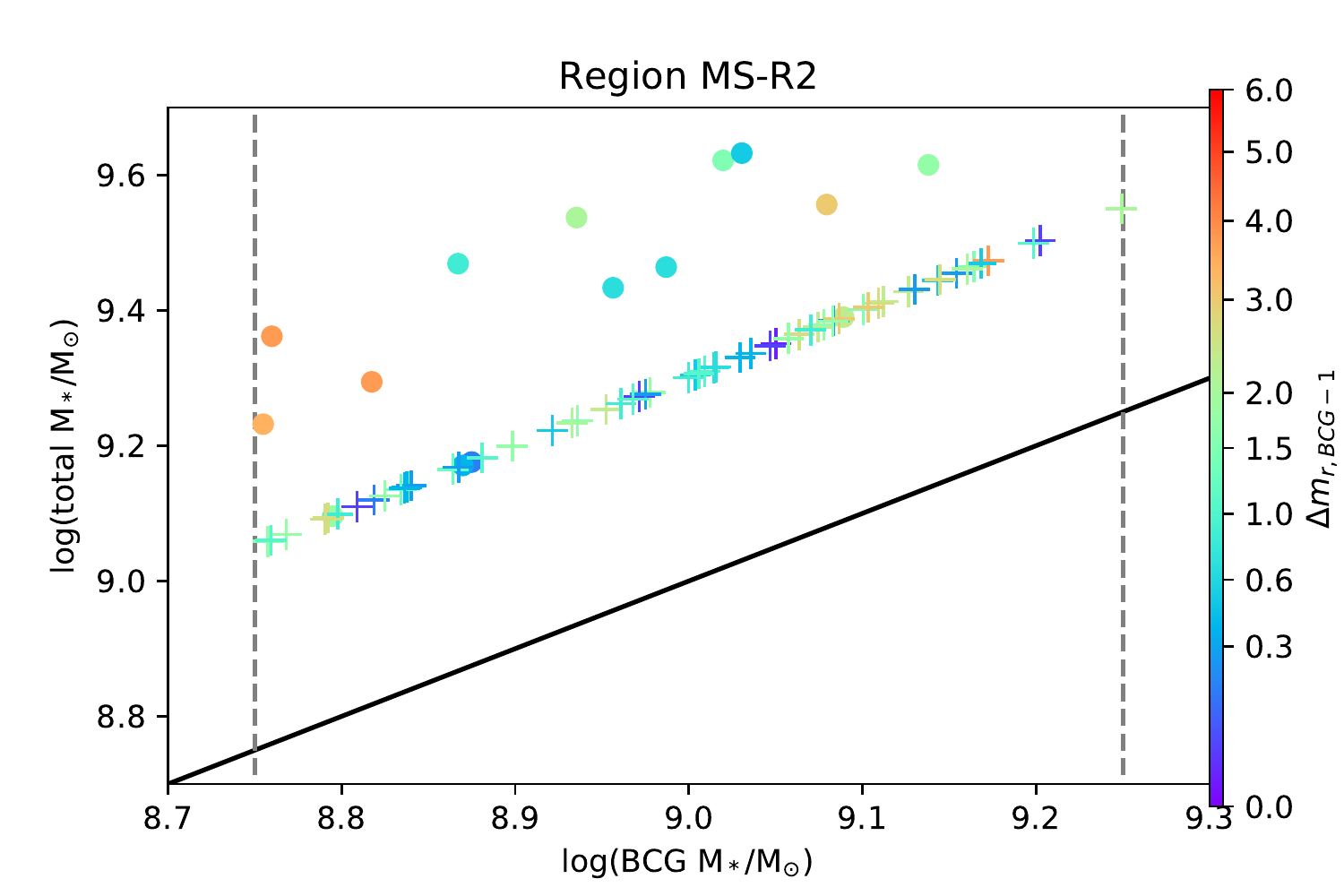}}}\\
{\mbox{\includegraphics[width=3.5truein]{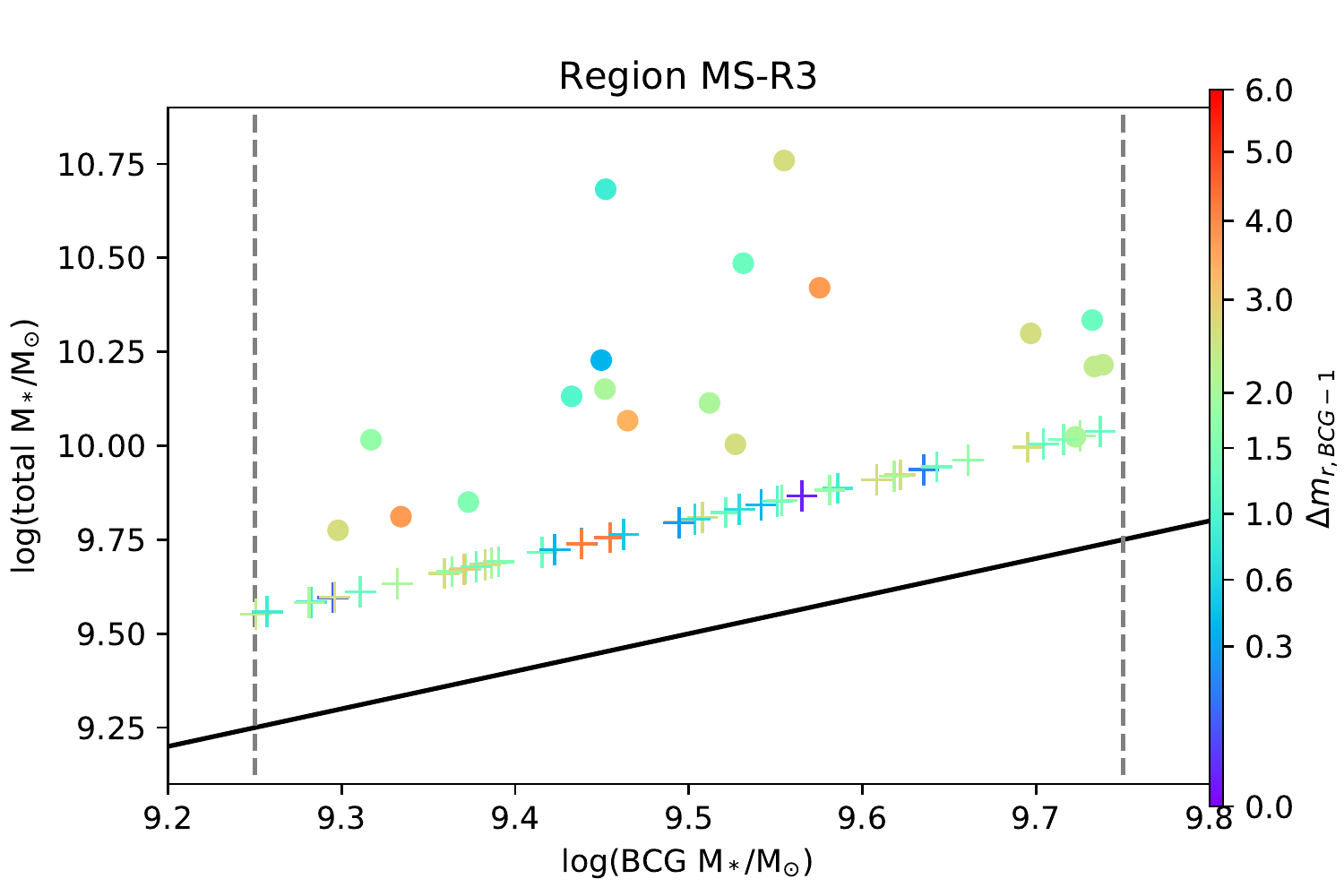}}}&
{\mbox{\includegraphics[width=3.5truein]{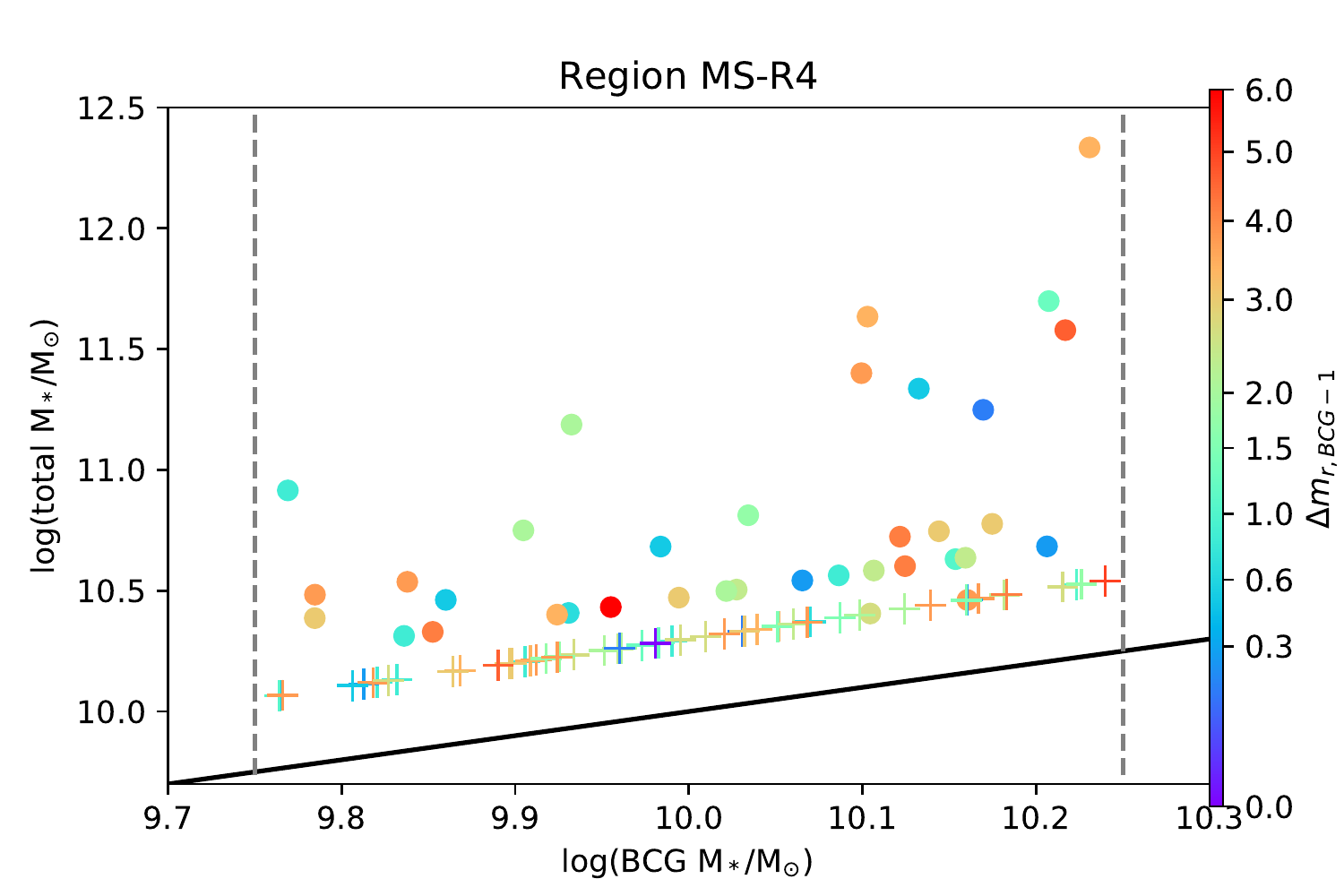}}}\\
{\mbox{\includegraphics[width=3.5truein]{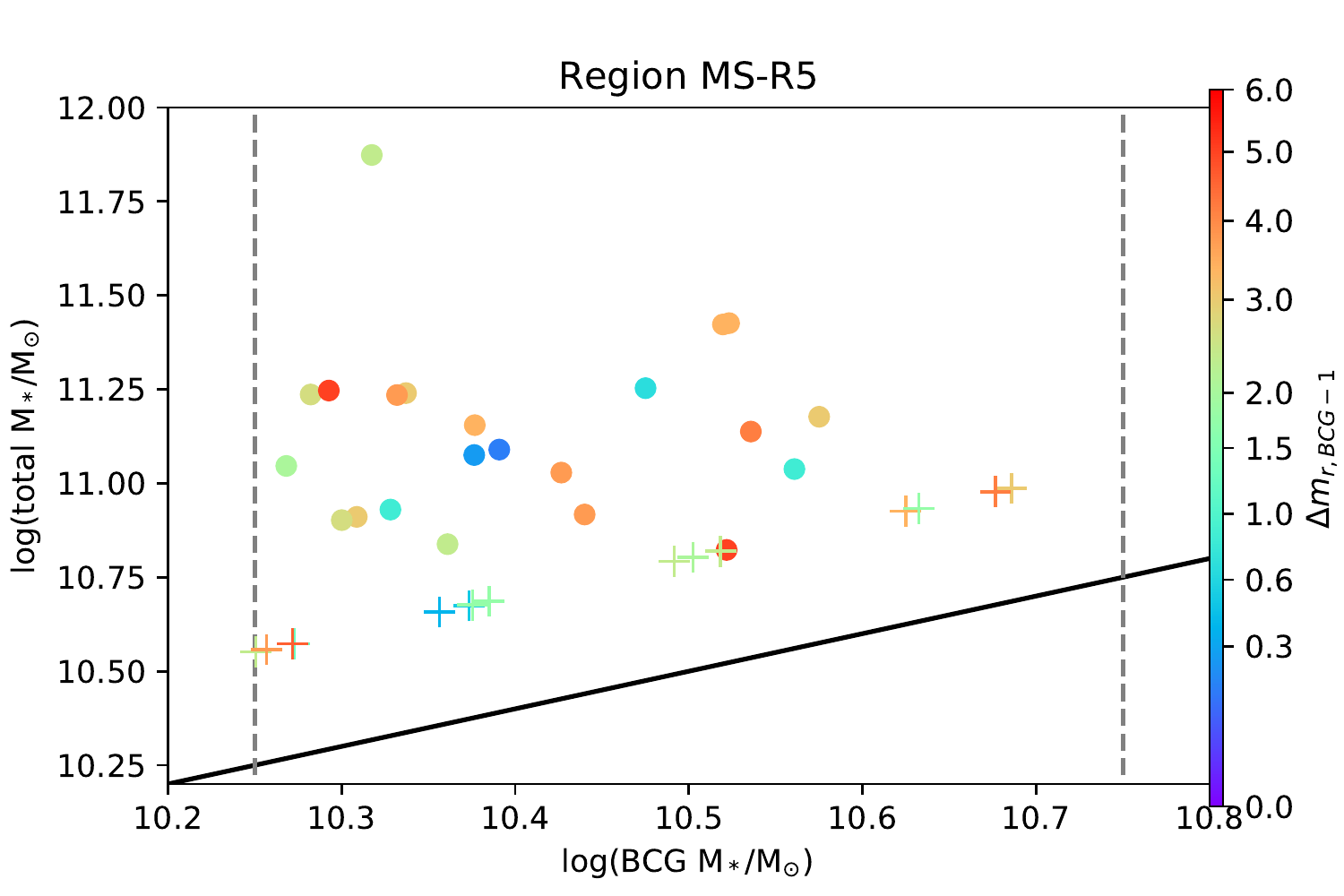}}}&\\
\end{tabular}
\end{center}
\caption{The total stellar masses of pairs (plus signs) and groups (filled circles) plotted against the stellar masses of their BCGs, for the 5 regions discussed in Section~\ref{sec:stack2}. The average stellar mass limits for each region (see Fig.~\ref{fig:mssfrms}) is marked by the dashed vertical grey lines in each panel. The bold black line is the 1:1 line, on which the isolated galaxies lie. The points are colour coded by the corresponding magnitude gap between the BCG and the brightest satellite in the halo.}
\label{fig:msvms}
\end{figure*}

In this section we aim to understand how the \hi\ mass -- halo mass relation might be built into the trends we observe in this study.
Our sample consists mostly of groups with low ($<5$) number of members, as well as pairs and isolated galaxies, for all of which estimating the dark matter halo mass is either not possible or subject to large errors when using velocity dispersions \citep{2011MNRAS.416.2640R}.
An alternative is to use abundance matching \citep{2010ApJ...717..379B,2013MNRAS.428.3121M}, which is inherently a statistical exercise and implicitly makes some important assumptions regarding how galaxies form and evolve within dark matter halos. 
We therefore avoid trying to explicitly estimate the dark matter halo masses, but use observed properties like \mst\ to qualitatively understand the variation in halo masses between categories and regions.

To start with, we check whether the dark matter halo masses of groups/pairs/isolated galaxies from similar regions in our study can be drastically different from category to category.
Stellar mass of the BCG is correlated with the dark matter halo mass, albeit with a large scatter \citep[e.g. see][]{2017MNRAS.470.2982L}.
Using luminosity information from additional member galaxies can provide better constraints on halo mass, and one such correction factor used is the difference in luminosity between the BCG and the brightest satellite galaxy \citep{2014ApJ...782...23S,2015ApJ...804...55L,2018ApJ...860....2G} -- the halo mass corresponding to a BCG stellar mass is {\it smaller} when the difference is {\it larger}.

We then explore if the units of different categories, but from similar regions in Section~\ref{sec:anares} potentially belong to very different halo masses.
In order to do this, we decided to focus on the regions from Section~\ref{sec:stack1} (Fig.~\ref{fig:mssfrms}) which span smaller ranges in log($\langle$\mst$\rangle$) compared to the regions in Sections~\ref{sec:sfrms},\ref{sec:stack1} (Fig.~\ref{fig:sfrms}).
This is because, if the halo mass of a unit is vastly different from its $\langle$\mst$\rangle$, and if such difference varies with category, the variation of the difference with category will be most apparent if we focussed on narrow ranges of \mst.
In Fig.~\ref{fig:msvms}, for each of the five regions from Section~\ref{sec:stack1} (Fig.~\ref{fig:mssfrms}), we separately plot the total stellar masses of the groups/pairs against the stellar mass of the respective BCG.
The points are colour coded according to the difference in $r$-magnitude between the BCG and the brightest satellite, which translates to the logarithm of the ratio of the corresponding luminosities when divided by a factor of 2.5.
The magnitude difference is also indicative of how evolved the BCG is compared to the other member galaxies, and therefore it is no surprise that groups/pairs with larger magnitude difference increases as we move to higher $\langle$\mst$\rangle$ regions, similar to the qualitative trend seen in Fig.~\ref{fig:mssfrms} in terms of the median colours of groups/pairs.

In Fig.~\ref{fig:msvms}, for each region the stellar masses of the isolated galaxies, which we can take to be the BCGs of their single-galaxy halos, lie between the dashed lines marking the boundaries of the corresponding region in each panel.
The BCG stellar masses for the groups/pairs for each region seem to also lie within these bounds for all regions, implying the halo masses are similar for groups/pairs/isolated galaxies lying within similar regions of the log($\langle$\mst$\rangle$) -- log($\langle$SFR$\rangle$) plane as are defined for our study.
From figure 3 in \citet{2017MNRAS.470.2982L} we can see that for most of our sample with BCG stellar masses $< {\rm 10^{10.5} M_{\odot}}$ the correction to halo masses based on the luminosity difference between the BCG and the brightest satellite is minimal. 
For groups/pairs in our sample with the highest BCG stellar masses, given the large luminosity differences in them the correction to halo masses will actually be negative.
Thus it seems clear that groups or pairs do not occupy significantly different (more massive) dark matter halos compared to isolated galaxies, when considering units lying in similar regions from our study.

Given the large range in \mst\ spanned by the regions in our study though, we lose sensitivity to the variation of \mhi\ with halo mass.
For example, our high \mst\ regions from Fig.~\ref{fig:sfrms} (R2,R4,R5) span almost the entire halo mass range covered by \citet[][see their Fig. 2]{2020ApJ...894...92G}, using the low redshift stellar mass -- halo mass function from \citet{2013MNRAS.428.3121M} as reference.
We are sensitive to much lower \mst\ and halo masses as the completeness limit for the GAMA group catalog is about two magnitudes ($m_r \sim$19.5) fainter than the SDSS group catalog used in \citet[][$m_r \sim$17.7]{2020ApJ...894...92G}. 
From our results we can see that when averaged over this large range in halo masses (at the higher end of halo masses), groups/pairs/isolated galaxies have similar gas fractions.
What our study does though, is divides the halos in terms of their position with relative to the SFMS, and reveals that there is a distinct pattern of decreasing gas fraction as we move away from the SFMS into the quenched regime.
This can be the primary cause of the scatter on the halo mass relation \citep[see also][for other potential causes of the scatter -- spin-parameter, fraction of mass in satelllites, feedback from active galactic nuclei]{2020MNRAS.498...44C}.

Similarly only the high \mst\ end of the low \mst\ regions from Fig.~\ref{fig:sfrms} (R1,R3) will be covered by the trends shown in Fig.2 from \citet{2020ApJ...894...92G}.
As discussed before, we find a distinct trend for our low \mst\ regions -- the gas fraction increases significantly as we move from isolated galaxies to pairs to groups.
This might be an effect of the trend of increasing \mhi\ with halo richness for decreasing halo masses seen in \citet{2020ApJ...894...92G}, becoming more pronounced at lower halo masses not covered in \citet{2020ApJ...894...92G}.
In the future using deeper \hi\ surveys like DINGO we would be able to disentangle these trends, by measuring the variation of \mhi\ or gas fraction as a function of both halo mass and offset from the SFMS.

\subsection{Comparison with simulations}
\label{sec:sim}

\begin{figure*}
\begin{center}
\begin{tabular}{cc}
{\mbox{\includegraphics[width=3.3truein]{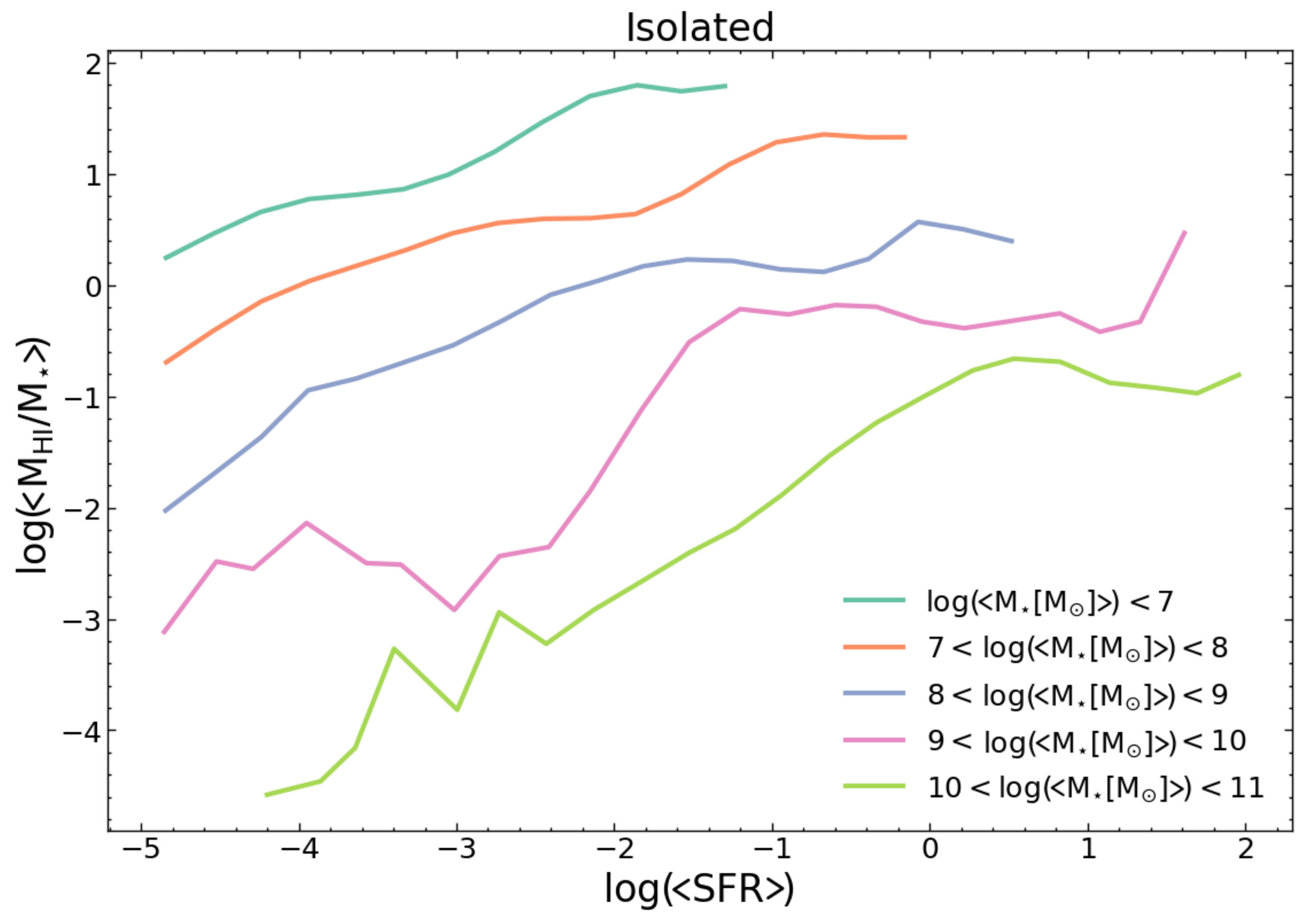}}}&
{\mbox{\includegraphics[width=3.3truein]{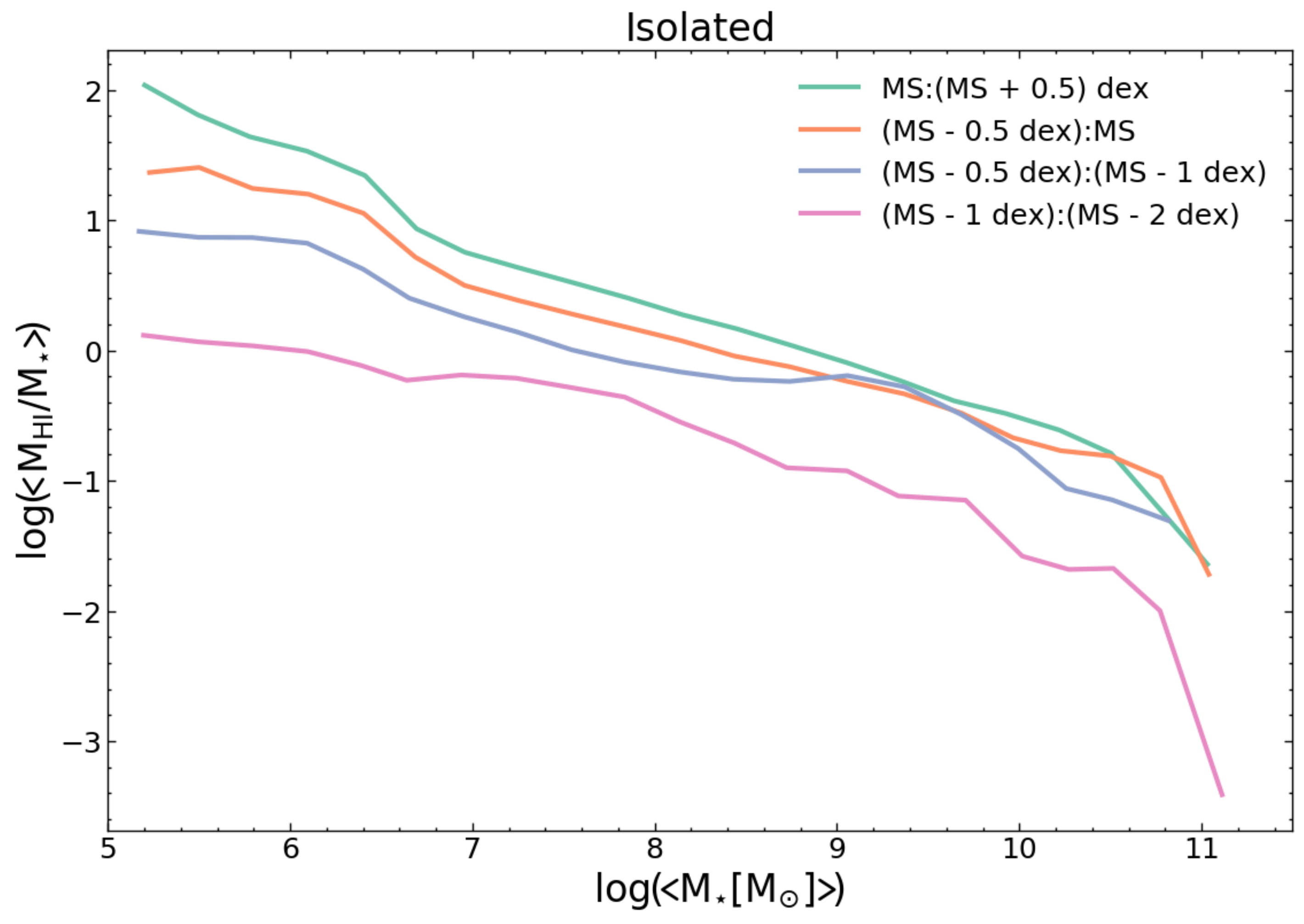}}}\\
{\mbox{\includegraphics[width=3.3truein]{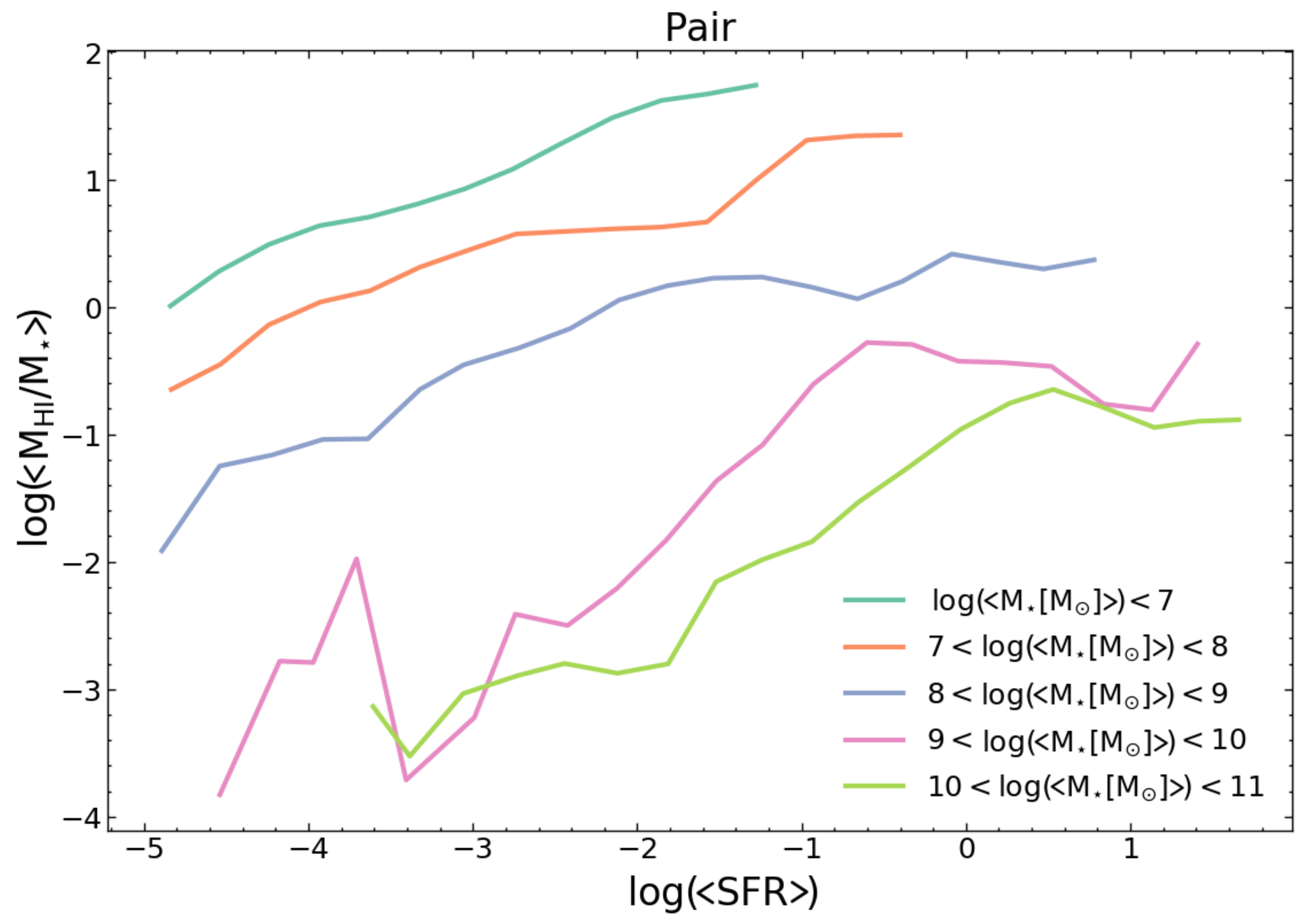}}}&
{\mbox{\includegraphics[width=3.3truein]{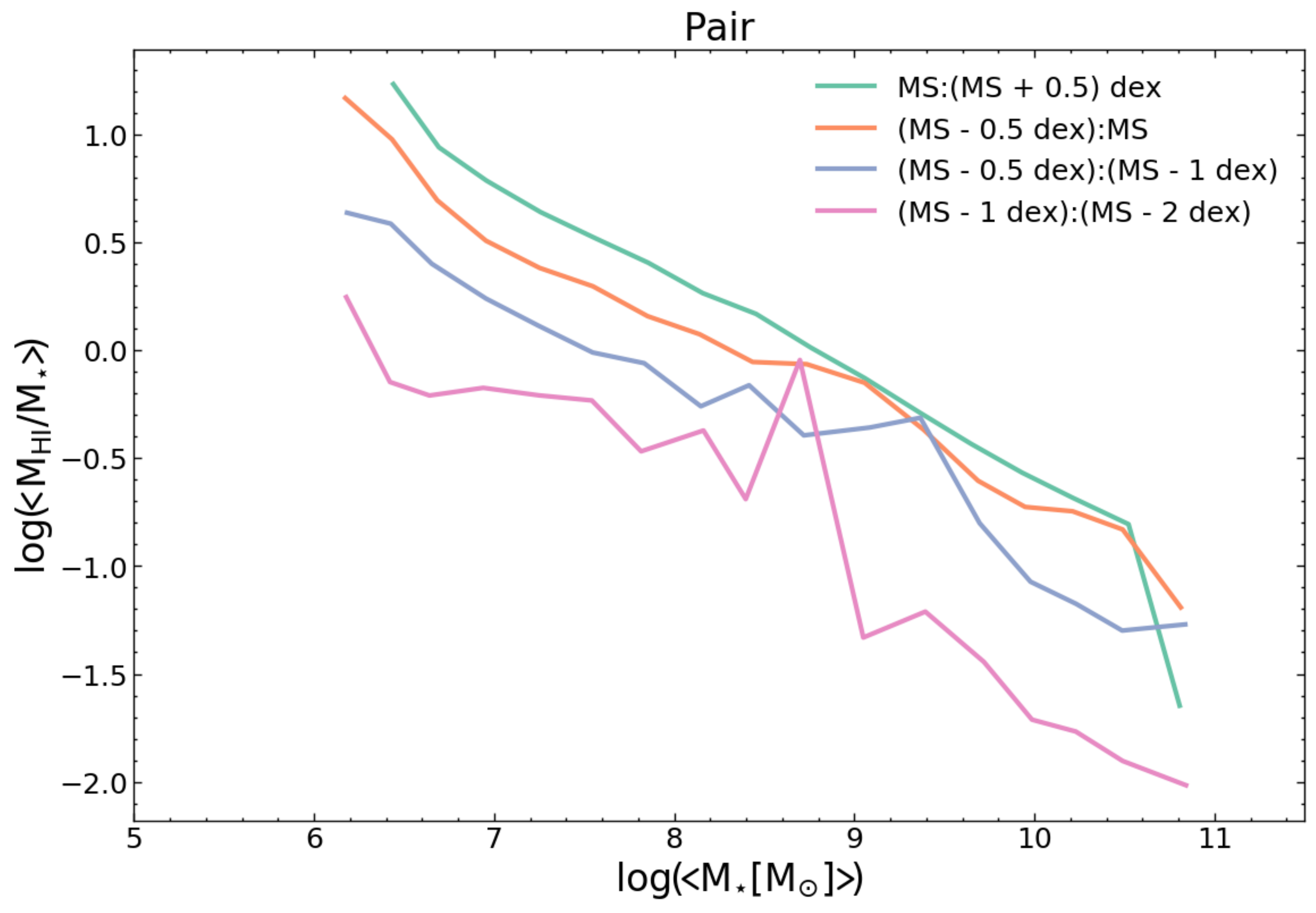}}}\\
{\mbox{\includegraphics[width=3.3truein]{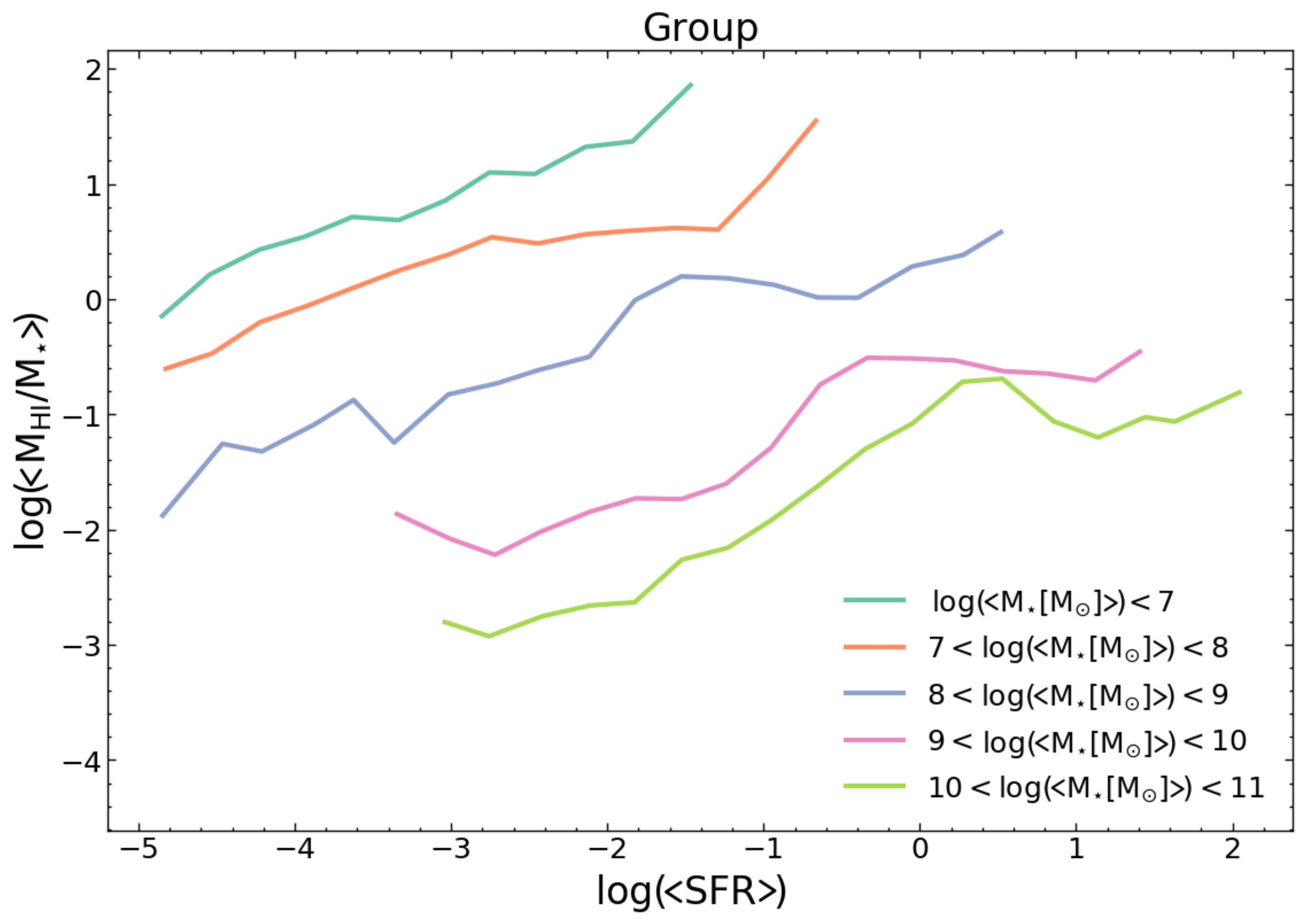}}}&
{\mbox{\includegraphics[width=3.3truein]{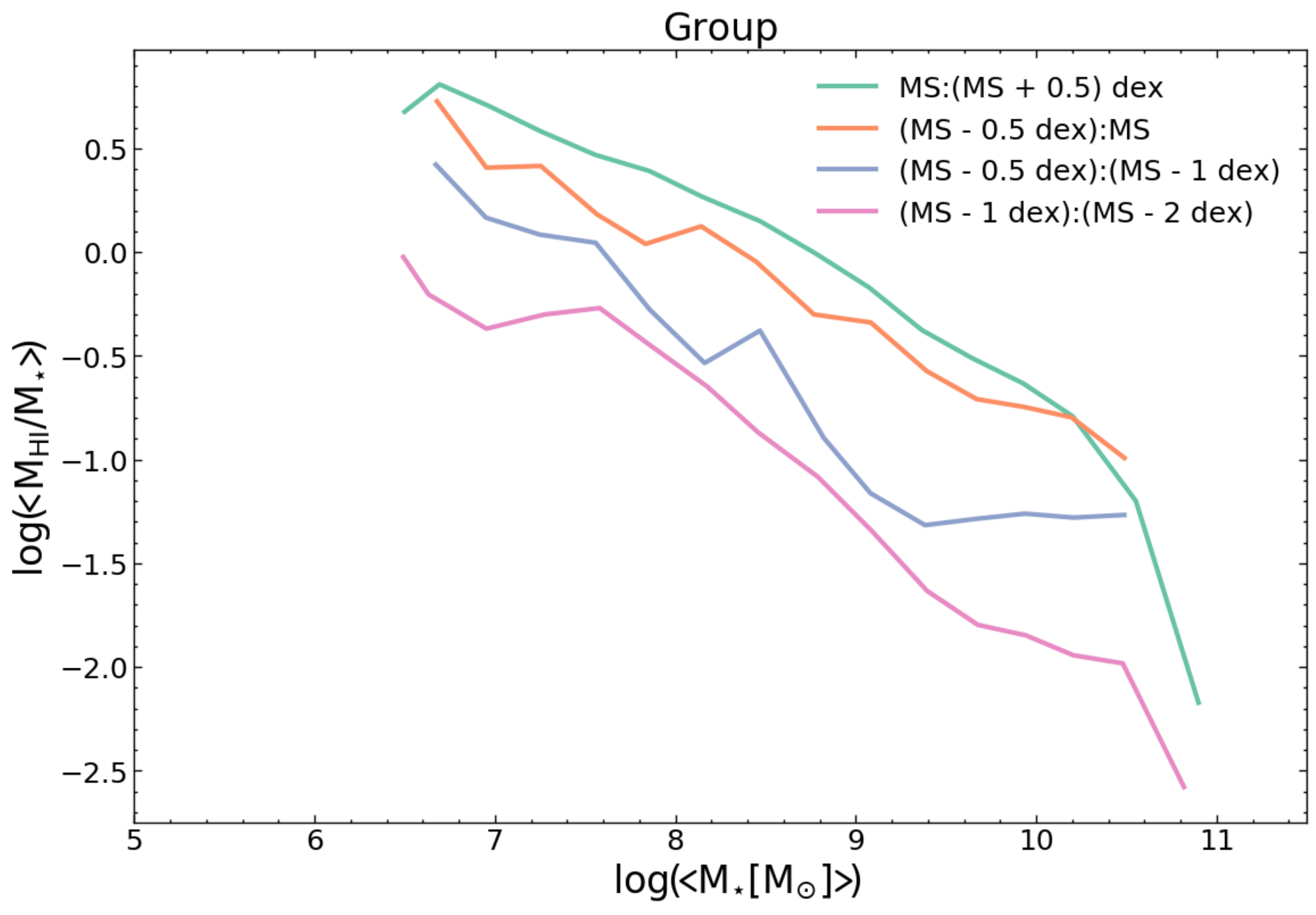}}}\\
\end{tabular}
\end{center}
\caption{Results derived using the full lightcone produced using the {\sc shark} simulations (see text for details). Variation of $\langle f_{HI} \rangle$ for isolated galaxies/pairs/groups (top to bottom row) plotted against, {\it left column}: their $\langle$SFR$\rangle$ when their $\langle$\mst$\rangle$ lies within specified cuts, {\it right column}: their $\langle$\mst$\rangle$ when they lie within specified cuts on the log($\langle$\mst$\rangle$) -- log($\langle$SFR$\rangle$) plane which are constant offsets from the SFMS.} 
\label{fig:sim}
\end{figure*}

\begin{figure*}
\begin{center}
\begin{tabular}{cc}
{\mbox{\includegraphics[width=3.3truein]{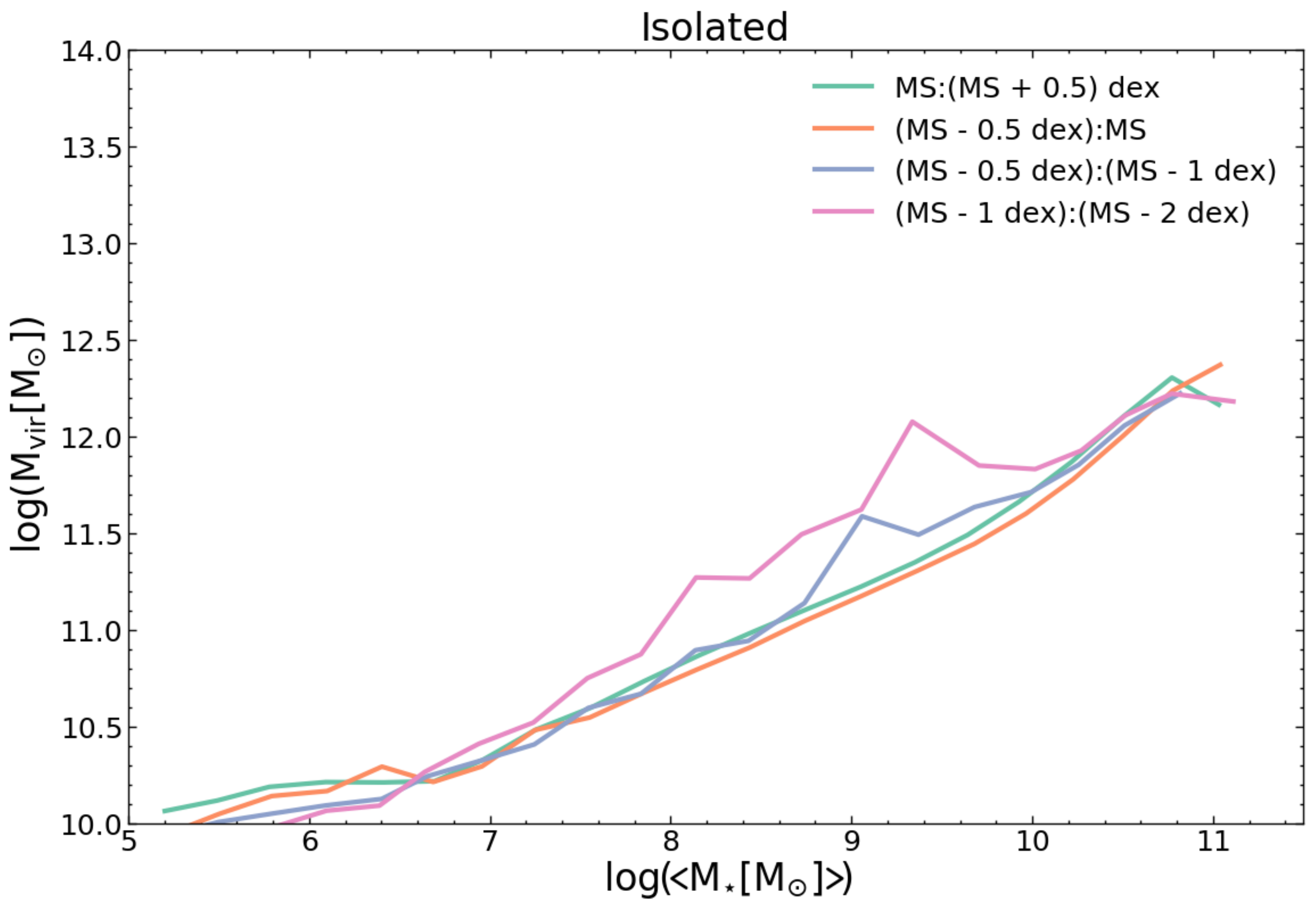}}}&
{\mbox{\includegraphics[width=3.3truein]{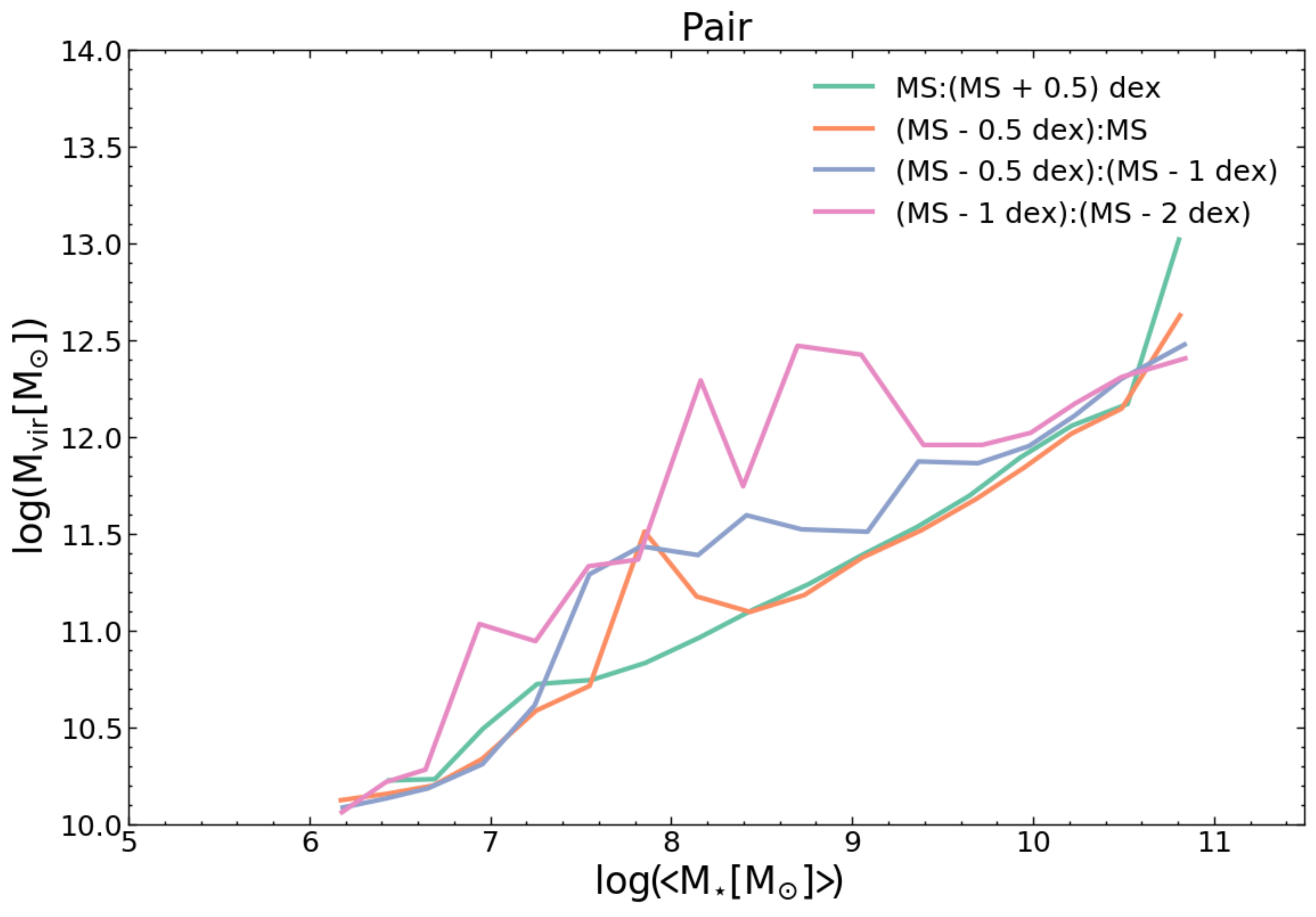}}}\\
{\mbox{\includegraphics[width=3.3truein]{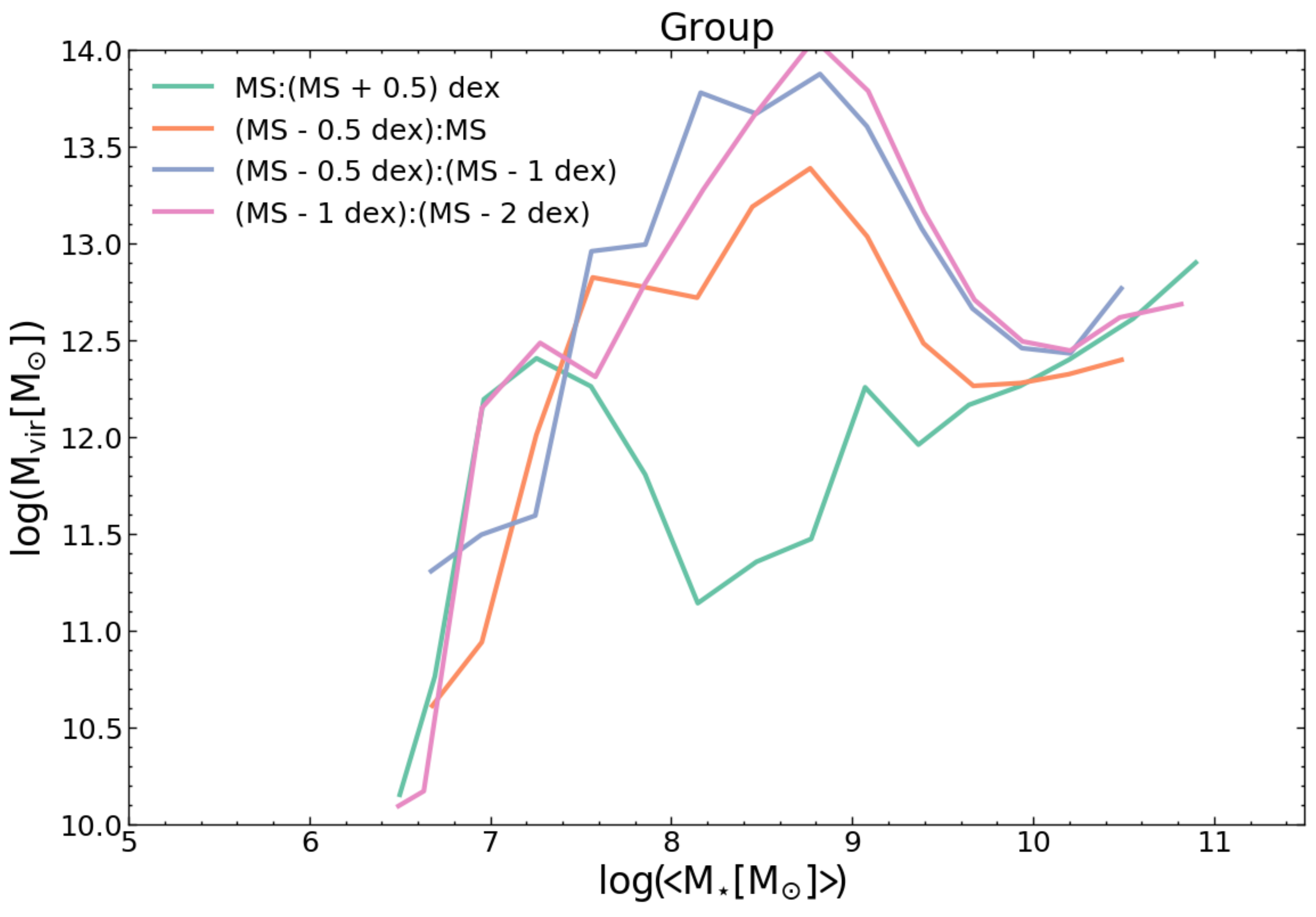}}}&\\
\end{tabular}
\end{center}
\caption{Results derived using the full lightcone produced using the {\sc shark} simulations (see text for details). Variation of the halo virial masses for isolated galaxies (top left), pairs (top right), and groups (bottom left) plotted against their $\langle$\mst$\rangle$ when they lie within specified cuts on the log($\langle$\mst$\rangle$) -- log($\langle$SFR$\rangle$) plane which are constant offsets from the SFMS.} 
\label{fig:sim}
\end{figure*}

Finally, we use the outputs from {\sc shark}, the open-source semi-analytic model (SAM) of galaxy formation \citep{2018MNRAS.481.3573L} to check how our observed results compare with what is expected based on our present understanding of galaxy formation and evolution. 
The models and parameters used to determine the \hi\ content of galaxies are the {\sc shark} defaults described in \citet{2018MNRAS.481.3573L} and applied in \citet{2019MNRAS.488.5898C,2020MNRAS.498...44C,2021arXiv210212203C}.
In these studies it has been shown that {\sc shark} is able to reproduce the observed \hi\ mass function, \hi--stellar mass scaling relation, \hi\ mass and velocity width distributions observed in ALFALFA, and \hi\ clustering.

We use groups/pairs/isolated galaxies from the full lightcone of {\sc shark} \citep{2021arXiv210212203C} to study the variation of their $f_{HI}$ with various quantities.
The constructed lightcone has a redshift range of z~$= 0 - 1$, area of $\sim$6900 deg$^2$, and contains all galaxies with \mst$\geq 10^5 {\rm M_{\odot}}$.
Optical magnitudes for galaxies within the lightcone are calculated from SEDs generated using {\sc prospect} \citep{2020MNRAS.495..905R} and {\sc viperfish} \citep{2019MNRAS.489.4196L}.
Note that when we apply GAMA magnitude cuts before deriving the trends, we end up with similar trends only curtailed at the low \mst\ and SFR regime.
{\sc shark} does not account for the presence of any intergalactic medium, and replenishment of the gas reservoir of galaxies happen when they are the central galaxy and another galaxy becomes a satellite.
In the simulation when a galaxy becomes a satellite, its hot halo gas is stripped and added to the reservoir of the central galaxy, where it cools and settles on the central. 
The satellite, on the other hand, is not provided with a hot-halo gas reservoir to get gas from.
Also, it is cut-off from the cosmic accretion, and so it eventually quenches.

It is worthwhile to note that in general SAMs, including {\sc shark}, assume that matter accreting onto halos is characterised by the universal baryon fraction. 
In most SAMs, baryons can be ejected from the halo and are reincorporated afer a timescale that depends on the growth rate of halos \citep[e.g.][]{2013MNRAS.431.3373H,2018MNRAS.481.3573L}. 
This typically leads to halos less massive than the Milky-Way mass having lower baryon fractions than the universal value, which is the ones more massive halos tend to. 
This is quite different to what is seen in some hydrodynamical simulations, in which the baryon fraction converges into the universal value at halo masses closer to galaxy cluster scales, and in which the matter accreting to halos can be severely baryon depleted due to the large-scale effects of stellar feedback \citep[e.g.][]{2020MNRAS.498.1668W}.
Given these caveats involved when trying to account for baryons ejected from galactic halos using simulations, we use {\sc shark} here specifically to try and understand the galaxy-based results from Section~\ref{sec:pilot}.

In Fig.~\ref{fig:sim}, we plot the variation of $\langle f_{HI} \rangle$ for groups/pairs/isolated galaxies against either their $\langle$SFR$\rangle$ when their $\langle$\mst$\rangle$ lies within specified cuts at different \mst s, or against their $\langle$\mst$\rangle$ when they lie within specified cuts on the log($\langle$\mst$\rangle$) -- log($\langle$SFR$\rangle$) plane which are constant offsets from the SFMS.
Overall for all units the simulations we reproduce the following trends seen in our results from Section~\ref{sec:anares}: (i) a decrease in $\langle f_{HI} \rangle$ with increasing $\langle$\mst$\rangle$ when moving parallel to the SFMS, (ii) a decrease in $\langle f_{HI} \rangle$ at a fixed $\langle$\mst$\rangle$ when moving from above the SFMS to below the SFMS, (iii) an increase in $\langle f_{HI} \rangle$ with $\langle$SFR$\rangle$ at a fixed $\langle$\mst$\rangle$, and (iv) a decrease in $\langle f_{HI} \rangle$ at a fixed $\langle$SFR$\rangle$ as one moves to higher $\langle$\mst$\rangle$ values.
When comparing the $\langle f_{HI} \rangle$ values for different units for the same $\langle$\mst$\rangle$ and $\langle$SFR$\rangle$, we find that it decreases as we move from isolated galaxies$\rightarrow$pairs$\rightarrow$groups.
This is exactly what we expect when focussing on galaxies as in Section~\ref{sec:pilot}.
Thus the semi-analytical model {\sc shark} is not equipped yet to resolve the origins of the \hi\ in the intergalactic space of groups/pairs with low $\langle$\mst$\rangle$.

Also plotted in Fig.~\ref{fig:sim} is the variation of the halo virial masses for groups/pairs/isolated galaxies against their $\langle$\mst$\rangle$ when they lie within specified cuts on the log($\langle$\mst$\rangle$) -- log($\langle$SFR$\rangle$) plane which are constant offsets from the SFMS.
These results take us beyond what is discussed in Section~\ref{ssec:halo}, and provide estimates of the actual halo masses to map our $\langle$\mst$\rangle$ onto.
It is interesting to note that the mapping from $\langle$\mst$\rangle$ to the virial mass of the halo becomes markedly non-linear as we move below and away from the SFMS into the quenched regime, especially for groups and pairs.
For isolated galaxies and pairs on or above the SFMS in the log($\langle$\mst$\rangle$) -- log($\langle$SFR$\rangle$) plane, the halo virial masses are comparable at a given $\langle$\mst$\rangle$.
This also seems to be true for groups lying above the SFMS in the log($\langle$\mst$\rangle$) -- log($\langle$SFR$\rangle$) plane, for log$(\frac{\langle M_* \rangle}{M_{\odot}}) \geq$ 8.
This is the $\langle$\mst$\rangle$ regime in which our results from Section~\ref{sec:anares} are derived.
And from Section~\ref{sec:stack1} we know that units above the SFMS have the highest gas fractions and likely dominate the measured gas fractions along the SFMS in Sections~\ref{sec:stack2} and \ref{sec:pilot}.
Thus our finding that the $\langle f_{HI} \rangle$, when considering the \hi\ in member galaxies only, is very similar for groups/pairs/isolated galaxies for a given $\langle$\mst$\rangle$ might be true because our measurements are dominated by star-forming groups/pairs/isolated galaxies with very similar halo virial masses.

\section{Summary and Conclusions}

The aim of this empirical study is to relate the total atomic gas (\hi) content of groups and pairs of galaxies with their total stellar masses (\mst) and star formation rates (SFR), and compare these relations to the corresponding relations for isolated galaxies, and thus gain insight into the conversion of gas to stars in the denser environments of groups and pairs.
For our work we use fields from the GAMA survey, with the \hi\ data from two different sources: (i) archival data from the ALFALFA survey which partially overlap three GAMA equatorial fields, and (ii) DINGO pilot survey data of the GAMA G15 field.
The environmental and galaxy parameters used in the study are from the GAMA survey, whose exquisite sensitivity allows us to include low mass galaxies and groups in our study.

We decided against measuring the variation of the gas content of the units in our study, i.e. groups/pairs/isolated galaxies, against their respective halo masses because of (i) the expected large variation in scales in terms of halo masses as we moved from isolated galaxies to groups; (ii) the large uncertainty in the inferred halo masses for isolated galaxies, pairs of galaxies, and galaxy groups with small number of members - i.e. most of the units used in our study.
In order to normalize the properties against which the gas content was to be measured and compared across the different units, we considered the log($\langle$\mst$\rangle$) -- log($\langle$SFR$\rangle$), and split the plane into sub-regions based on the z$\sim 0$ SFMS of individual galaxies, for all units have substantial populations along and around the SFMS in the respective log($\langle$\mst$\rangle$) -- log($\langle$SFR$\rangle$) planes.
The normalized gas quantity that we measure in each sub-region is the mean \hi\ gas fraction $\langle f_{HI} \rangle$ (where $f_{HI}=\frac{M_{HI}}{M_{*}}$), through flux-weighted stacking of the $f_{HI}$ of the individual units.

We determine the $f_{HI}$ of the groups and pairs in slightly different ways for ALFALFA and the DINGO data, considering the low spatial resolution of ALFALFA data, and the higher resolution but relative shallowness of the DINGO pilot data.
For ALFALFA the $f_{HI}$ spectra of groups and pairs are measured by extracting \hi\ spectra over the entire groups or pair areas and dividing by the total \mst\ of all member galaxies.
For DINGO the $f_{HI}$ spectra of groups and pairs are measured by co-adding \hi\ spectra of individual member galaxies, and then dividing by the total \mst\ of all members galaxies. 

We find that (i) $\langle f_{HI} \rangle$ for all units decreases along the SFMS with increasing $\langle M_{*} \rangle$,
(ii) $\langle f_{HI} \rangle$ for all units decreases as one moves from above the SFMS to below it at any given $\langle M_{*} \rangle$. 
(iii) $\langle f_{HI} \rangle$ values are also found to be comparable for units residing the same sub-region of the log($\langle$\mst$\rangle$) -- log($\langle$SFR$\rangle$) plane, with two major caveats: (a) when the gas content is measured by summing over the gas contents of member galaxies (DINGO data), for $\langle f_{HI} \rangle$ s at any given $\langle M_{*} \rangle$, isolated galaxies have slightly higher values compared to groups (and pairs for higher $\langle$\mst$\rangle$).
The $\langle f_{HI} \rangle$ s are comparable across units when the gas content of groups and pairs are measured over the entire group or pair areas (ALFALFA data) for higher $\langle$\mst$\rangle$ bins (log$(\frac{\langle M_* \rangle}{M_{\odot}}) >$ 9.5), but maybe with groups having slightly higher values compared to isolated galaxies.
More importantly though, (b) for bins with log$(\frac{\langle M_* \rangle}{M_{\odot}}) \lesssim$ 9.5, $\langle f_{HI} \rangle$ is higher for groups compared to isolated galaxies (with those for pairs being intermediate), and significantly so when log$(\frac{\langle M_* \rangle}{M_{\odot}}) \lesssim$ 9.0).

The evidence seems to suggest that low average/total \mst\ groups (and pairs) contain substantial amounts of \hi\ not associated with already detected member galaxies, and likely situated in the intergalactic space either as dark clouds or associated with faint undetected galaxies.
Even though our sample is dominated by low multiplicity groups (N$_{FoF}$ = 3 or 4) in the lowest $\langle$\mst$\rangle$ bins, the trend does not seem to be driven by multiplicity.
We compare our results with predictions from the semi-analytical model of galaxy formation SHARK, which reproduces our results from galaxy-based \hi\ spectra, suggesting that in denser galaxy environments galaxies on average are less gas rich.
Comparison with the SHARK predictions also throws up the intriguing result that groups, pairs and isolated galaxies used in our study which lie on or above the SFMS in their respective log($\langle$\mst$\rangle$) -- log($\langle$SFR$\rangle$) plane, i.e. the units which dominate the measured $\langle f_{HI} \rangle$ at a given $\langle$\mst$\rangle$, might have very similar dark matter halo masses across unit types for any given $\langle$\mst$\rangle$. 

The possibility though of the presence of large amounts of atomic hydrogen in the intergalactic space of low mass groups is the unique result to emerge from this study, which demands further observational exploration.
We hope to confirm this result in the future using data from the DINGO survey alone -- as more data are collected enabling the measurement of the gas content over the entire areas of groups and pairs, and not just from individual galaxies separable due to the high spatial resolution.

\section*{Acknowledgements}

We would like to thank the ALFALFA team for providing us access to their data cubes.
SR would like to thank Barbara Catinella and Luca Cortese for helpful dicsussions.

The Australian SKA Pathfinder is part of the Australia Telescope National Facility which is funded by the Commonwealth of Australia for operation as a National Facility managed by CSIRO. This scientific work uses data obtained from the Murchison Radio-astronomy Observatory (MRO), which is jointly funded by the Commonwealth Government of Australia and State Government of Western Australia. The MRO is managed by the CSIRO, who also provide operational support to ASKAP. We acknowledge the Wajarri Yamatji people as the traditional owners of the Observatory site. The work was supported by the Pawsey supercomputing centre through the use of advanced computing resources.

This paper includes archived data obtained through the CSIRO ASKAP Science Data Archive, CASDA (\url{https://www.doi.org/10.25919/2424-6m26}). 

GAMA is a joint European-Australasian project based around a spectroscopic campaign using the Anglo-Australian Telescope. The GAMA input catalog is based on data taken from the Sloan Digital Sky Survey and the UKIRT Infrared Deep Sky Survey. Complementary imaging of the GAMA regions is being obtained by a number of independent survey programmes including GALEX MIS, VST KiDS, VISTA VIKING, WISE, Herschel-ATLAS, GMRT and ASKAP providing UV to radio coverage. GAMA is funded by the STFC (UK), the ARC (Australia), the AAO, and the participating institutions. The GAMA website is \url{http://www.gama-survey.org/}. Based on observations made with ESO Telescopes at the La Silla Paranal Observatory under programme ID 179.A-2004. Based on observations made with ESO Telescopes at the La Silla Paranal Observatory under programme ID 177.A-3016. 

This research was supported by the Australian Research Council Centre of Excellence for All Sky Astrophysics in 3 Dimensions (ASTRO  3D), through project number CE170100013.
LJMD and ASGR and acknowledge support from the Australian Research Councils Future Fellowship scheme (FT200100055 and FT200100375 respectively).
RK acknowledges support from the Bundesministerium fuer Bildung und Forschung (BMBF) award 05A20WM4.
AHW is supported by an European Research Council Consolidator Grant (No. 770935). 

This research made use of Astropy,\footnote{\url{http://www.astropy.org}} a community-developed core Python package for Astronomy \citep{2013A&A...558A..33A,2018AJ....156..123A}. 

\bibliography{sambit}{}
\bibliographystyle{aasjournal}

\appendix

\section{Stacked spectra}

\figsetstart
\figsetnum{17}
\figsettitle{Stacked spectra}

\figsetgrpstart
\figsetgrpnum{17.1}
\figsetgrptitle{ALFALFA isolated galaxies, 5$+$2 regions}
\figsetplot{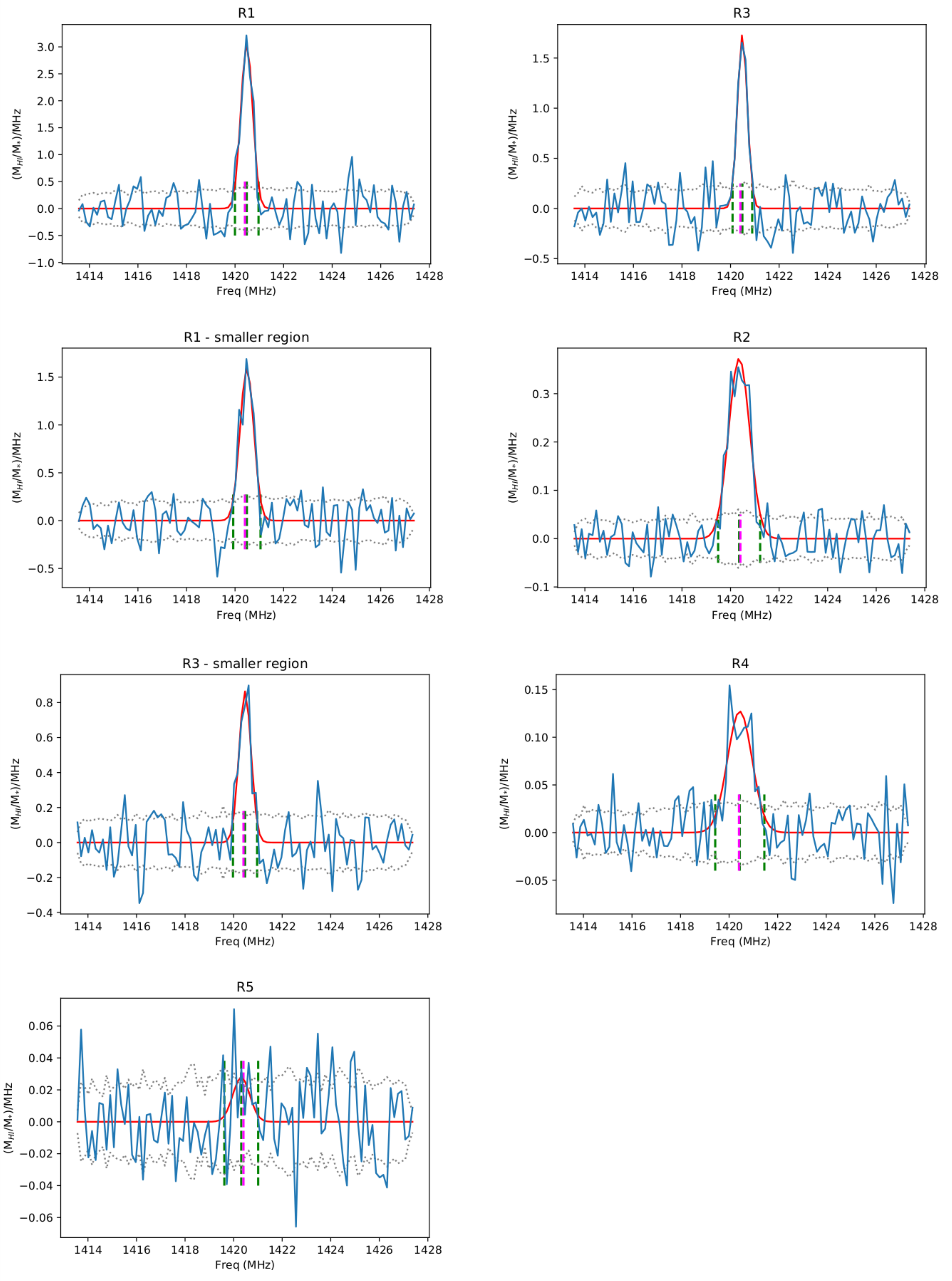}
\figsetgrpnote{Flux-weighted mean stacks of $f_{HI}$ of the 5$+$2 regions for {\bf isolated galaxies} discussed in Sections~3.2, 3.3 in blue. The Gaussian fit to each spectrum is shown in red. The green vertical dashed lines mark the centre of the Gaussian and $\pm {\rm 2} \sigma$ positions, while the magenta dashed line marks the rest frequency of the {\rm H}{\sc i} 21 cm line. The grey dotted lines mark the standard deviation per channel calculated using 1000 bootstrapped spectra drawn randomly from the original set of spectra (see Section~3.3 for details). The complete figure set (13 images) is available in the online journal.}
\figsetgrpend

\figsetgrpstart
\figsetgrpnum{17.2}
\figsetgrptitle{ALFALFA pairs, 5 regions}
\figsetplot{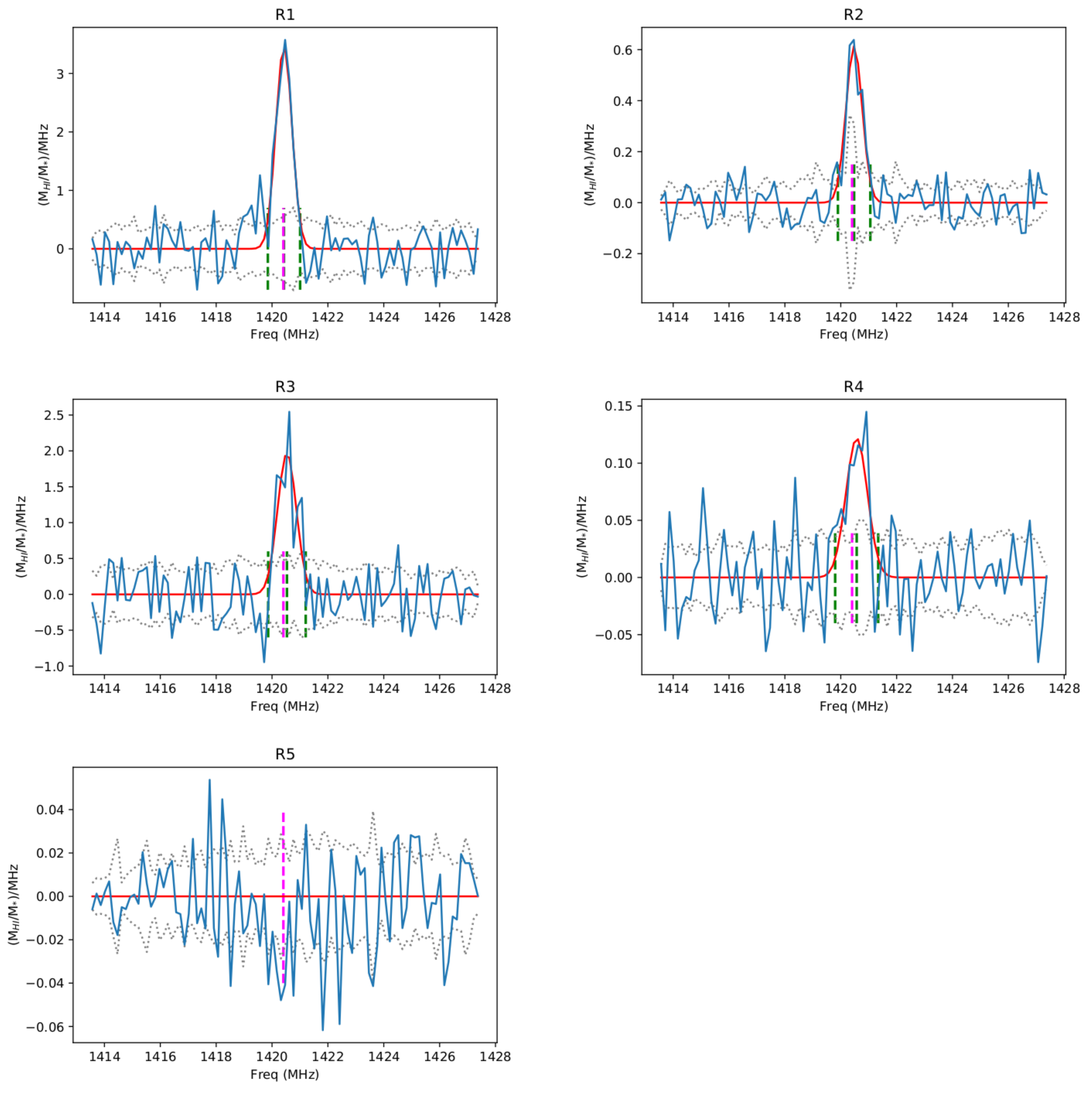}
\figsetgrpnote{Flux-weighted mean stacks of $f_{HI}$ of the 5 regions discussed in Sections~3.2, 3.3 in blue, for {\bf pairs of galaxies}.  The Gaussian fit to each spectrum is shown in red. The green vertical dashed lines mark the centre of the Gaussian and $\pm {\rm 2} \sigma$ positions, while the magenta dashed line marks the rest frequency of the {\rm H}{\sc i} 21 cm line. The grey dotted lines mark the standard deviation per channel calculated using 1000 bootstrapped spectra drawn randomly from the original set of spectra (see Section~3.3 for details).}
\figsetgrpend

\figsetgrpstart
\figsetgrpnum{17.3}
\figsetgrptitle{ALFALFA groups, 5 regions}
\figsetplot{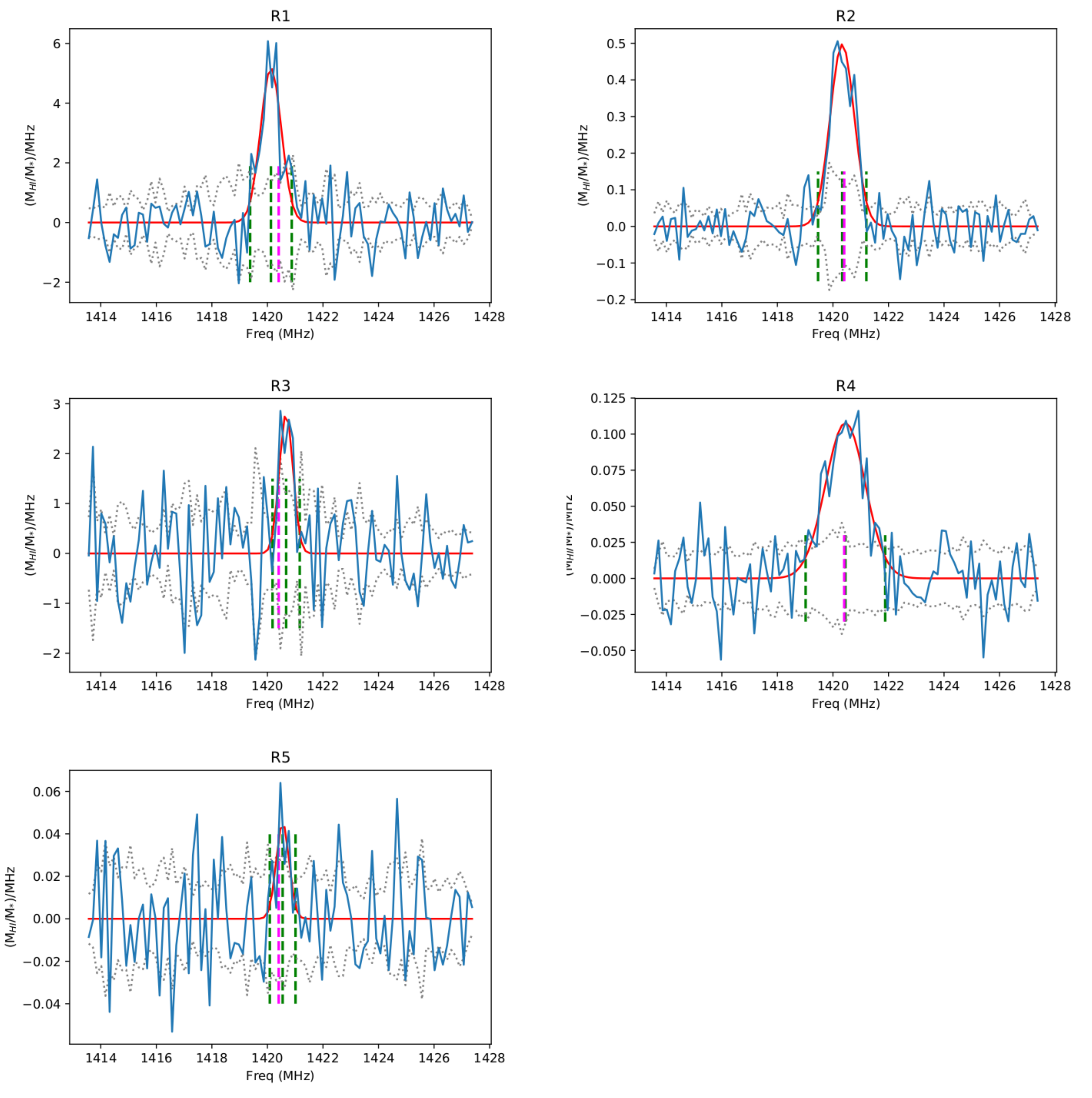}
\figsetgrpnote{Flux-weighted mean stacks of $f_{HI}$ of the 5 regions discussed in Sections~3.2, 3.3 in blue, for {\bf groups of galaxies}. The Gaussian fit to each spectrum is shown in red. The green vertical dashed lines mark the centre of the Gaussian and $\pm {\rm 2} \sigma$ positions, while the magenta dashed line marks the rest frequency of the {\rm H}{\sc i} 21 cm line. The grey dotted lines mark the standard deviation per channel calculated using 1000 bootstrapped spectra drawn randomly from the original set of spectra (see Section~3.3 for details).}
\figsetgrpend

\figsetgrpstart
\figsetgrpnum{17.4}
\figsetgrptitle{ALFALFA isolated galaxies, 5 SFMS regions}
\figsetplot{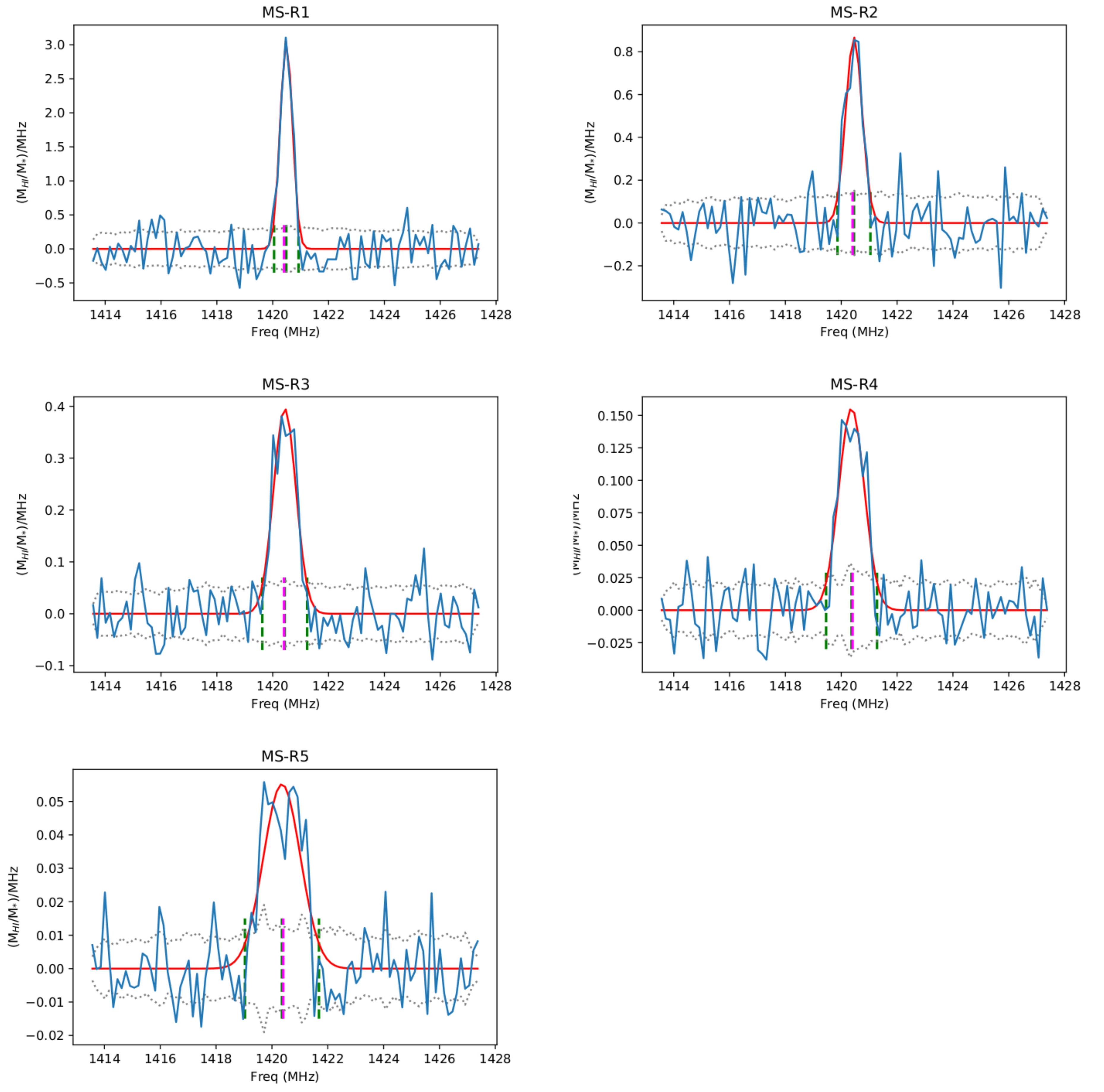}
\figsetgrpnote{Flux-weighted mean stacks of $f_{HI}$ of the 5 SFMS regions discussed in Section~3.4 in blue, for {\bf isolated galaxies}. The Gaussian fit to each spectrum is shown in red. The green vertical dashed lines mark the centre of the Gaussian and $\pm {\rm 2} \sigma$ positions, while the magenta dashed line marks the rest frequency of the {\rm H}{\sc i} 21 cm line. The grey dotted lines mark the standard deviation per channel calculated using 1000 bootstrapped spectra drawn randomly from the original set of spectra (see Section~3.4 for details).}
\figsetgrpend

\figsetgrpstart
\figsetgrpnum{17.5}
\figsetgrptitle{ALFALFA pairs, 5 SFMS regions}
\figsetplot{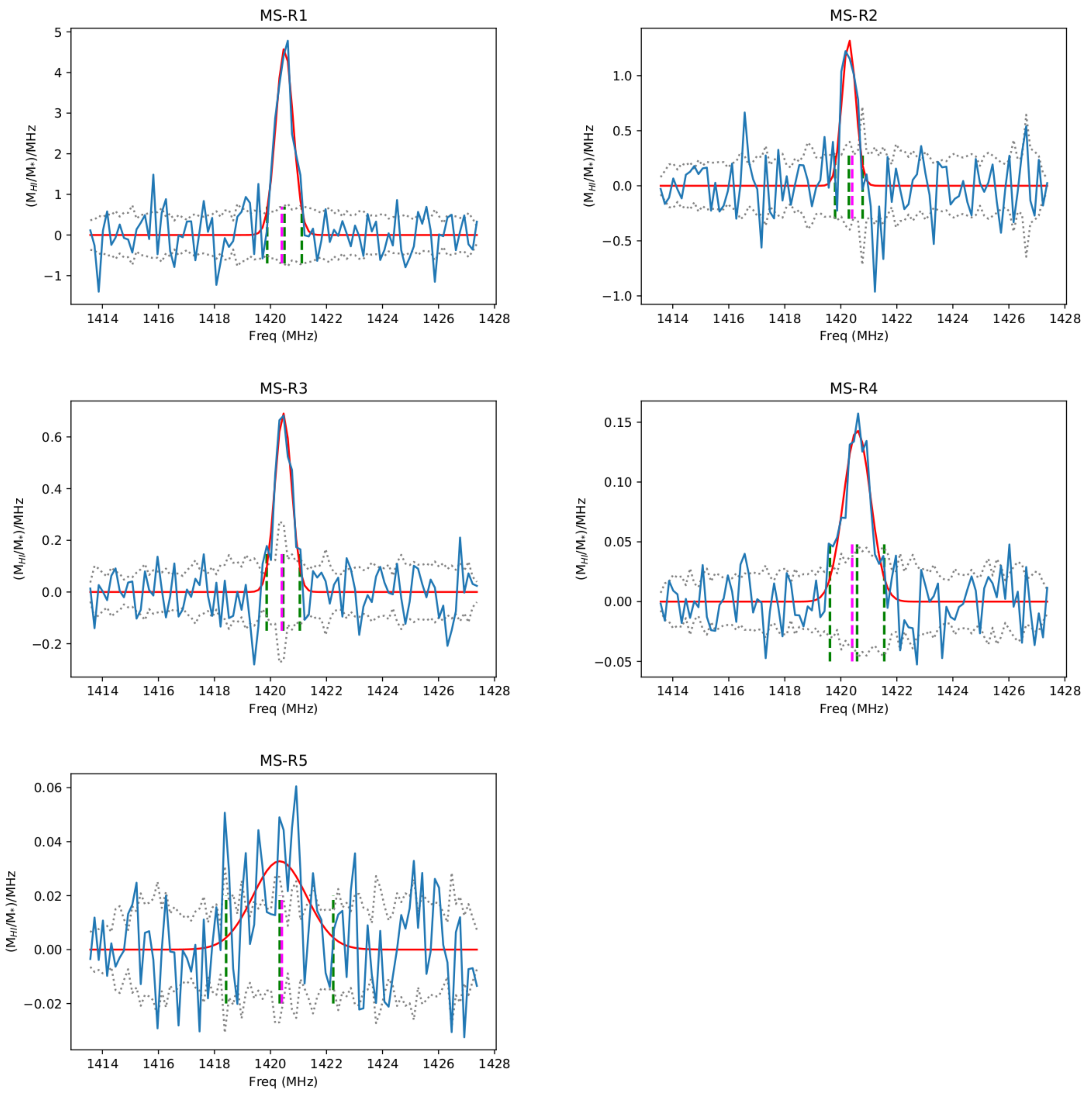}
\figsetgrpnote{Flux-weighted mean stacks of $f_{HI}$ of the 5 SFMS regions discussed in Section~3.4 in blue, for {\bf pairs of galaxies}. The Gaussian fit to each spectrum is shown in red. The green vertical dashed lines mark the centre of the Gaussian and $\pm {\rm 2} \sigma$ positions, while the magenta dashed line marks the rest frequency of the {\rm H}{\sc i} 21 cm line. The grey dotted lines mark the standard deviation per channel calculated using 1000 bootstrapped spectra drawn randomly from the original set of spectra (see Section~3.4 for details).}
\figsetgrpend

\figsetgrpstart
\figsetgrpnum{17.6}
\figsetgrptitle{ALFALFA groups, 5 SFMS regions}
\figsetplot{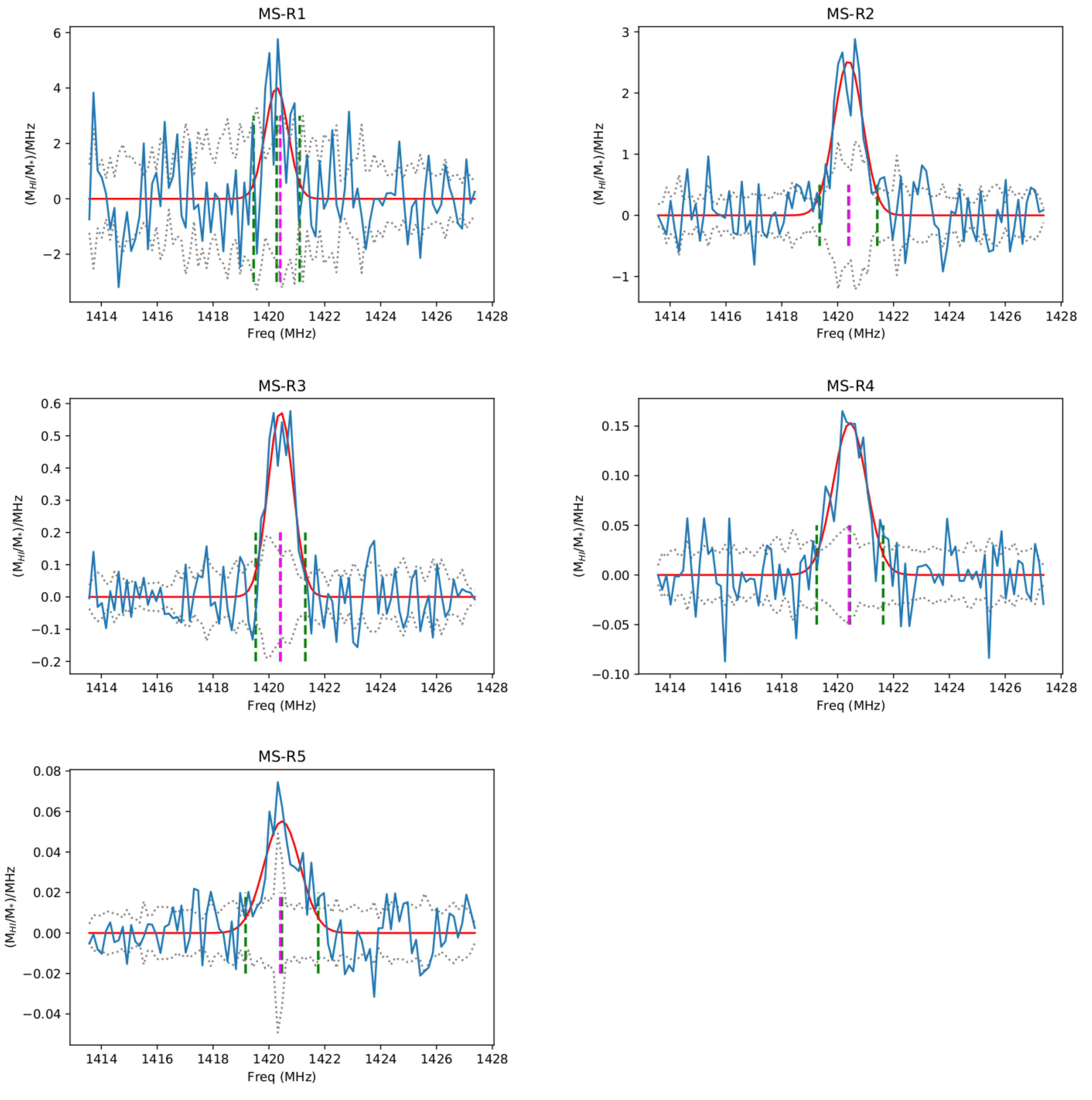}
\figsetgrpnote{Flux-weighted mean stacks of $f_{HI}$ of the 5 SFMS regions discussed in Section~3.4 in blue, for {\bf groups of galaxies}. The Gaussian fit to each spectrum is shown in red. The green vertical dashed lines mark the centre of the Gaussian and $\pm {\rm 2} \sigma$ positions, while the magenta dashed line marks the rest frequency of the {\rm H}{\sc i} 21 cm line. The grey dotted lines mark the standard deviation per channel calculated using 1000 bootstrapped spectra drawn randomly from the original set of spectra (see Section~3.4 for details).}
\figsetgrpend

\figsetgrpstart
\figsetgrpnum{17.7}
\figsetgrptitle{DINGO isolated galaxies, 3 SFMS regions}
\figsetplot{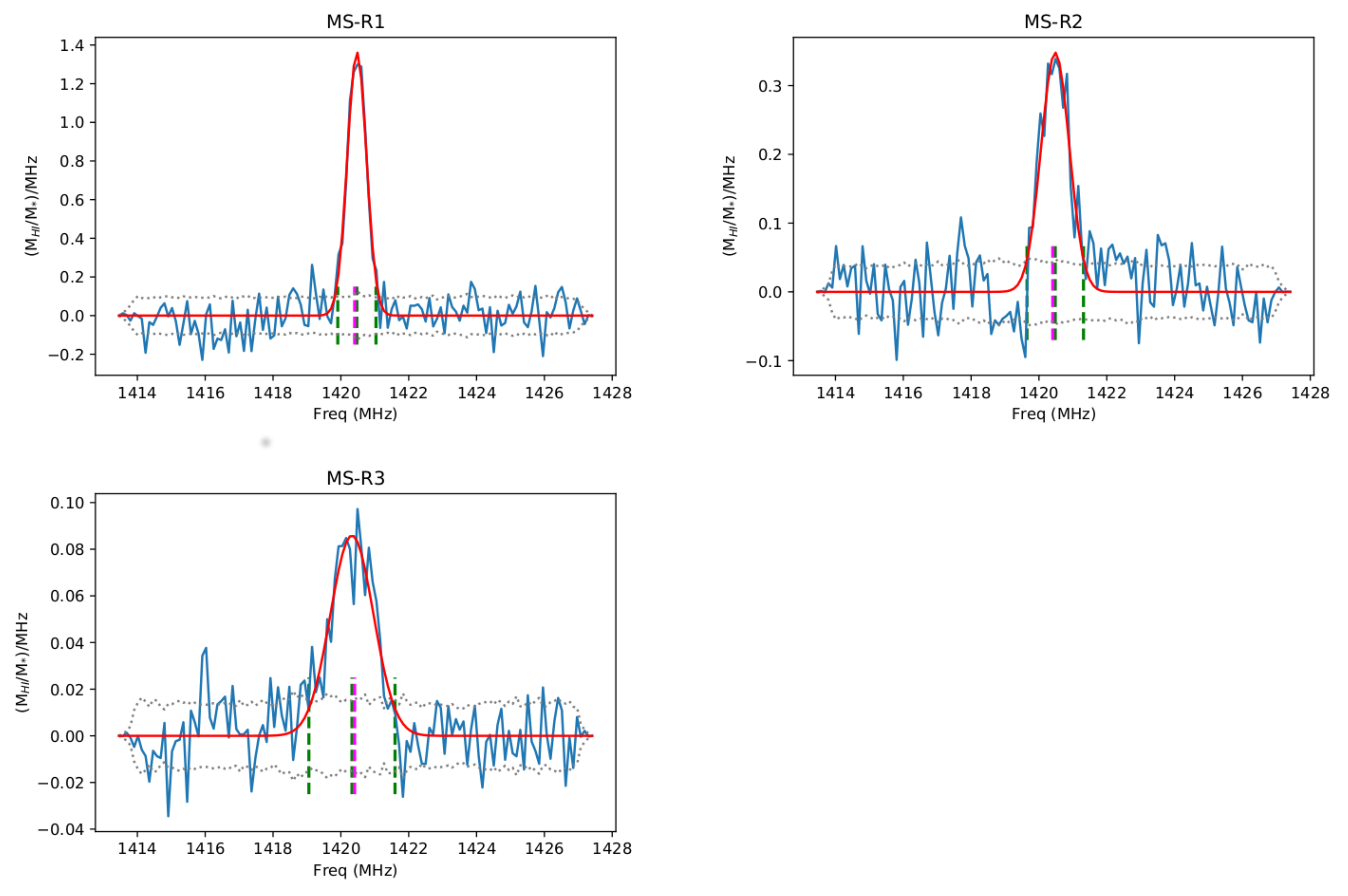}
\figsetgrpnote{Flux-weighted mean stacks of $f_{HI}$ of the 3 SFMS regions discussed in Section~3.5 in blue, for {\bf isolated galaxies}. The Gaussian fit to each spectrum is shown in red. The green vertical dashed lines mark the centre of the Gaussian and $\pm {\rm 2} \sigma$ positions, while the magenta dashed line marks the rest frequency of the {\rm H}{\sc i} 21 cm line. The grey dotted lines mark the standard deviation per channel calculated using 1000 bootstrapped spectra drawn randomly from the original set of spectra (see Section~3.5 for details).}
\figsetgrpend

\figsetgrpstart
\figsetgrpnum{17.8}
\figsetgrptitle{DINGO pairs, 3 SFMS regions}
\figsetplot{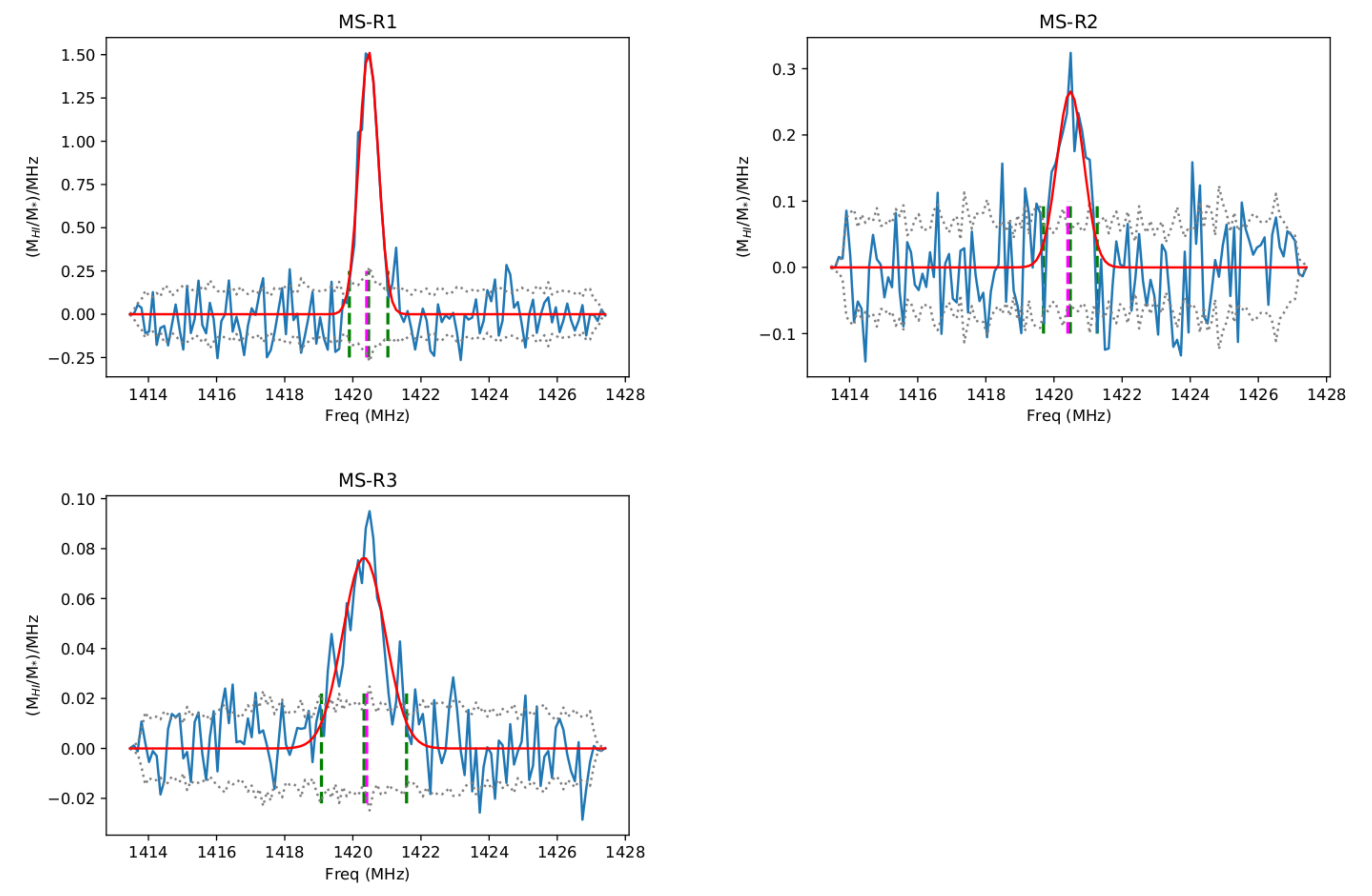}
\figsetgrpnote{Flux-weighted mean stacks of $f_{HI}$ of the 3 SFMS regions discussed in Section~3.5 in blue, for {\bf pairs of galaxies}. The Gaussian fit to each spectrum is shown in red. The green vertical dashed lines mark the centre of the Gaussian and $\pm {\rm 2} \sigma$ positions, while the magenta dashed line marks the rest frequency of the {\rm H}{\sc i} 21 cm line. The grey dotted lines mark the standard deviation per channel calculated using 1000 bootstrapped spectra drawn randomly from the original set of spectra (see Section~3.5 for details).}
\figsetgrpend

\figsetgrpstart
\figsetgrpnum{17.9}
\figsetgrptitle{DINGO groups, 3 SFMS regions}
\figsetplot{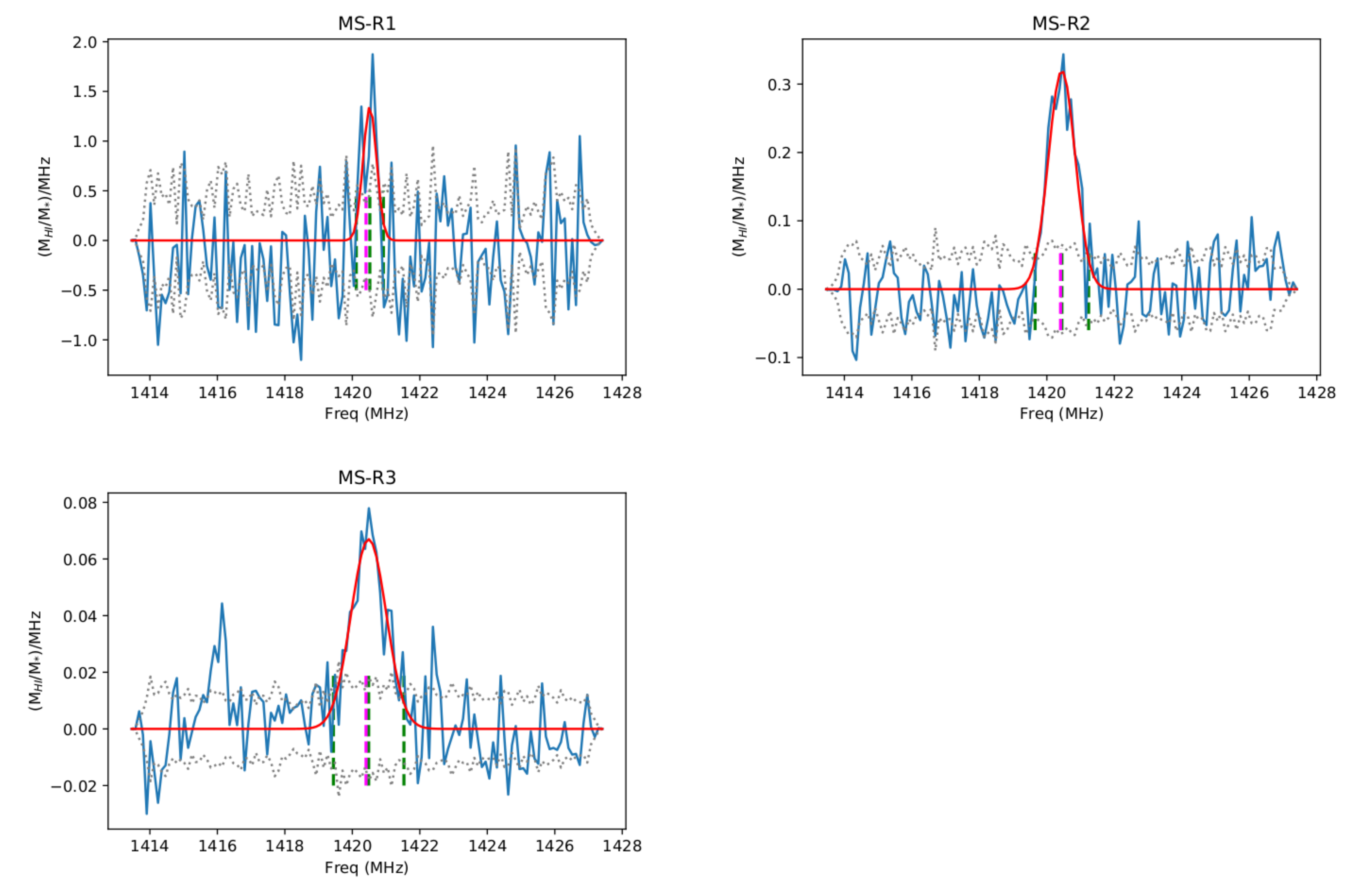}
\figsetgrpnote{Flux-weighted mean stacks of $f_{HI}$ of the 3 SFMS regions discussed in Section~3.5 in blue, for {\bf groups of galaxies}. The Gaussian fit to each spectrum is shown in red. The green vertical dashed lines mark the centre of the Gaussian and $\pm {\rm 2} \sigma$ positions, while the magenta dashed line marks the rest frequency of the {\rm H}{\sc i} 21 cm line. The grey dotted lines mark the standard deviation per channel calculated using 1000 bootstrapped spectra drawn randomly from the original set of spectra (see Section~3.5 for details).}
\figsetgrpend

\figsetgrpstart
\figsetgrpnum{17.10}
\figsetgrptitle{ALFALFA isolated galaxies, 3 SFMS regions}
\figsetplot{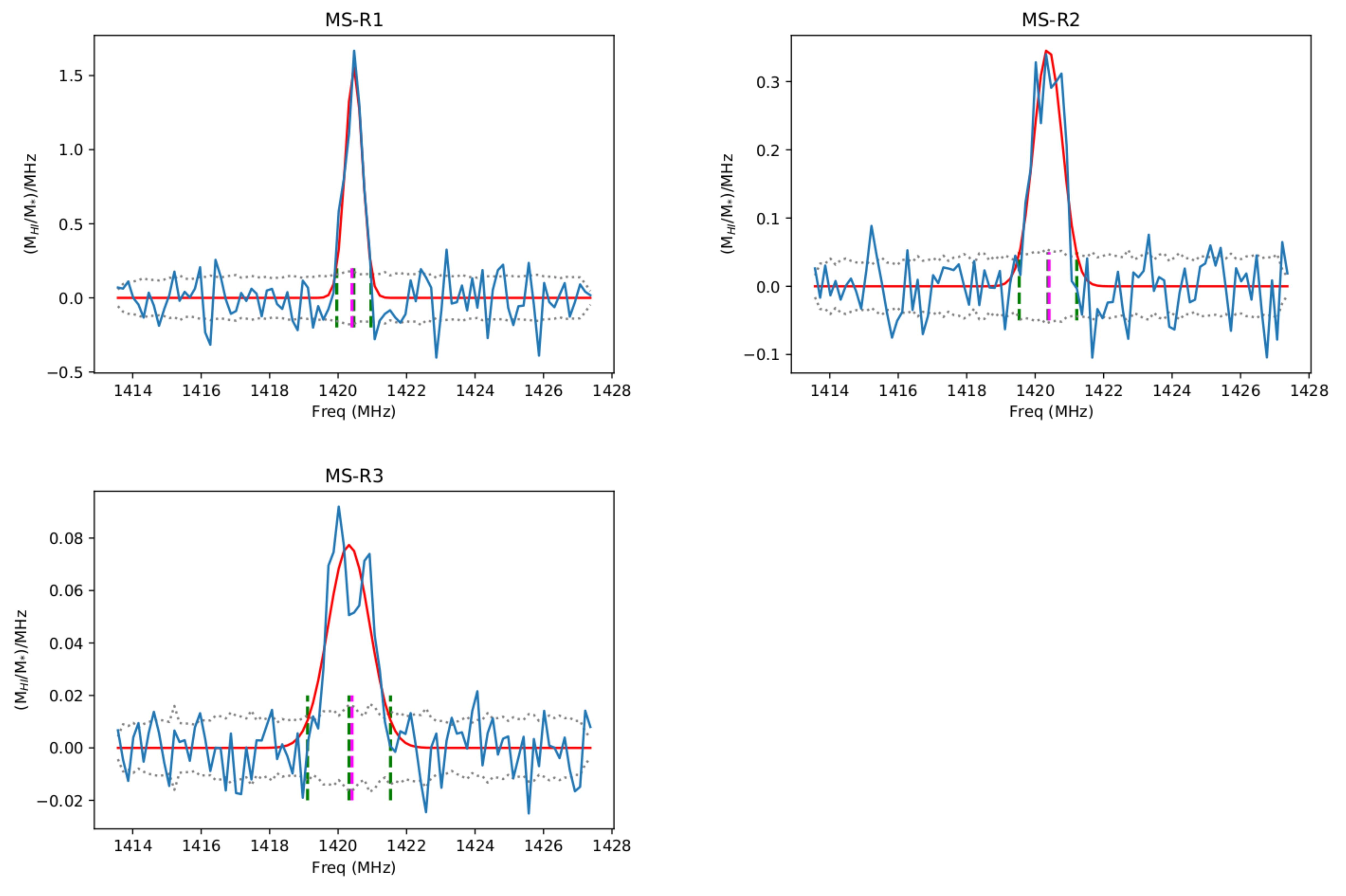}
\figsetgrpnote{Flux-weighted mean stacks of $f_{HI}$ of the 3 SFMS regions discussed in Section~3.5 but for the ALFALFA-GAMA overlap sample in blue, for {\bf isolated galaxies}. The Gaussian fit to each spectrum is shown in red. The green vertical dashed lines mark the centre of the Gaussian and $\pm {\rm 2} \sigma$ positions, while the magenta dashed line marks the rest frequency of the {\rm H}{\sc i} 21 cm line. The grey dotted lines mark the standard deviation per channel calculated using 1000 bootstrapped spectra drawn randomly from the original set of spectra (see Section~3.5 for details).}
\figsetgrpend

\figsetgrpstart
\figsetgrpnum{17.11}
\figsetgrptitle{ALFALFA pairs, 3 SFMS regions}
\figsetplot{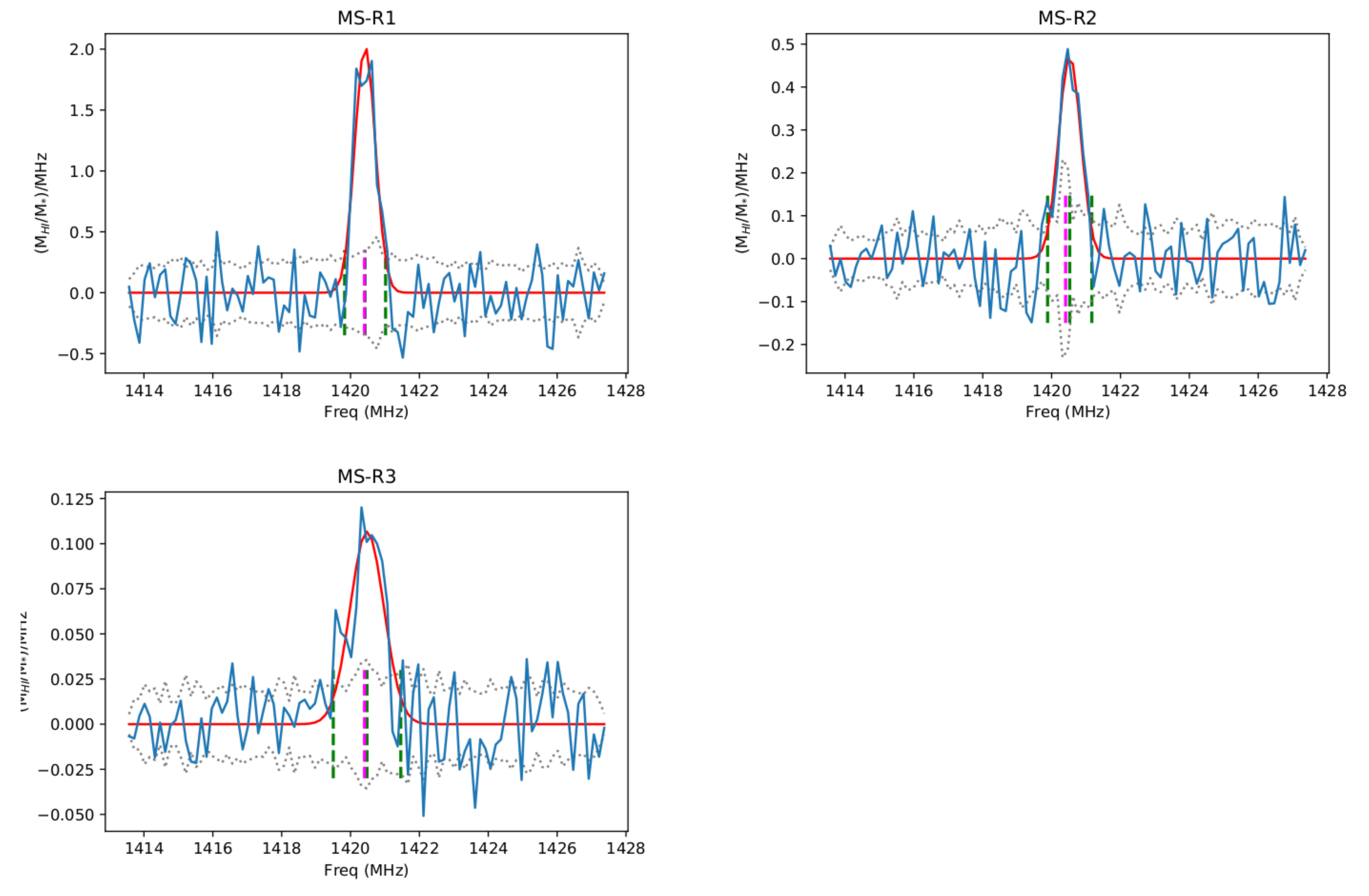}
\figsetgrpnote{Flux-weighted mean stacks of $f_{HI}$ of the 3 SFMS regions discussed in Section~3.5 but for the ALFALFA-GAMA overlap sample in blue, for {\bf pairs of galaxies}. The Gaussian fit to each spectrum is shown in red. The green vertical dashed lines mark the centre of the Gaussian and $\pm {\rm 2} \sigma$ positions, while the magenta dashed line marks the rest frequency of the {\rm H}{\sc i} 21 cm line. The grey dotted lines mark the standard deviation per channel calculated using 1000 bootstrapped spectra drawn randomly from the original set of spectra (see Section~3.5 for details).}
\figsetgrpend

\figsetgrpstart
\figsetgrpnum{17.12}
\figsetgrptitle{ALFALFA groups, 3 SFMS regions}
\figsetplot{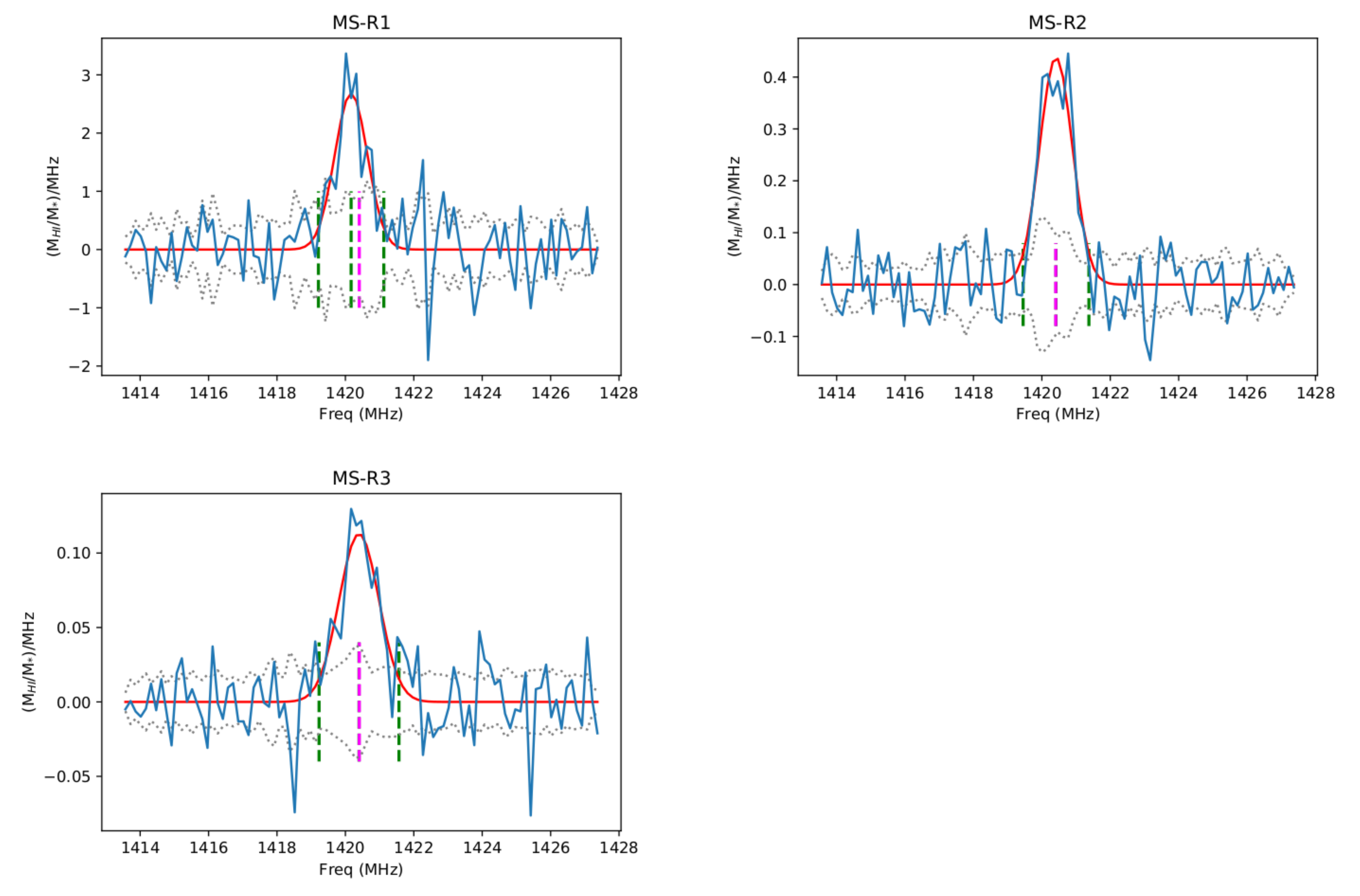}
\figsetgrpnote{Flux-weighted mean stacks of $f_{HI}$ of the 3 SFMS regions discussed in Section~3.5 but for the ALFALFA-GAMA overlap sample in blue, for {\bf groups of galaxies}. The Gaussian fit to each spectrum is shown in red. The green vertical dashed lines mark the centre of the Gaussian and $\pm {\rm 2} \sigma$ positions, while the magenta dashed line marks the rest frequency of the {\rm H}{\sc i} 21 cm line. The grey dotted lines mark the standard deviation per channel calculated using 1000 bootstrapped spectra drawn randomly from the original set of spectra (see Section~3.5 for details).}
\figsetgrpend

\figsetgrpstart
\figsetgrpnum{17.13}
\figsetgrptitle{ALFALFA groups by multiplicity, 2 SFMS regions}
\figsetplot{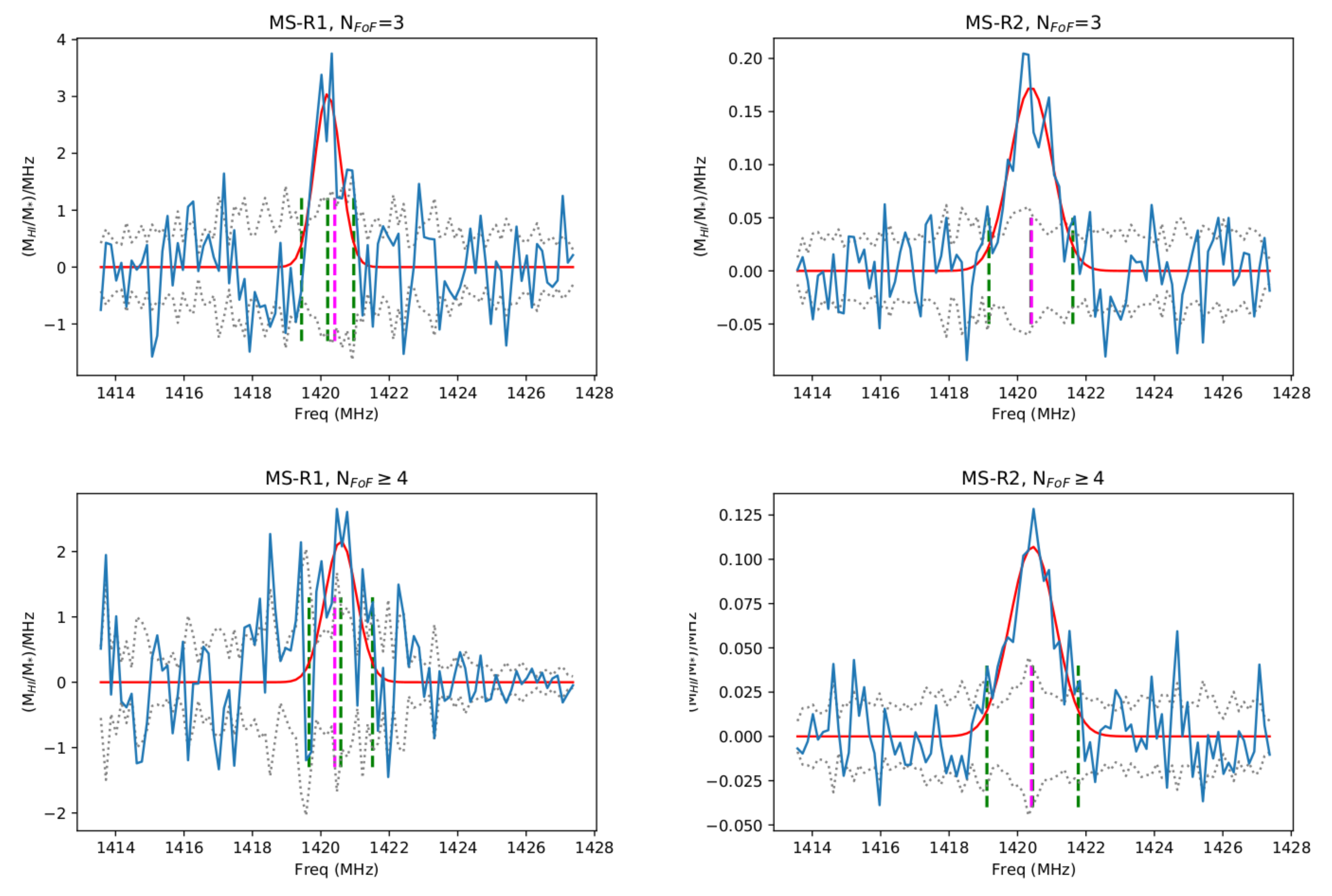}
\figsetgrpnote{Flux-weighted mean stacks of $f_{HI}$ of the 2 SFMS regions discussed in Section~4.1 in blue, for {\bf groups of galaxies} for the two different categories of groups defined by their multiplicity. The Gaussian fit to each spectrum is shown in red. The green vertical dashed lines mark the centre of the Gaussian and $\pm {\rm 2} \sigma$ positions, while the magenta dashed line marks the rest frequency of the {\rm H}{\sc i} 21 cm line. The grey dotted lines mark the standard deviation per channel calculated using 1000 bootstrapped spectra drawn randomly from the original set of spectra (see Section~4.1 for details).}
\figsetgrpend

\figsetend

\begin{figure}[h!]
\figurenum{17.1}
\plotone{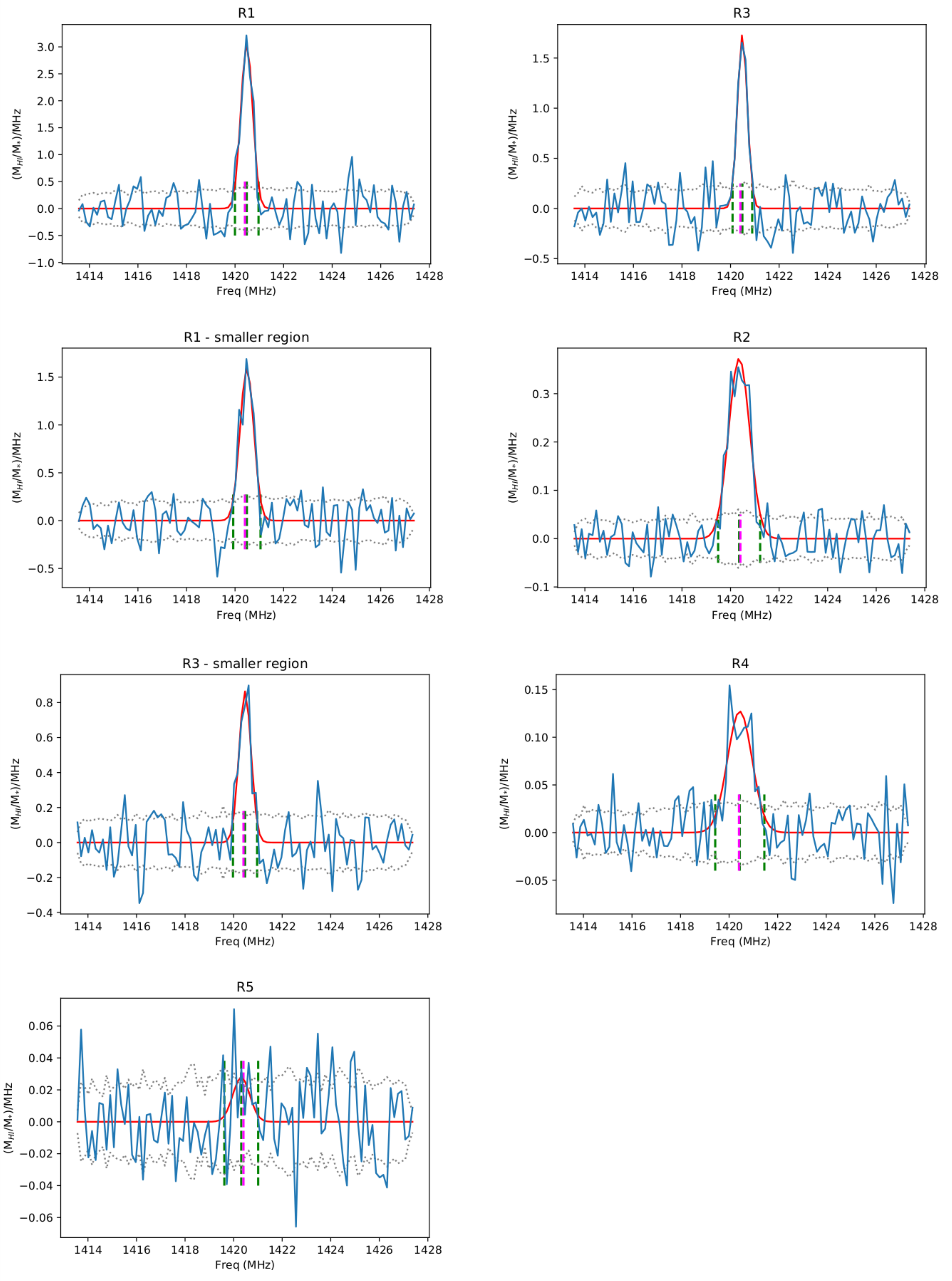}
\caption{Flux-weighted mean stacks of $f_{HI}$ of the 5$+$2 regions for {\bf isolated galaxies} discussed in Sections~\ref{sec:sfrms},\ref{sec:stack1} in blue. The Gaussian fit to each spectrum is shown in red. The green vertical dashed lines mark the centre of the Gaussian and $\pm {\rm 2} \sigma$ positions, while the magenta dashed line marks the rest frequency of the \hi\ 21 cm line. The grey dotted lines mark the standard deviation per channel calculated using 1000 bootstrapped spectra drawn randomly from the original set of spectra (see Section~\ref{sec:stack1} for details). The complete figure set (13 images) is available in the online journal.}
\end{figure}

\section{Results for ALFALFA-GAMA sample with regions same as DINGO G15 sample}
\label{app:3reg}

\begin{table}[h!]
\begin{center}
\caption{Measured values of different parameters for units from the ALFALFA-GAMA overlap sample within the three main-sequence regions from Section~\ref{sec:pilot}.
See Table~\ref{tab:5reg} caption for description of the tabulated quantities.}
\label{tab:apval}
\begin{tabular}{lccccc}
\hline
\hline
Unit			& Region MS-R1 			& Region MS-R2 			& Region MS-R3 \\
				& log(${\rm \frac{\langle M_* \rangle}{M_{\odot}}}$)	& log(${\rm \frac{\langle M_* \rangle}{M_{\odot}}}$)	& log(${\rm \frac{\langle M_* \rangle}{M_{\odot}}}$) \\
\hline
Isolated 			& 8.69$\pm$0.27			& 9.57$\pm$0.17			& 10.20$\pm$0.20 \\
Pairs 		 		& 8.84$\pm$0.27			& 9.58$\pm$0.18			& 10.09$\pm$0.21 \\
Groups 		 		& 8.87$\pm$0.26			& 9.71$\pm$0.17			& 10.17$\pm$0.19 \\
\hline         
\hline                           
Unit 			& Region MS-R1 				& Region MS-R2 				& Region MS-R3 \\
				& log(${\rm \frac{\langle SFR \rangle}{M_{\odot} yr^{-1}}}$)	& log(${\rm \frac{\langle SFR \rangle}{M_{\odot} yr^{-1}}}$)	& log(${\rm \frac{\langle SFR \rangle}{M_{\odot} yr^{-1}}}$) \\
\hline
Isolated 			& $-$0.89$\pm$0.39			& $-$0.33$\pm$0.37			& $-$0.08$\pm$0.40 \\
Pairs 				& $-$0.73$\pm$0.41			& $-$0.30$\pm$0.33			& $-$0.17$\pm$0.42 \\
Groups 				& $-$0.71$\pm$0.34			& $-$0.38$\pm$0.43			& $-$0.34$\pm$0.40 \\
\hline    
\hline 
Unit 			& Region MS-R1 			& Region MS-R2 			& Region MS-R3 \\
				& ${\rm \langle \frac{M_{HI}}{M_{*}} \rangle}$	& ${\rm \langle \frac{M_{HI}}{M_{*}} \rangle}$	& ${\rm \langle \frac{M_{HI}}{M_{*}} \rangle}$ \\
\hline
Isolated 			& 0.98$\pm$0.07			& 0.35$\pm$0.03			& 0.11$\pm$0.01	 \\
Pairs 				& 1.45$\pm$0.17			& 0.36$\pm$0.06			& 0.12$\pm$0.02	 \\
Groups 				& 3.09$\pm$0.70			& 0.51$\pm$0.07			& 0.15$\pm$0.02	 \\    
\hline    
\hline   
Unit 			& Region MS-R1 			& Region MS-R2 			& Region MS-R3 \\
				& $\Delta v_{HI,\pm 2 \sigma}$	& $\Delta v_{HI,\pm 2 \sigma}$	& $\Delta v_{HI,\pm 2 \sigma}$\\
				& ${\rm (km~s^{-1})}$    		& ${\rm (km~s^{-1})}$			& ${\rm (km~s^{-1})}$ \\
\hline
Isolated			& 211				& 355				& 512 \\
Pairs				& 251				& 271				& 414 \\
Groups				& 404				& 406				& 493 \\
\hline
\hline
\end{tabular}
\end{center}
\end{table}

\end{document}